\documentclass[mnsc,nonblindrev]{informs3-nologo}

\newif\ifdebug
\debugtrue

\usepackage[utf8]{inputenc}

\usepackage{amsmath}
\usepackage{amssymb}
\usepackage{amsfonts}
\usepackage{commath}
\usepackage{dsfont,upgreek,marvosym,stmaryrd,bbold,bm,euscript}
\usepackage{graphicx}
\usepackage{verbatim}
\usepackage{url,hyperref}
\usepackage{mathtools}
\usepackage{tikz}
\usepackage{xifthen}
\usepackage{xspace}
\usepackage{float}
\usepackage{wrapfig}
\usepackage{subfig}
\usepackage[inline]{enumitem}
\graphicspath{{fig/}}

\definecolor{myblue}{RGB}{0,0,0}
\definecolor{myred}{RGB}{223,21,32}
\definecolor{myfuch}{HTML}{CA2C92}
\definecolor{mygreen}{RGB}{70,124,13}

\ifdebug

	\newenvironment{changed}[0]{\color{myblue}}{}
\fi

\renewcommand{\alpha}[0]{\upalpha}
\renewcommand{\beta}[0]{\upbeta}
\renewcommand{\lambda}[0]{\uplambda}
\renewcommand{\sigma}[0]{\upsigma}

\newcommand{\const}[0]{\ensuremath{\operatorname{const}}}

\newcommand{\T}[0]{\ensuremath{\mathbb{T}}}
\newcommand{\Nout}[0]{\ensuremath{\EuScript{N}^+}}
\newcommand{\Nin}[0]{\ensuremath{\EuScript{N}^-}}
\newcommand{\dout}[0]{\ensuremath{d^+}}
\newcommand{\din}[0]{\ensuremath{d^-}}
\newcommand{\eout}[0]{\ensuremath{e^+}}
\newcommand{\lout}[0]{\ensuremath{\lambda^+}}
\newcommand{\lbarout}[0]{\ensuremath{\widebar{\lambda}^+}}
\newcommand{\sigout}[0]{\ensuremath{\sigma^+}}
\newcommand{\rout}[0]{\ensuremath{\uprho^+}}
\newcommand{\douthat}[0]{\ensuremath{\widehat{d}^+_{1, i}}}
\newcommand{\Fjmihat}[0]{\ensuremath{\widehat{F}_{2, j}^{-i}}}

\newcommand{\from}[0]{\ensuremath{\leftarrow}}
\DeclareMathOperator{\E}{\mathbb{E}}

\DeclareMathOperator{\Var}{\operatorname{Var}}
\DeclareMathOperator*{\pr}{\mathbb{P}}
\renewcommand{\qed}[0]{\nobreak\hfill\quad\hbox{$\blacksquare$}}  % \Halmos

\input{widebar.tex}

\newcommand{\mysetminusD}{\hbox{\tikz{\draw[line width=0.6pt,line cap=round] (3pt,0) -- (0,6pt);}}}
\newcommand{\mysetminusT}{\mysetminusD}
\newcommand{\mysetminusS}{\hbox{\tikz{\draw[line width=0.45pt,line cap=round] (2pt,0) -- (0,4pt);}}}
\newcommand{\mysetminusSS}{\hbox{\tikz{\draw[line width=0.4pt,line cap=round] (1.5pt,0) -- (0,3pt);}}}
\renewcommand{\setminus}{\mathbin{\mathchoice{\mysetminusD}{\mysetminusT}{\mysetminusS}{\mysetminusSS}}}

\newcommand{\ecbigwidth}[0]{0.24in}
\newcommand{\ecwidth}[0]{0.14in}
\newcommand{\ecswidth}[0]{0.12in}

\newcommand{\ecsempty}[0]{\big\rvert_{\includegraphics[width=\ecswidth]{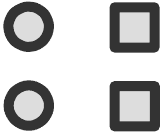}}}
\newcommand{\eccone}[0]{\raisebox{-0.1em}{\includegraphics[width=\ecwidth]{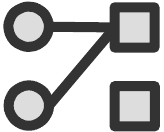}}}
\newcommand{\ecconebig}[0]{\raisebox{-0.1em}{\includegraphics[width=\ecbigwidth]{fig-2x2-chain-equil-cand-2-cone-subs.pdf}}}
\newcommand{\ecscone}[0]{\big\rvert_{\includegraphics[width=\ecswidth]{fig-2x2-chain-equil-cand-2-cone-subs.pdf}}}
\newcommand{\ecsscone}[0]{_{\operatorname{cone}}}
\newcommand{\ecpara}[0]{\raisebox{-0.1em}{\includegraphics[width=\ecwidth]{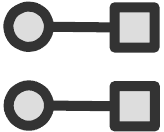}}}
\newcommand{\ecparabig}[0]{\raisebox{-0.1em}{\includegraphics[width=\ecbigwidth]{fig-2x2-chain-equil-cand-3-para-subs.pdf}}}
\newcommand{\ecspara}[0]{\big\rvert_{\includegraphics[width=\ecswidth]{fig-2x2-chain-equil-cand-3-para-subs.pdf}}}
\newcommand{\ecsspara}[0]{_{\operatorname{para}}}
\newcommand{\eczee}[0]{\raisebox{-0.1em}{\includegraphics[width=\ecwidth]{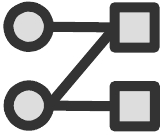}}}
\newcommand{\ecszee}[0]{\big\rvert_{\includegraphics[width=\ecswidth]{fig-2x2-chain-equil-cand-4-zee-subs.pdf}}}
\newcommand{\ecsszee}[0]{_{\operatorname{zee}}}
\newcommand{\ecsszeeone}[0]{_{\operatorname{zee-1}}}
\newcommand{\ecsszeetwo}[0]{_{\operatorname{zee-2}}}
\newcommand{\ecfull}[0]{\raisebox{-0.1em}{\includegraphics[width=\ecwidth]{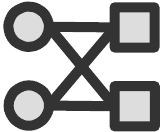}}}
\newcommand{\ecfullbig}[0]{\raisebox{-0.1em}{\includegraphics[width=\ecbigwidth]{fig-2x2-chain-equil-cand-5-full-subs.pdf}}}
\newcommand{\ecsfull}[0]{\big\rvert_{\includegraphics[width=\ecswidth]{fig-2x2-chain-equil-cand-5-full-subs.pdf}}}
\newcommand{\ecssfull}[0]{_{\operatorname{full}}}

\newcommand{\vsa}[0]{\vspace{-0.1in}}

\ifdebug
	\SingleSpacedXII
\else
	\OneAndAHalfSpacedXI
\fi

\usepackage{natbib}
\bibpunct[, ]{(}{)}{,}{a}{}{,}
\def\bibfont{\small}

\TheoremsNumberedThrough
\ECRepeatTheorems
\EquationsNumberedThrough

\MANUSCRIPTNO{}

\begin{document}

% Outcomment only when entries are known. Otherwise leave as is and default values will be used.
%\setcounter{page}{1}
%\DOI{10.1287/xxxx.0000.0000}%

\RUNAUTHOR{Amelkin and Vohra}

\RUNTITLE{Strategic Formation and Reliability of Supply Chain Networks}

\TITLE{
    Strategic Formation and Reliability\\of Supply Chain Networks\\[0.1in]
    {\small \emph{(Working Paper, \today)}}
}

\ARTICLEAUTHORS{%
\AUTHOR{Victor Amelkin}
\AFF{Warren Center for Network and Data Sciences, Department of Electrical and Systems Engineering\\University of Pennsylvania, Philadelphia, PA 19104, \EMAIL{vctr@seas.upenn.edu}, \URL{https://victoramelkin.com}}
\AUTHOR{Rakesh Vohra}
\AFF{Warren Center for Network and Data Sciences, Department of Electrical and Systems Engineering, Department of Economics\\University of Pennsylvania, Philadelphia, PA 19104, \EMAIL{rvohra@seas.upenn.edu}, \URL{https://sites.google.com/site/quaerereverum9/}}
}

\ABSTRACT{%
Supply chains are the backbone
of the global economy. Disruptions to them can be costly. Centrally managed supply chains invest in ensuring their resilience. Decentralized supply chains, however, must rely upon the self-interest of their individual components to maintain the resilience of the entire chain.

We examine the incentives that independent self-interested agents have in forming a  resilient supply chain network in the face of production disruptions and competition. In our model, competing suppliers are subject to yield uncertainty (they deliver less than ordered) and congestion (lead time uncertainty or, ``soft'' supply caps). Competing retailers must decide which suppliers to link to based on both price and reliability. 
In the presence of yield uncertainty only, the resulting supply chain networks are sparse. Retailers concentrate their links on a single supplier, counter to the idea that they should mitigate yield uncertainty by diversifying their supply base. This happens because retailers benefit from supply variance. It suggests that competition will amplify output uncertainty.
When congestion is included as well, the resulting networks are denser and resemble the 
bipartite expander graphs that have been proposed in the supply chain literature,
\begin{changed}
thereby, providing the first example of endogenous formation of resilient supply chain networks,
without resilience being explicitly encoded in payoffs.
\end{changed}
Finally, we show that a supplier's investments in improved yield can make it worse off. This happens because high production output saturates the market, which, in turn  lowers prices and profits for participants.
}


\maketitle
{\small
(The most recent version of this paper is available at {\normalfont\footnotesize\url{https://victoramelkin.com/pub/supply-chains/}}.)
}

\clearpage
\tableofcontents
\clearpage

\section{Introduction}

 Supply chains are the backbone
of the global economy. Disruptions to them can be costly. They happen because the individual components of  the
 chain are subject to   yield uncertainty (a supplier comes up short on the ordered
product quantity) as well as lead time uncertainty (clients of overly congested suppliers experience
delivery delays)~\citep{snyder2016or}. The degree of
uncertainty can be large. \citet{bohn1999economics}, for example, suggest that disk drive manufacturer Seagate experiences production yields as low as 50\%. Centrally managed and controlled supply chains invest a great deal in mitigating these disruptions. Decentralized supply chains, however, must rely upon the self-interest of their individual components to maintain the resilience of the entire chain. 

Cursory reflection suggests that the incentives of an individual component in the chain should align with the chain as a whole. A supplier, for example, will be rewarded with greater business if it invests in reducing the possibility of it being disrupted relative to its competitors. However, a supplier's customers can also hedge against disruption by multi-sourcing~\citep{cachon2008matching,chopra2016supply,tomlin2006value}.
Thus, a potential customer may prefer to source from many \begin{changed}low-cost\end{changed} unreliable suppliers rather than a few highly reliable but costlier suppliers. Cursory reflection also ignores the impact on output prices that would result from reducing the frequency of disruptions. If prices adjust to clear markets, increased throughput may result in lower prices. Thus, one must compare the profits earned from high volumes with low margins with those generated from lower volumes but with higher margins. It is not obvious which will dominate. 

In this paper we examine the strategic formation of a two-tier\footnote{Our model straightforwardly generalizes to the
case of an arbitrary number of tiers.} supply chain network  by \emph{independent} self-interested
agents. Retailers occupy the first tier  and suppliers the second tier. The price at which trade takes place between tiers is set to clear the market.  Retailers decide which suppliers to source from. There is also a cost
for linking to a supplier.\footnote{According to~\citet{cormican2007supplier}, it takes, on average, six months to a year to qualify a new supplier.}

Every agent present in the supply chain is subject to yield uncertainty which affects their capacity.\footnote{To quote Yossi Sheffi: ``The essence of most disruptions is a reduction in capacity and, therefore, inability to meet demand.''} It is modeled as a Bernoulli random variable. Yield uncertainty of this kind can arise from the nature of the production process (e.g., farming); it can also arise from disruptions like a natural disaster or a union strike.
The resulting random output of each tier is distributed among agents in the downstream tier in proportion to their
demands, following the proportional rationing rule~\citep{rong2017bullwhip,cachon1999equilibrium}.  Every supplier in the chain is also subject to congestion, and the resulting congestion costs are borne by the retailers. Congestion in our model has at least two interpretations. One is a delay cost associated with lead time uncertainty.  The second is a ``soft'' supply constraint.

We are interested in whether a \emph{decentralized} supply chain network resilient to disruptions will form \emph{endogenously} in the presence of competition and different types of uncertainty.

The \emph{three major findings} of our analyses are as follows:

$\vartriangleright$
With only yield uncertainty and no congestion,
\emph{retailers create a sparse network, with a single link per retailer, and concentrate
links on a single supplier}. This generalizes  to the case of more than two tiers where the corresponding 
supply chain network is  almost a chain. Link concentration runs counter to
 the
common wisdom about the benefits of multi-sourcing.
 \emph{Link concentration helps retailers secure low upstream
prices} in the presence of high upstream yield, and low expenditures in case of the target upstream
supplier's failure. Therefore, retailers benefit from supply variance. It suggests that
competition  can amplify output uncertainty.
% presented with two quantity-controlled markets, the buyer may have an incentive to buy from the more uncertain market.
The network formed in our model is dramatically different from the ones that are \emph{assumed} in the existing literature. \citet{bimpikis2019supply}, for example, assume
that, in the presence of yield uncertainty only, a $k$-tier supply chain
network will take the form of a complete $k$-partite graph.  
		
$\vartriangleright$
In the presence of  yield uncertainty \emph{and} congestion, the network formed is \emph{sparse, yet well-connected} resembling an expander
graph. Similar objects have been shown to have good resilience properties in the context of centrally organized supply chains, see for example, \citet{chou2011process}.
Congestion, unlike yield uncertainty, encourages retailers to
split their demand across several suppliers to lower congestion costs.
\begin{changed}
Thus, our work provides the first example of the endogenous formation of resilient supply
chain networks, without an explicit concern for resilience being encoded in the payoffs.
\end{changed}

$\vartriangleright$
\emph{Yield uncertainty and congestion have fundamentally different implications for supply chains.}
In the presence of yield uncertainty only, each supplier has a unilateral incentive to increase its average yield. 
With both yield uncertainty \emph{and} congestion, a unilateral reduction in
congestion costs unconditionally benefits that supplier, but increasing mean yield could make a supplier worse off! This is because high yield results in market saturation, which leads to low prices
and profits for market members.

The rest of the paper is organized as follows. The next section discusses prior work. The subsequent section 
introduces notation. Sec.~\ref{sec:supchain-formation} describes the model of strategic formation of supply
chains with costly links, competition, and 
yield uncertainty only. Sec.~\ref{sec:supchain-formation-w-congestion} augments the previous model with
congestion, and provides its comprehensive characterization. Sec.~\ref{sec:general-model-w-congest-2x2}
focuses on the case of a small two-tier supply chain. Sec.~\ref{sec:general-model-w-congest-analysis}
provides a limited set of results for the general two-tier case. Finally, in
Sec.~\ref{sec:hetero-suppliers}, we describe the quality-investment behavior of
competing heterogeneous suppliers and the qualitative differences between
yield uncertainty and congestion / lead time uncertainty.

\section{Prior Work}
\label{sec:prior-work}
Our model has three features:
\begin{enumerate}
	\item strategic network formation,
	\item disruptions, and
	\item competition.
\end{enumerate}
In the extensive literature on supply chain networks, one will find models that possess some,
but not all three features, with the exception of one recent model of~\citet{amelkin2019yield}.
Table~\ref{tbl:prior-work-summary} categorizes a sample of recent related works. A detailed
comparison with prior work on network formation in supply chains follows.

\newcommand{\yes}[0]{\small\ensuremath{\bm{+}}}
\newcommand{\no}[0]{\small\ensuremath{\bm{-}}}
\newcommand{\yesno}[0]{\small\ensuremath{\bm{\pm}}}
\begin{table}
	\centering
    \footnotesize
	\begin{tabular}[ht]{|l|>{\centering\arraybackslash}p{1.3in}|>{\centering\arraybackslash}p{1.3in}|>{\centering\arraybackslash}p{1.3in}|}
		\hline
		\phantom{xxxxxxx}\textbf{Paper~{\textbackslash}~Feature} & \textbf{Network~Formation} & \textbf{Disruptions} & \textbf{Competition} \\ \hline
		Present work & \yes & \yes & \yes \\ \hline \hline
		\citet{amelkin2019yield} & \yes & \yes & \yes \\ \hline
		\citet{bimpikis2018multisourcing} & \yes & \yes & \no \\ \hline
		\citet{ang2016disruption} & \yes & \yesno & \no \\ \hline
		\citet{tang2011supplier} & \yesno & \yes & \yesno \\ \hline
		\hline
		\citet{chod2019supplier} & \no & \yesno & \yes \\ \hline
		\citet{bimpikis2019supply} & \no & \yes & \yes \\ \hline
		\citet{fang2015managing} & \no & \yes & \yes \\ \hline
		\citet{adida2011supply} & \no & \yes & \yes \\ \hline
		\citet{babich2007competition} & \no & \yes & \yes \\ \hline
		\citet{carr2005competition} & \no & \no & \yes \\ \hline
		\citet{bernstein2005decentralized} & \no & \yes & \yes \\ \hline
		\cite{anupindi1993diversification} & \no & \yes & \no \\ \hline
		\hline
		\citet{kotowski2018trading}* & \yes & \yesno & \yes \\ \hline
		\citet{ambrus2014consumption}* & \no & \yes & \no \\ \hline
		\citet{deo2009cournot}* & \no & \yes & \yesno \\ \hline
		\citet{kranton2001theory}* & \yes & \yesno & \yes \\ \hline
	\end{tabular}
	\caption{
	Prior work summary.
	Partial presence (\yesno) of disruption means that they are present only in a part of
	the system, e.g., in the upstream tier of suppliers~\citep{ang2016disruption} or the tier of buyers / retailers~\citep{chod2019supplier}, while in our model every agent can
	be disrupted. Also, none of the mentioned works considers anything analogous to our
	congestion or lead time uncertainty. Absence (\no) of competition implies fixed prices.	
	Works marked with an asterisk are not about supply chains.
	}
	\label{tbl:prior-work-summary}
\end{table}

There are many papers that study supply chains in the presence of competition. Examples are~\citet{carr2005competition} and~\citet{fang2015managing} which use Cournot competition, while~\citet{chod2019supplier} uses Bertrand competition. Some also incorporate disruptions, such as~\citet{deo2009cournot}, and~\citet{babich2007competition}. However, none of considers
endogenous network formation.

A significantly smaller set of papers compare supply chain performance across different network structures, but do not consider endogenous formation. Within this stream two papers are closely related to ours. The first is~\citet{bimpikis2019supply}. We share the same price
formation process in every tier of the supply chain and the same production
model with Bernoulli yield. Our supply chain network, however, is endogenously formed while
theirs is exogenously fixed to be a
complete $k$-partite graph. Our results cannot be deduced from the model of~\citet{bimpikis2019supply}.  
In particular, our analysis in Sec.~\ref{sec:supchain-formation}
demonstrates that the complete $k$-partite networks assumed in~\citet{bimpikis2019supply} need not arise
endogenously in their setting.

The second work is~\citet{tang2011supplier},
with two competing reliable retailers sourcing  from two non-competing suppliers subject to yield uncertainty that is correlated. Because supplier prices are fixed exogenously, the focus is on how order quantities change with the sourcing decisions of retailers. This paper compares outcomes across possible  networks, but \begin{changed}does not\end{changed} consider all possible networks. In particular,  the case when retailers
single-source from the \emph{same} supplier is excluded. In one of our models, this configuration arises in equilibrium.

\citet{tang2011supplier} also study the
effect of an exogenously given correlation between supplier yields. In our model, supplier yields are independent of each other. However, when different
retailers in our model have overlap in their supplier bases, it results in an implicit correlation of
their production outputs. This implicit correlation influences network formation
in our model.

Papers that consider endogenous supply chain network formation can be numbered on a single hand.
\citet{amelkin2019yield} consider an endogenous supply chain network formation model, with
yield uncertainty and competition. In their model, suppliers, having uncertain i.i.d.
supplies, strategically announce wholesale prices to retailers, and the retailers, then,
compete to sell the product to consumers. There are two major 
differences between their model and ours. 
\begin{changed}
 In \cite{amelkin2019yield}, wholesale prices are set in advance while in the present paper they are set to clear the market as in \cite{bimpikis2019supply}. This seems particularly relevant to supply chains where prices are determined by a spot market, for example, the memory chip industry\footnote{\url{https://www.dramexchange.com}}~\citep{bimpikis2019supply} and
            the US corn, soybeans, wheat, and tobacco industries~\citep{mendelson2007strategic}. Even in markets where prices are set by contract, they may be pegged to an index, such as the average spot price on an agreed upon date. The equilibrium outcomes under a spot price are radically different. The second difference is that, in~\citet{amelkin2019yield},
retailers are quantity-takers, that is, they have unconstrained demands, while in our model,
there is a specific demand originating at the consumer tier, which, then, ``propagates''
through the supply chain network based on how the latter is structured. These two differences
produce different network outcomes in equilibrium.
            \end{changed}

Another work capturing endogenous network formation is~\citet{ang2016disruption}, who consider a supply chain comprised
of a manufacturer issuing orders to two suppliers, who, in turn, are linked to two
higher-level suppliers. The manufacturer issues contracts
incorporating quantities and prices, thereby, affecting sourcing decisions of intermediary suppliers. 
In this model only top-tier suppliers fail, while every agent can fail in our model. 
Further, unlike our paper, there is no competition: top-tier sourcing costs---different
for reliable (higher) and unreliable (lower) suppliers---are exogenously fixed; so is the
price at which the manufacturer sells a unit of product to consumers.

\cite{ang2016disruption} are interested in how the intermediate suppliers decide between
single- vs. multi-sourcing decisions. In their paper, the manufacturer, primed by a deterministic
exogenous consumer demand issues price-quantity contracts to the intermediary suppliers, and each of the intermediaries
decides upon how much to order from each of the top-tier suppliers at the fixed prices. Thus, sourcing
decisions of intermediate suppliers  are the result of strategic price-setting by the
manufacturer. 
In our model, prices are
formed via competition in every tier of the supply chain. Thus, the formed networks we observe
are a result of competition in all tiers.

\citet{ang2016disruption} characterize
optimal sourcing strategies (optimal order quantities) in a network where intermediaries
source from different top-tier suppliers (V-shaped network), and in the network where they source from the
same top-tier supplier (diamond-shaped network). They also propose and analyze a
game where intermediate suppliers strategically decide upon sourcing
from the top tier, and characterize its pure equilibria, which happen to be the diamond-
and V-shaped networks. V-shaped networks are equilibria in our network formation game as well, but we find other equilibria as well (see
Sec.~\ref{sec:general-model-w-congest-2x2}). It is important to emphasize that V-shaped
equilibria are absent in our model when yield uncertainty is the \emph{only}
disruption present (see Sec.~\ref{sec:supchain-formation}). Hence, if competition was incorporated into~\citet{ang2016disruption}, it would eliminate the V-shaped equilibria.

\begin{changed}
    The presence of competition in our work (and its absence in~\citet{ang2016disruption})
    is crucial, as competition---additionally augmented with congestion in our model---largely drives
    strategic link formation in the supply chain network.
\end{changed}

Finally, this  paper is related to the literature on strategic network formation, including
the work on buyer-seller networks of~\citet{kranton2001theory}, trading networks with
intermediaries of~\citet{kotowski2018trading}, and risk-sharing networks
of~\citet{ambrus2014consumption}. However, the multipartite structure of our networks,
the mechanics of our models, and the conclusions we arrive at are both different and distant from the ones in these works.

\section{Preliminaries and Notation}
\label{sec:prelims}

In this section, we introduce notation\begin{changed}---summarized in Table~\ref{tbl:notation-summary}---\end{changed}and  several useful definitions.

\textbf{Supply Chain:} A supply chain is a multi-tier network comprised of $T$ tiers of agents---also known as firms
or suppliers---where tier $t$ is denoted with $\T_t = \{1, 2, \dots\}$. Most of our modeling
efforts will target 2-tier supply chains, in which $T = 2$, $|\T_1| = n > 1$, $|\T_2| = m > 1$.
The agents in tier $\T_1$ are referred to as \emph{retailers}, who sell product
to consumers;
higher-tier agents are \emph{suppliers}. When using notation independent of the total number of tiers,
we may also refer to any agent in any tier of the supply chain as a supplier. Implicitly present is tier
$\T_0$ of consumers, and another tier $\T_{T + 1}$ corresponding to the raw material
market\footnote{The consumer and the raw material producer tiers will actually
be represented by a single meta-consumer and meta-raw material producer, respectively.}.
The product will flow from higher numbered tiers (\emph{upstream}) to lower numbered tiers
(\emph{downstream}). Throughout this paper, we assume that downstream agents
strategically link to upstream agents in the adjacent tier. All retailers
are linked to consumers, and all top-level suppliers are  linked
to raw materials producers.

\textbf{Demands:}
Each retailer will experience a fixed consumer demand of $D > 0$.
By considering equal consumer demand distribution over the retailers, we make sure that
the retailers differ only in what suppliers they link to. $\Delta = n D$ is the
total consumer demand. More generally, we denote by $D_{t, i} \in \mathbb{R}_+$ the demand
experienced by agent $i \in \T_t$, indicating the amount of product collectively requested
from supplier $i$ by downstream agents from tier $\T_{t - 1}$.

\textbf{Prices:} Each tier $\T_1, \dots, \T_{T+1}$ of the supply chain consists of agents competing to supply agents in the adjacent downstream tier. Within a tier, ``supply'' is the \emph{realized} total quantity present in the tier. The realized quantity
may be lower than the demanded quantity due to upstream production failures.
Denote the supply of $i \in \T_t$ by $S_{t, i}$, and the total supply of tier $\T_t$ by
\begin{align}
    S_t = \sum_{i \in \T_t}{S_{t, i}}.
    \label{eq:tier-supply}
\end{align}

The market price $p_t$ per unit of output of tier $t$ is set so as to ``clear'' the market, i.e.,
\begin{align}
	p_t = \Delta - S_t.
	\label{eq:market-price}
\end{align}

\textbf{Production:} Each supplier, having
received some product quantity, supplies the same amount of product downstream---provided there are downstream agents linked to this supplier---with probability $\lambda \in (0, 1)$,
and fails to produce any output with the complementary probability $(1 - \lambda)$. We exclude $\lambda \in \{0,1\}$ to avoid trivialities. If  $\lambda = 0$,
the agents clearly cannot make a profit. Setting $\lambda = 1$ entails the same degenerate outcome, the reasons for which are
given in Theorem~\ref{thm:empty-equil-unique-when-no-failures}.
Raw material producers never fail.  

We use $R_{t, i} \in \mathbb{R}_+$ to denote the \emph{realized demand} of supplier
$i \in \T_t$, that is, how much supplier $i$ receives from upstream suppliers in 
$\T_{t + 1}$ in response to $i$'s demand of $D_{t, i}$. The \emph{production success indicator}
$\omega_{t, i} \sim \operatorname{Bernoulli}(\lambda)$ is a random variable that
indicates whether supplier $i \in \T_t$ has succeeded in producing output. 

\textbf{Network:}
All suppliers together with their links comprise the \emph{network} underlying the supply
chain. Let $\Nin_{t, i} \subseteq \T_{t - 1}$ and $\Nout_{t, i} \subseteq \T_{t + 1}$
denote in- and out-neighborhoods of supplier $i \in \T_t$, that is, the sets of suppliers that
source product from $i$ or that $i$ sources product from, respectively. Thus, the network
formed by suppliers of tier $\T_t$ is $g_t = (\Nout_{t, 1}, \dots, \Nout_{t, n_t})$, with
$g = g_1$ being used in the analysis of two-tier chains. We also define
$g_t^{-i} = (\Nout_{t, 1}, \dots, \Nout_{t, i-1}, \Nout_{t, i+1}, \dots \Nout_{t, n_t})$, and
$g^{-i} = g_1^{-i}$ for two-tier chains. In- and out-degrees of
$i$ are $\din_{t, i} = |\Nin_{t, i}|$ and $\dout_{t, i} = |\Nout_{t, i}|$, respectively.
$\dout_{i \cap i'} = |\Nout_{t, i} \cap \Nout_{t, i'}|$ stands for the size of out-neighborhood
overlap of suppliers $i, i' \in \T_t$. Additionally, we introduce the effective out-degree
$\eout_{t, i} = \sum_{j \in \Nout_{t, i}}{\omega_{t + 1, j}}$ of supplier $i \in \T_t$, that
measures the number of its out-neighbors who successfully produced output. We will say
that a supplier is \emph{active} if its in- and out-degrees are both positive; other suppliers
are inactive (and they cannot possibly earn profits due to their inability to either buy or sell).
By $\T_t^a \subseteq \T_t$ we will denote the \emph{subset of active suppliers} of tier $\T_t$,
and the \emph{number of active suppliers} is $n_t^a = |\T_t^a| \begin{changed}\leq n_t\end{changed}$.

Finally, we introduce the following expressions useful in the analysis of our models:
\begin{align}
	\rout_{t, i} &= \sum_{i' \in \T_t^a \setminus \{ i \}}{
		\frac{
			\dout_{i \cap i'}
		}{
			\dout_{t, i'}
		}
	}
	=
	\sum_{i' \in \T_t^a \setminus \{ i \}}{
		\frac{
			|\Nout_{t, i'} \cap \Nout_{t, i}|
		}{
			\dout_{t, i'}
		}
	}, \label{eq:rho} \\
	F_{t, j} &= \sum\limits_{ i \in \Nin_{t, j} \setminus \{ i \}}{
		\tfrac{1}{\dout_{t - 1, i}}
	}, \label{eq:eff}\\
	F^{-i}_{t, j} &= \sum\limits_{ i' \in \Nin_{t, j} \setminus \{ i \} }{
		\tfrac{1}{\dout_{t - 1,i'}}
	}. \label{eq:eff-mi}
\end{align}
Here, $\rout_{t, i}$ measures the aggregate relative extent to which out-neighborhoods of
active suppliers in tier $\T_t$ overlap with the out-neighborhood of supplier $i$, or, less
formally, how well supplier $i$ is ``embedded'' in its tier.  Thus, we will refer to $\rout$ as
the \emph{degree of overlap of $i$ with its peer suppliers}. As suppliers
distribute their demand uniformly over out-neighborhoods, $F_{t, j}$ quantifies
(scaled) \emph{congestion at supplier $j \in \T_t$} (where ``congestion'' is understood with
respect to the demand coming from downstream agents), and $F_{t, j}^{-i}$ measures the
same quantity excluding the impact of supplier $i$. Seemingly different, $\rout_{t, i}$ and
$F_{t + 1, j}^{-i}$ are actually closely related, as shown in the next Lemma.
\begin{lemma}[About $\rout_{t, i}$ and $F_{t + 1, j}^{-i}$]
	\begin{align*}
		\rout_{t, i} = \sum\limits_{j \in \Nout_{t, i}}{ F_{t + 1, j}^{-i}}.
	\end{align*}
\label{thm:rho-f-lemma}
\end{lemma}
\proof{Proof of Lemma~\ref{thm:rho-f-lemma}:}
	\begin{align*}
	\rout_{t, i} &= \sum_{i' \in \T_t^a \setminus \{ i \}}{
			\frac{
				\dout_{i \cap i'}
			}{
				\dout_{t, i'}
			}
		}
		=
		\sum_{i' \in \T_t^a \setminus \{ i \}}{
			\frac{
				|\Nout_{t, i'} \cap \Nout_{t, i}|
			}{
				\dout_{t, i'}
			}
		}
		=
		\sum_{i' \in \T_t^a \setminus \{ i \}}{
		\sum_{j \in \Nout_{t, i'} \cap \Nout_{t, i}}{
			\frac{
				1
			}{
				\dout_{t, i'}
			}
		}}
		=
		\sum_{i \to j \from i' \neq i}{
			\frac{
				1
			}{
				\dout_{t, i'}
			}
		}\\
		&=
		\sum_{j \in \Nout_{t, i}}{
		\sum_{i' \in \Nin_{t + 1, j} \setminus \{i\}}{
			\frac{
				1
			}{
				\dout_{t, i'}
			}
		}}
		= \sum_{j \in \Nout_{t, i}}{ F_{t + 1, j}^{-i} }.
	\end{align*}
	\qed
\endproof

When dealing with two-tier supply chains, we call the underlying network
\emph{left-regular} if all retailers have identical out-degrees, and \emph{right-regular} if all
suppliers have identical in-degrees.

We restrict attention to pure strategy Nash equilibrium.

\begin{table}[ht]
    \small
    \centering
    \begin{tabular}{|c|l|}
        \hline
        	$\T_t$ &  $\{ 1, \dots, n_t \}$ -- tier $t \in \{1, \dots, T\}$ of the supply chain; $|\T_1| = n$, $|\T_T| = m$, $|\T_t = n_t|$\\ \hline
        	$D_{t, i}$ & demand exerted upon supplier $i \in \T_t$ by downstream suppliers from $\T_{t-1}$\\ \hline
		    $D$ & $D_{1,i} = \const$ -- consumer demand per retailer\\ \hline
        	$R_{t, i}$ & realized demand of supplier $i \in \T_t$; $R_{t, i} \leq D_{t, i}$\\ \hline
        	$\Delta$ & $n D$ -- total consumer demand\\ \hline
        	$S_{t, i}$ & supply delivered by supplier $i \in \T_t$ to downstream suppliers sourcing from $i$\\ \hline
        	$S_t$ & $\sum_{i \in \T_t}{S_{t, i}}$ -- total supply of tier $\T_t$\\ \hline        	
        	$\omega_{t, i}$ & $\omega_{t, i} \sim \operatorname{Bernoulli}(\lambda)$ -- production success indicator of supplier $i \in \T_t$\\ \hline
        	$\lambda$ & $\pr\{\omega_{t, i} = 1\} \in (0, 1)$ -- production success likelihood\\ \hline
        	$\Nout_{t, i}$ & $\Nout_{t, i} \subseteq \T_{t+1}$ -- out-neighborhood of supplier $i \in \T_t$; ($\Nin_{t, i} \subseteq \T_{t - 1}$ -- in-neighborhood)\\ \hline
        	$g_t$ & $(\Nout_{t, 1}, \dots, \Nout_{t, n_t})$ -- network between tiers $\T_t$ and $\T_{t + 1}$; when $T = 2$, $g = g_1$\\ \hline
        	$\dout_{t, i}$ & $|\Nout_{t, i}|$ -- out-degree of supplier $i \in \T_t$; ($\din_{t, i}$ -- in-degree)\\ \hline
        	$\dout_{t, i \cap i'}$ & $|\Nout_{t, i} \cap \Nout_{t, i'}|$ -- number of out-neighbors that $i, i' \in \T_t$ share\\ \hline
        	$\eout_{t, i}$ & $\sum_{j \in \Nout_{t, i}}{\omega_{t + 1, j}}$ -- effective out-degree of supplier $i \in \T_t$\\ \hline
        	$\T^a_t$ & $\{i \in \T_t \mid \din_{t, i} \cdot \dout_{t, i} > 0 \}$ -- subset of active suppliers in tier $\T_t$\\ \hline
        	$n^{a}_t$ & $|\T^a_t|$ -- number of active suppliers in tier $\T_t$; $0 \leq n_t^a \leq n_t$\\ \hline
        	$p_t$ & market price of a unit of product at which tier $\T_t$ sells downstream\\ \hline
        	$\rout_{t,i}$ & aggregate relative extent of overlap of out-neighborhoods in $\T_t$ with $\Nout_{t, i}$\\ \hline
        	$F_{t,j}$ & congestion at supplier $j \in \T_t$\\ \hline
        	$F_{t,j}^{-i}$ & $F_{t,j}$ excluding the contribution of $i \in \T_{t - 1}$\\ \hline
        	$c$ & constant cost of linking to an upstream supplier\\ \hline
        	$\gamma$ & constant congestion cost\\ \hline
        	$\pi_{t,i}$ & payoff of supplier $i \in \T_t$\\ \hline
    \end{tabular}
    \caption{Notation summary.}
\label{tbl:notation-summary}
\end{table}

\section{Strategic Formation of Supply Chain Networks Without Congestion}
\label{sec:supchain-formation}

In this section, we consider a supply chain model, similar to~\cite{bimpikis2019supply}, augmented with strategic link formation. Our 
equilibrium networks will differ from the networks exogenously imposed in~\cite{bimpikis2019supply}.

At a high-level, the two-tier version of our model is as follows. In a supply chain, consumers
and raw material producers are connected via two tiers---retailers (linked to consumers) and
suppliers (linked to the raw material producers). Only \begin{changed}retailers are strategic in that they
% (i) plan what price per unit of product to announce to consumers; as well as (ii)
decide which suppliers to source product from.\end{changed} In each tier of the supply chain, prices are determined via market clearing. When planning, retailers take into account production
failures that may occur at any agent present in the system. Retailers pay a constant cost for each link they create.
Each supplier is capable of delivering any amount of product---conditional upon
production success in the chain---regardless of the collective demand retailers exert upon it.

\subsection{Model Without Congestion}
\label{sec:model-no-congest}

\begin{figure}[ht!]
	\centering
	\includegraphics[width=0.5\linewidth]{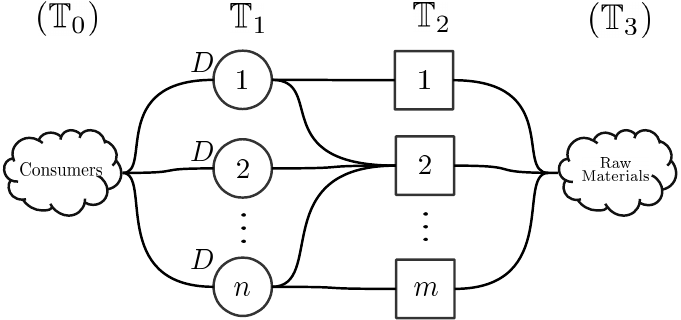}
	\caption{
		Two-tier supply chain with $n$ retailers, $m$ suppliers, and two implicitly present
		tiers of consumers and raw material producers. Links are
		directed from downstream to upstream agents, and appear only between adjacent tiers.
	}
	\label{fig:two-tier-chain}
\end{figure}

Let us consider a two-tier ($T = 2$) supply chain model---illustrated in
Fig.~\ref{fig:two-tier-chain}---in which tier $\T_1$ consists of $n$ retailers, all linked to consumers,
tier $\T_2$ consists of $m$ suppliers, all linked to raw material producers. It is up to the
retailers in $\T_1$ to decide which suppliers in $\T_2$ to link to.

\subsubsection{Demands}
Demand in the supply chain originates in the consumer tier, and, propagates
up the chain. \begin{changed}Consumers exert a fixed demand of $D$ units of product per
retailer.\footnote{\begin{changed}This is similar to the model of
\citet{ang2016disruption}, where the manufacturer is primed with a fixed consumer demand.\end{changed}}
\end{changed}
Consumer demand across retailers is equal consistent with the absence of differentiation. 

Each agent $i \in \T_t$ allocates its demand $D_{t, i}$ equally among its out-neighborhood. Thus, each upstream agent in $i$'s out-neighborhood receives an order for 
$D_{t, i} / \dout_{t, i}$ units (which is true even for consumers, as long as tier $\T_0$ is represented by
a single meta-consumer, with demand \begin{changed}$\Delta = n D$\end{changed}), so the demand of a supplier $j$ in tier $\T_2$ is
\begin{align}
	D_{2, j} = \sum\limits_{i \in \Nin_{2, j}}{
		\frac{D_{1, i}}{\dout_{1, i}}
	} = D \sum\limits_{i \in \Nin_{2, j}}{
		\frac{1}{\dout_{1, i}}
	},
	\label{eq:tier-2-demand}
\end{align}
where $\Nin_{2, j}$ is the in-neighborhood of supplier $j \in \T_2$, and $\dout_{1, i}$ is the
out-degree of retailer $i \in \T_1$; or, more generally,
\begin{align}
	D_{t, j} = \sum\limits_{i \in \Nin_{t, j}}{
		\frac{D_{t - 1, i}}{\dout_{t - 1, i}}
	}.
	\label{eq:demand}
\end{align}
Agent $i \in \T_t$, experiencing demand $D_{t, i}$, orders that exact amount of product upstream. An agent could order more than this to mitigate the uncertainty with upstream supply. \begin{changed}However, as discussed
at the end of Sec.~\ref{sec:no-congest-nash-equil-char}, allowing agents to strategically set order
quantities will not affect the equilibria networks arising in the model without congestion.\end{changed}

The demand formation process just described is equivalent to~\citet{bimpikis2019supply} who ``prime'' the supply chain with a fixed price for raw materials. Under the market clearing assumption this \begin{changed}translated\end{changed} into demand for raw materials, and the latter
propagates through the supply chain, \begin{changed}eventually, determining consumer demand\end{changed}. In our model, demand propagates in the opposite direction---from the consumers towards the raw material producers.

\subsubsection{Production, Failures, and Supplies}
Having received up to $D_{t, i}$ units of product from upstream, agent $i$ passes all it receives from the upstream tier to the  downstream tier with \emph{production
success probability} $\lambda \in (0, 1)$, or fails to do so with the complementary probability
$(1 - \lambda)$.

To analyze production failures, we introduce random variables
$\omega_{t, i} \sim \operatorname{Bernoulli}(\lambda)$---\emph{production success indicators}---indicating
whether production at agent $i \in \T_t$ succeeds.  Production failures at different
suppliers are independent, so $\omega_{t, i}$ are i.i.d. Using these random variables, we define
realized demand $R_{t, i}$ of agent $i \in \T_t$---the amount of product delivered to this agent
by upstream suppliers in response to its demand $D_{t, i}$ and along its out-links $\Nout_{t, i}$---where each supplier allocates its available product (whose quantity may be lower than
the one requested from the supplier due to disruptions) over its downstream agents
proportionally\footnote{Proportional allocation is a widely used scheme; for 
justification, see, for example~\cite{rong2017bullwhip}} to the latters' demands:
\begin{align}
	R_{t, i} &= \sum\limits_{j \in \Nout_{t, i}}{
		\underbrace{\left(
			\omega_{t + 1, j}
			\frac{ R_{t + 1, j} }{ D_{t + 1, j} }
		\right)}_{
			\substack{\text{produced share of}\\\text{supply of $j \in \T_{t + 1}$}}
		}
		\underbrace{
			\frac{ D_{t, i} }{ \dout_{t, i} }
		}_{
			\substack{\text{amount $i$}\\\text{requested}\\\text{from $j$}}
		}
	}, \label{eq:realized-demand}\\
	R_{2, j} &= D_{2, j}, \label{eq:realized-supplier-demand}
\end{align}
where the second expression implies never-failing raw material producers in the two-tier model.
Substituting~\eqref{eq:realized-supplier-demand} into~\eqref{eq:realized-demand},
we get the realized amount that a retailer receives is
\begin{align}
	R_{1, i} &= D \frac{ \sum_{j \in \Nout_{1, i}}{\omega_{2, j}} }{ \dout_{1, i} }
		= D \frac{\eout_{1, i}}{\dout_{1, i}},
	\label{eq:realized-retailer-demand}
\end{align}
where
\begin{align}
	\eout_{1, i} = \sum_{j \in \Nout_{1, i}}{\omega_{2, j}}
	\label{eq:effective-outdeg}
\end{align}
is the \emph{effective out-degree} of $i$,
that is, the number of $i$'s out-neighbors whose production succeeded.
We define supply $S_{t, i}$ of agent $i \in \T_t$ to its downstream customers as
\begin{align}
	S_{t, i} = \omega_{t, i} R_{t, i}.
\label{eq:supply}
\end{align}

Market clearing at every pair of adjacent tiers of the supply chain
translates into the following equality
\begin{align}
	S_{t + 1} = \Big(\sum\limits_{j \in \T_{t + 1}}{S_{t + 1, j}}\Big) = \sum\limits_{i \in \T_t}{
		R_{t, i}
	},
	\label{eq:market-clearance}
\end{align}
that is, we assume that the entire amount of product $S_{t+1}$ supplied by the upstream suppliers
$j \in \T_{t +1}$ is consumed by downstream suppliers $i \in \T_t$, as expressed via $\sum_i{R_{t, i}}$.

\subsubsection{Payoffs {\color{myblue}and Prices}}
The payoff $\pi_{t, i}$ of agent $i \in \T_t$ is as follows:
\begin{align}
	\pi_{t, i} &=
		\underbrace{ S_{t, i} \cdot p_t }_{ \substack{\text{selling}\\\text{downstream}} }
		\underbrace{-~R_{t, i} \cdot p_{t + 1}}_{ \substack{\text{buying}\\\text{upstream}} }
		\underbrace{-~c \cdot \dout_{t, i}}_{ \substack{\text{linking}\\\text{cost}} }, \label{eq:payoff}\\
	\pi_{T, i} &= S_{T, i} \cdot p_T - R_{T, i} \cdot p_{T + 1}, \label{eq:payoff-top-tier}
\end{align}
where $c \geq 0$ is a fixed linking cost, and
\begin{align}
    p_t = \Delta - S_t \tag{\ref{eq:market-price}}
\end{align}
is the market price in tier $\T_t$
being a function of the total output $S_t$ of that tier ($p_{T + 1}$ is the market price of raw materials).
This cost can be interpreted as the expense a retailer \begin{changed}incurs\end{changed} to establish a relationship with a new supplier. In a two-tier
model, suppliers are not strategic\footnote{In Sec.~\ref{sec:hetero-suppliers} we allow suppliers to invest in their reliability.} and, hence,
do not pay for their links to the raw material producers. From~\eqref{eq:payoff}, it is clear
that an agent's payoff depends upon how agents are interlinked, how much product agent
$i$ requests from upstream suppliers, as well as the random  production failures.

The price formation mechanism in~\eqref{eq:market-price} implies
that every active supplier in tier $\T_t$ contributes to the tier's output $S_t$ and,
hence, to the market price $p_t$. This can be justified by assuming that negotiations
happen \emph{ex ante},
without any pre-existing relationships between buyers and sellers.
\begin{changed}
    A link corresponds to a contract, according to which a buyer
    promises to buy up to a given quantity of product from the
    supplier at the market price which is determined after 
    production failures are realized and the total product
    quantity in the upstream market is established. This can be
    implemented, for example, through a price-matching clause in
    the contract, based on which suppliers would be discouraged
    to deviate from the market price, thereby---and due to the
    complete information assumption---establishing a single market
    price for the upstream market.

    Each retailer in tier $\T_1$ experiences a fixed consumer demand of $D$ units. However, the actual amount supplied by each retailer will, because of production failures, be less than $D$. If $S_1$ is the total realized supply at the retailer level, the price paid per unit by consumers will be given by $p_1 = \Delta - S_1$. When no production failures have occurred and, consequently, $S_1 = \Delta$, then, the retailers sell
    at the (zero) marginal cost.
    
    This price formation mechanism is identical to~\citet{bimpikis2019supply}, with one cosmetic difference.
    In~\citet{bimpikis2019supply}, the
    supply chain is ``bootstrapped'' with a fixed price for raw materials,
    which, then, translates into demand for raw materials and propagates through
    the supply chain under market clearance, until it reaches consumers; while
    in our model, the supply chain is bootstrapped with consumer demand, which
    propagates up the chain, eventually, determining the market price for
    raw materials. Mathematically, both pricing mechanism are identical.
    \citet{bimpikis2019supply} interpret this pricing model---where prices are
    not fixed in advance in a contract---as a spot market, which is a common
    mechanism occurring ``in a number of real-world supply chains, such as
    those for semiconductors and microelectronics'' (see also~\citet{mendelson2007strategic}).
\end{changed}

\begin{changed}
Finally, we note that the payoff~\eqref{eq:payoff} does not include an explicit penalty
for product under-delivery (in addition to the lost opportunity itself). While such
penalties can be included in a contract, they are complex to implement and are rarely
used~\citep{ang2016disruption,hwang2015simple}.
\end{changed}

\subsubsection{Network Formation Game Without Congestion}

The major qualitative difference between our model without congestion from the model considered
in~\cite{bimpikis2019supply} is that the agents in our model are allowed to choose their links. We model the agents'
link formation behavior as a one-shot network formation game. We describe the game for the 
two-tier model but it generalizes to the multi-tier case.

\begin{definition}[Strategic Network Formation Game Without Congestion]
In a two-tier supply chain, every retailer is considered a player, with payoff~\eqref{eq:payoff},
and whose pure strategy is its out-neighborhood $\Nout_{1, i}$, that is, which upstream suppliers
in $\T_2$ to link to. The retailers (or, all the
strategic agents in tiers $1, \dots, T - 1$ in the multi-tier case) simultaneously decide upon
their pure strategies, rationally maximizing their expected payoffs.
\label{def:game}
\end{definition}

We will be interested in pure strategy Nash equilibria of this game, defined with respect to
arbitrary unilateral deviations in a standard fashion as follows.
\begin{definition}[Pure Strategy Nash Equilibrium]
	$g^* = ({\Nout_{1, i}}^*)_{i \in \T_1}$ is a \emph{pure strategy Nash equilibrium} of the
	network formation game without congestion if for any retailer $i \in \T_1$ and for any
	$\Nout_{1, i}$, it holds that
	$$
		\E[\pi_{1, i}(
			\Nout_{1, i}, ({\Nout_{1, i'}}^*)_{i' \in \T_1 \setminus \{i\}}
		)]
		\leq
		\E[\pi_{1, i}(g^*)].
	$$
\label{def:equilibrium}
\end{definition}

Throughout this work, whenever we refer to an equilibrium, we mean pure strategy Nash
equilibrium.

\subsection{Analysis}
\label{sec:analysis-no-congest}

In this section we
characterize the pure strategy Nash equilibria of the network formation game without congestion.

\subsubsection{Expected Payoffs}

The first step in the analysis is to obtain the expected payoff of a retailer, based on
equation~\eqref{eq:payoff-top-tier}.

\begin{proposition}[Retailer's Expected Payoff]
For an active retailer $i \in \T_1$ (with $\dout_{1, i} > 0$),
\begin{align}
	\E[ \pi_{1, i} ]
		&= \lambda (1 - \lambda) D (\lambda D ((1 + \lambda) n_1^a - \lambda) - \Delta)
			+ \lambda (1 - \lambda)^2 D^2 \frac{ 1 + (1 + \lambda)\rout_{1, i} }{ \dout_{1, i} }
			- c \dout_{1, i},
	\label{eq:expected-retailer-payoff}
\end{align}
where $\lambda$ is the production success likelihood, $D$ is the consumer demand per
retailer, $n_1^a \leq n$ is the number of active retailers in tier $\T_1$,
$\Delta = n D$ is the total consumer demand, and
$$
	\rout_{1, i} = \sum_{i' \in \T_1 \setminus \{i\}}{\dout_{1, i' \cap i}~/~\dout_{1, i'}} = \sum_{i' \in \T_1 \setminus \{i\}}{|\Nout_{1, i} \cap \Nout_{1, i'}|~/~\dout_{1, i'}}
$$
measures the extent of overlap between the out-neighborhood of retailer $i$ and those of $i$'s
peers $i' \in \T_1$, $i' \neq i$. For a retailer $i$ having no out-links, $\E[ \pi_{1, i} ] = 0$.
\label{thm:expected-retailer-payoff}
\end{proposition}
\proof{Proof of Proposition~\ref{thm:expected-retailer-payoff}:}
	From equation~\eqref{eq:payoff}, we have
	\begin{align*}
		\pi_{1, i} &= S_{1, i} p_1 - R_{1, i} p_2 - c \dout_{1, i}
				= (\text{from~\eqref{eq:supply} and~\eqref{eq:market-price}})
				= \omega_{1, i} R_{1, i} (\Delta - S_1) - R_{1, i} (\Delta - S_2) - c \dout_{1, i}\\[0.15in]
			&= (\text{from~\eqref{eq:tier-supply} and~\eqref{eq:market-clearance}})
				= \omega_{1, i} R_{1, i} (\Delta - \sum\limits_{i' \in \T_1^a}{\omega_{1, i'} R_{1, i'}}) - R_{1, i} (\Delta - \sum\limits_{i' \in \T_1^a}{R_{1, i'}}) - c \dout_{1, i}\\
			&= R_{1, i} \Big(\sum\limits_{i' \in \T_1^a}{(1 - \omega_{1, i} \omega_{1, i'}) R_{1, i'} - (1 - \omega_{1, i}) \Delta}\Big) - c \dout_{1, i},
	\end{align*}
	where $\T_1^a \subseteq \T_1$ is a subset of active retailers that have at least one out-link each.
	To compute expectation of the obtained expression for $\pi_{1, i}$, let us first compute
	expectations of its components.
	\begin{align*}
		\E[R_{1, i}] &= (\text{from }~\eqref{eq:realized-retailer-demand})
			= \E\Big[ D \frac{\eout_{1, i}}{\dout_{1, i}} \Big]
			= \frac{D}{\dout_{1, i}} \E[ \eout_{1, i} ]
			= (\text{from }~\eqref{eq:effective-outdeg})
			= \frac{D}{\dout_{1, i}} \E\Big[ \sum\limits_{j \in \Nout_{1, i}}{\omega_{2, j}} \Big]\\
			&= \frac{D}{\dout_{1, i}} \sum\limits_{j \in \Nout_{1, i}}{\E[\omega_{2, j}]}
			= \frac{D}{\dout_{1, i}} \sum\limits_{j \in \Nout_{1, i}}{\lambda}
			= \frac{D}{\dout_{1, i}} \lambda \dout_{1, i} = \lambda D.
	\end{align*}
	\begin{align*}
		\E[R_{1, i} R_{1, i'}] &= (\text{from~\eqref{eq:realized-retailer-demand}})
			= \E\Big[ \Big( D \frac{\eout_{1, i}}{\dout_{1, i}} \Big) \Big( D \frac{\eout_{1, i'}}{\dout_{1, i'}} \Big) \Big]
			= \frac{D^2}{\dout_{1, i}\dout_{1, i'}} \E[ \eout_{1, i} \eout_{1, i'} ]\\[0.15in]
			& = \frac{D^2}{\dout_{1, i}\dout_{1, i'}} \E\Big[ \Big( \sum\limits_{j \in \Nout_{1, i}}{\omega_{2, j}} \Big) \Big( \sum\limits_{j' \in \Nout_{1, i'}}{\omega_{2, j'}} \Big) \Big]\\
			&= (\text{as $\omega_{t, i}$ are i.i.d., and $\E[X^2] = \E^2[X] + \Var[X]$})\\[0.1in]
			&= \frac{D^2}{\dout_{1, i}\dout_{1, i'}} \Big(
				\sum\limits_{j \in \Nout_{1, i}}{
					\sum\limits_{j' \in \Nout_{1, i'}}{
						\E[\omega_{2, j}] \E[\omega_{2, j'}]
					}
				}
				+
				\sum\limits_{j \in \Nout_{1, i} \cap \Nout_{1, i'}}{
					\Var[\omega_{2, j}]
				}
				\Big)\\
			&= \frac{D^2}{\dout_{1, i}\dout_{1, i'}} \Big(
				\sum\limits_{j \in \Nout_{1, i}}{
					\sum\limits_{j' \in \Nout_{1, i'}}{
						\lambda^2
					}
				}
				+
				\sum\limits_{j \in \Nout_{1, i} \cap \Nout_{1, i'}}{
					\lambda(1- \lambda)
				}
				\Big)
				= \frac{D^2}{\dout_{1, i}\dout_{1, i'}} \Big(
					\lambda^2 \dout_{1, i} \dout_{1, i'}
					+ \lambda(1- \lambda) \dout_{1, i \cap i'}
				\Big)\\
			&= \lambda D^2 \Big(
				\lambda + \frac{1 - \lambda}{\dout_{1, i}} \cdot \frac{\dout_{1, i \cap i'}}{\dout_{1, i'}}
			\Big),
	\end{align*}
	where $\dout_{1, i \cap i'} = | \Nout_{1, i} \cap \Nout_{1, i'} |$. In particular, when $i' = i$,
	\begin{align*}
			\E[R_{1, i}^2] = \lambda D^2 \Big(
				\lambda + \frac{1 - \lambda}{\dout_{1, i}} \cdot \frac{\dout_{1, i \cap i}}{\dout_{1, i}}
			\Big)
			= \lambda D^2 \Big(
				\lambda + \frac{1 - \lambda}{\dout_{1, i}} \cdot \frac{\dout_{1, i}}{\dout_{1, i}}
			\Big)			
			= \lambda D^2 \Big(
				\lambda + \frac{1 - \lambda}{\dout_{1, i}}
			\Big).
	\end{align*}
	Having computed expectations of expressions involving realized demands, we can now return to
	the computation of expectation of retailer payoff.
	\begin{align*}
		\E[\pi_{1, i}] &= \E\Big[
			R_{1, i} \Big(\sum\limits_{i' \in \T_1^a}{(1 - \omega_{1, i} \omega_{1, i'}) R_{1, i'}
				- (1 - \omega_{1, i}) \Delta}\Big) - c \dout_{1, i}
		\Big] = \text{(as $\omega_{t, i}$ and $R_{t+1,j}$ are indep.)}\\
		&=
			(1 - \E[\omega_{1, i}^2]) \E[R_{1, i}^2]
			+ \sum\limits_{\substack{i' \in \T_1^a, i' \neq i}}{
				(1 - \E[\omega_{1, i} \omega_{1, i'}])
				\E[R_{1, i} R_{1, i'}]
			}
			- \Delta (1 - \E[\omega_{1, i}]) \E[R_{1, i}]
			- c \dout_{1, i}\\
		&=
			(1 - \lambda) \lambda D^2 \Big(
				\lambda + \frac{1 - \lambda}{\dout_{1, i}}
			\Big)
			+ (1 - \lambda^2) \sum\limits_{\substack{i' \in \T_1^a, i' \neq i}}{
				\lambda D^2 \Big(
					\lambda + \frac{1 - \lambda}{\dout_{1, i}} \cdot \frac{\dout_{1, i \cap i'}}{\dout_{1, i'}}
				\Big)
			}
			- \Delta (1 - \lambda) \lambda D
			- c \dout_{1, i}\\
		&=
			\lambda^2 D^2	(1 - \lambda)
			+ \frac{\lambda(1 - \lambda)^2 D^2}{\dout_{1, i}}
			+ \lambda^2 (1 - \lambda^2) D^2 (|\T_1^a| - 1)\\
			&\phantom{xxxx} + \frac{\lambda (1 - \lambda) (1 - \lambda^2) D^2}{\dout_{1, i}}
				\sum\limits_{i' \in \T_1^a \setminus \{ i \}}{
					\frac{\dout_{1, i' \cap i}}{\dout_{1, i'}}
				}
			- \lambda (1- \lambda) \Delta D - c \dout_{1, i}\\
		&=
			\lambda^2 D^2	(1 - \lambda)
			+ \frac{\lambda(1 - \lambda)^2 D^2}{\dout_{1, i}}
			+ \lambda^2 (1 - \lambda^2) D^2 (n_1^a - 1)
			+ \frac{\lambda (1 - \lambda) (1 - \lambda^2) D^2}{\dout_{1, i}}
				\rout_{1, i}\\
			&\phantom{=} - \lambda (1- \lambda) \Delta D - c \dout_{1, i}\\
		&= \lambda (1 - \lambda) D (\lambda D ((1 + \lambda) n_1^a - \lambda) - \Delta)
			+ \lambda (1 - \lambda^2) D^2 \frac{ 1 + (1 + \lambda) \rout_{1, i} }{\dout_{1, i}}
			- c \dout_{1, i}.
	\end{align*}
	\qed
\endproof

\subsubsection{Bounding Costs}
\label{sec:bounding-costs-no-congest}

To prevent trivial equilibrium outcomes such as an empty network, we need to ensure that costs are not excessive.

\begin{assumption}[Bounding Costs for Network Formation Without Congestion]
If the number of suppliers is
	at least as large as the number of retailers, that is, $m \geq n$, then, the network with parallel
	links---in which every retailer maintains a single link, pointing to an exclusive supplier yields each retailer  positive expected payoff.
\label{asm:parallel-network-feasibility-no-congestion}
\end{assumption}

\begin{figure}[ht!]
	\centering
	\includegraphics[width=0.5\linewidth]{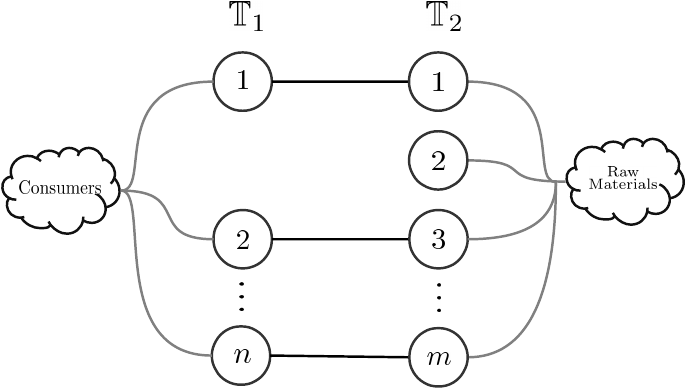}
	\caption{A network with parallel links. Model's parameters are assumed to be bounded, so
	that in such a network, if $m \geq n$, every retailer would earn
	positive expected payoff.}
	\label{fig:parallel-network}
\end{figure}

Assumption~\ref{asm:parallel-network-feasibility-no-congestion} states that the
model's parameters are such that the network with parallel links---illustrated in Fig.~\ref{fig:parallel-network}---is at least as good as the empty network. This network is the
 simplest and least cost---from the point of view
of link maintenance cost---network in which retailers can turn a profit.

\begin{proposition}[Bounding Costs in the Model Without Congestion]
	For the model without congestion,
	Assumption~\ref{asm:parallel-network-feasibility-no-congestion} holds if and only if
	\begin{align}
		c &< \lambda (1 - \lambda) (n - 1) (\lambda^2 + \lambda - 1) D^2, \label{eq:ub-c-no-congest}\\
		\lambda &> \frac{\sqrt{5} - 1}{2} \approx 0.618. \label{eq:lb-lambda-no-congest}
	\end{align}
	\label{thm:cost-bounds-no-congestion}
\end{proposition}
\proof{Proof of Proposition~\ref{thm:cost-bounds-no-congestion}:}
	From Proposition~\ref{thm:expected-retailer-payoff}, we know that the expected retailer
	payoff in a network with parallel links and enough suppliers ($m \geq n$)---shown in
	Fig.~\ref{fig:parallel-network}---is as follows.
	\begin{align*}
		\E[ \pi_{1, i} ]
		 &= \lambda (1 - \lambda) D (\lambda D ((1 + \lambda) n_1^a - \lambda) - \Delta)
			+ \lambda (1 - \lambda)^2 D^2 \frac{ 1 + (1 + \lambda)\rout_{1, i} }{ \dout_{1, i} }
			- c \dout_{1, i}\\
		&= \lambda (1 - \lambda) D (\lambda D ((1 + \lambda) n - \lambda) - \Delta)
			+ \frac{ \lambda (1 - \lambda)^2 D^2 }{ 1 }
			- c \cdot 1\\[0.1in]
		&= D^2 \cdot \lambda (1 - \lambda) ((n - 1) \lambda^2 + (n - 1)\lambda + 1)
			- D \cdot \lambda (1 - \lambda) \Delta - c\\[0.1in]
		&= (\Delta = n D) = \lambda (1 - \lambda) (n - 1) (\lambda^2 + \lambda - 1) D^2 - c.
		\end{align*}
		The proposition statement's requirement $\E[ \pi_{1, i} ] \geq 0$ immediately translates into the upper bound for the linking cost
		$$
			c < \lambda (1 - \lambda) (n - 1) (\lambda^2 + \lambda - 1) D^2.
		$$
		For this upper bound to be well-defined, however, it must be non-negative, since
		$c \geq 0$. It is non-negative as long as $\lambda^2 + \lambda - 1 \geq 0$,
		which holds iff $\lambda \in (\tfrac{\sqrt{5} - 1}{2}, 1)$.
		
		\qed
\endproof

\subsubsection{Nash Equilibria Characterization}
\label{sec:no-congest-nash-equil-char}
In our analysis of equilibria, we will first deal with an empty network equilibrium. While such
a network itself is not of particular interest to us, its being an equilibrium provides useful
insights into the model without congestion and the impact of the presence of multiple active
retailers upon the latters' profit-making.
\begin{theorem}[Empty Equilibrium Existence]
	If the linking cost $c > 0$, then an empty network is a pure strategy Nash equilibrium of
	the network formation game without congestion.
\label{thm:two-tier-empty-equil-no-congest}
\end{theorem}
\proof{Proof of Theorem~\ref{thm:two-tier-empty-equil-no-congest}:}
From~\eqref{eq:payoff}, we know
$$
\pi_{t, i} =
		\underbrace{ S_{t, i} \cdot p_t }_{ \substack{\text{selling}\\\text{downstream}} }
		\underbrace{-~R_{t, i} \cdot p_{t + 1}}_{ \substack{\text{buying}\\\text{upstream}} }
		\underbrace{-~c \cdot \dout_{t, i}}_{ \substack{\text{linking}\\\text{cost}} }.
$$
When only one retailer $i \in \T_1$ is active, $\dout_{1, i} > 0$, while all its peers have no
links,
\begin{align*}
	p_1 &= (\text{from~\eqref{eq:market-price}}) = \Delta - S_1 = (\text{from~\eqref{eq:tier-supply}}) = \Delta - S_{1, i} = (\text{from~\eqref{eq:supply}}) = \Delta - \omega_{1, i} R_{1, i},\\
	p_2 &= (\text{from~\eqref{eq:market-price}}) = \Delta - S_2 = (\text{from~\eqref{eq:market-clearance}}) = \Delta - R_{1, i},
\end{align*}
so,
\begin{align*}
	\pi_{1, i} &= S_{1, i} \cdot p_1 - R_{1, i} \cdot p_2 - c \cdot \dout_{1, i}
		= \omega_{1, i} R_{1, i} (\Delta - \omega_{1, i} R_{1, i})
			- R_{1, i} (\Delta - R_{1, i}) - c \dout_{1, i}\\[0.1in]
		&\leq R_{1, i} (\Delta - R_{1, i})
			- R_{1, i} (\Delta - R_{1, i}) - c \dout_{1, i} = - c \dout_{1, i} < 0
\end{align*}
as long as $\dout_{1, i} > 0$. Hence, no retailer would prefer to unilaterally deviate from
an empty network, making it an equilibrium.
\qed
\endproof

In words, Theorem~\ref{thm:two-tier-empty-equil-no-congest} states that
a single active retailer cannot create and exploit a gap between upstream and downstream prices
if no other active retailers are in the market. It is useful to note that a similar effect is present
when there are multiple active retailers and no production failures,
as Theorem~\ref{thm:empty-equil-unique-when-no-failures} states.
\begin{theorem}[Empty Equilibrium Uniqueness When Nobody Fails]
	If production never fails, that is, $\lambda = 1$, and $c > 0$, then, the empty network is the
	unique equilibrium of the network formation game without congestion.
% 	(while the
% 	corresponding consumer demand $D$ per retailer is arbitrary).
\label{thm:empty-equil-unique-when-no-failures}
\end{theorem}
\proof{Proof of Theorem~\ref{thm:empty-equil-unique-when-no-failures}:}
If $\lambda = 1$, the expected payoff of a retailer having positive out-degree $\dout_{1, i} > 0$
is as follows:
\begin{align*}
	\pi_{1, i} &= S_{1, i} \cdot p_1 - R_{1, i} \cdot p_2 - c \cdot \dout_{1, i}
		= (\text{from~\eqref{eq:market-price}})
	 	= S_{1, i} \cdot (\Delta - S_1) - R_{1, i} \cdot (\Delta - S_2) - c \cdot \dout_{1, i}\\[0.1in]
	 	&= (\text{from~\eqref{eq:supply}}) = \omega_{1, i} R_{1, i} \cdot (\Delta - S_1) - R_{1, i} \cdot (\Delta - S_2) - c \cdot \dout_{1, i} = (\text{since no failures})\\[0.1in]
	 	&= R_{1, i} \cdot (\Delta - n_1^a D) - R_{1, i} \cdot (\Delta - n_1^a D) - c \cdot \dout_{1, i}
	 	 	= - c \dout_{1, i} < 0.
\end{align*}
Thus, a retailer cannot have a positive out-degree at an equilibrium, making the empty
network---which is an equilibrium as per Theorem~\ref{thm:two-tier-empty-equil-no-congest}---a  unique equilibrium.
\qed
\endproof
Theorem~\ref{thm:empty-equil-unique-when-no-failures} easily generalizes to
a supply chain with an arbitrary number $T \geq 2$ of tiers. Theorem~\ref{thm:empty-equil-unique-when-no-failures} states that
production failures are essential for the agents' ability to make positive profit in the model.
The latter is the result of price formation through competition under market clearance, as well as due to our
stipulation that the agents function as ``repeaters'', at best reproducing the input quantity,
and not actually transforming the product and/or adding any value to it.

\begin{figure}[ht!]
	\centering
	\includegraphics[width=0.5\linewidth]{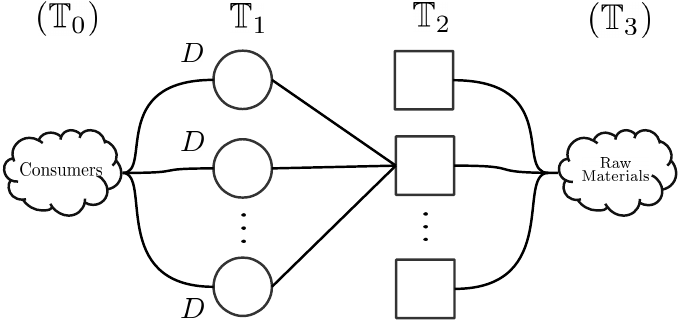}
	\caption{
		Pure strategy Nash equilibrium in supply chain network formation game without
		congestion.
	}
	\label{fig:two-tier-equil-no-congest}
\end{figure}

From now on we will be interested in non-trivial equilibria.
In the following Theorem~\ref{thm:two-tier-equil-no-congest}, we characterize non-trivial
equilibria of the supply chain network formation game without congestion. The networks from
these equilibria are illustrated in Fig.~\ref{fig:two-tier-equil-no-congest}.

\begin{theorem}[Non-empty Equilibria Characterization]
	In the supply chain network formation game without congestion (Definition~\ref{def:game}),
	under Assumption~\ref{asm:parallel-network-feasibility-no-congestion}, if $c > 0$,
	then 	a \emph{cone network}---in which every retailer $i \in \T_1$ maintains a single
	link, and all the retailers link to the same upstream supplier---is a pure strategy
	Nash equilibrium. This is a unique non-empty equilibrium of the game, up to supplier
	labeling.
	\label{thm:two-tier-equil-no-congest}
\end{theorem}
\proof{Proof of Theorem~\ref{thm:two-tier-equil-no-congest}}
	We first show that under the assumptions about the linking cost, cone networks are Nash equilibria, and, then, show their uniqueness.

	\emph{1) Cone networks are equilibria:}
	Consider one such network---as shown in Fig.~\ref{fig:two-tier-equil-no-congest}---where all the retailers maintain a single link each, linking to the same supplier.
	From~\eqref{eq:expected-retailer-payoff}, the expected payoff of retailer $i$ is
	\begin{align*}
		\E[ \pi_{1, i} ]
		&= \lambda (1 - \lambda) D (\lambda D ((1 + \lambda)n_1^a - \lambda) - \Delta)
			+ \lambda (1 - \lambda)^2 D^2 \frac{ 1 + (1 + \lambda)\rout_{1, i} }{ \dout_{1, i} }
			- c \dout_{1, i}.
	\end{align*}
	In a cone network, in the expression above,
	\begin{align*}
		n_1^a = n,\quad
		\dout_{1, i} = 1,\quad
		\rout_{1, i} = \sum_{i' \in \T_1^a \setminus \{i\}}{
			\frac{ \dout_{1, i' \cap i} }{ \dout_{1, i'} }
		} = \sum_{i' \in \T_1^a \setminus \{i\}}{
			1
		} = n_1^a - 1 = n - 1,
	\end{align*}
	so, in such a network,
	\begin{align*}
	\E[ \pi_{1, i} ]
		&= \lambda (1 - \lambda) D (\lambda D ((1 + \lambda) n - \lambda) - n D)
			+ \lambda (1 - \lambda)^2 D^2 (1 + (1 + \lambda) (n - 1))
			- c\\[0.1in]
		&=		
		\lambda (1 - \lambda) ((1 + \lambda) n - \lambda) D^2
		- \lambda (1 - \lambda) D^2 n
		- c\\[0.1in]
	    &= \lambda^2 (1 - \lambda) (n - 1) D^2 - c.
	\end{align*}
	This is non-negative as long as
	$$	
		c \leq \lambda^2 (1 - \lambda) (n - 1) D^2.
	$$
	Simultaneously, we have an upper bound~\eqref{eq:expected-retailer-payoff} on $c$
	from Proposition~\ref{thm:cost-bounds-no-congestion}, coming from the assumption about
	the feasibility of a network with parallel links
	$$
		c < \lambda (1 - \lambda) (n - 1) ( \lambda^2 + \lambda - 1 ) D^2 
	$$
	that holds in the considered region $\lambda \in (\tfrac{\sqrt{5} - 1}{2}, 1)$.
	The latter bound on $c$ is tighter for such $\lambda$ than the former one, so, under
	Assumption~\ref{asm:parallel-network-feasibility-no-congestion}, for the considered
	network we have
	$$
		\E[ \pi_{1, i} ] > 0.
	$$
	It is also clear that no agent strictly prefers to unilaterally
	deviate from that network: (i) by dropping a link, a retailer would change its positive expected
	payoff to zero expected payoff; (ii.a) $\rout_{1, i}$ is already at its maximum ($i$'s
	out-neighborhood completely overlaps with that of each of its peers, of all whom are active)
	and cannot be improved; and (ii.b) $\E[ \pi_{1, i} ]$ is strictly
	decreasing in $\dout_{1, i}$. Consequently, the considered network is a pure strategy
	Nash equilibrium.
	
	\emph{2) Cone networks are the only equilibria:}
	Now, we show that the cone networks---each corresponding to a different supplier $j \in \T_2$
	to whom every retailer links---are unique.
	
	 From Assumption~\ref{asm:parallel-network-feasibility-no-congestion} it is clear that if
	some---but not all---retailers have no links, the corresponding network
	is not an equilibrium---by assumption, these retailers would earn positive profit
	by maintaining a single link having no out-neighborhood overlap with other retailers; in an
	arbitrary network (rather than a network with parallel links and sufficiently many suppliers), the
	overlap can only increase a retailer's expected payoff. Thus, we are only concerned with proving
	that networks where every retailer is active are not equilibria unless it is a cone network.
	
	Now, assume a network, where every retailer is active, $\forall i \in \T_1: \dout_{1, i} > 0$,
	yet the retailer degree sequence is non-uniform. We show that for a retailer $i$ such that
	$\dout_{1, i} > 1$, one can always find a link in $\Nout_{1, i}$ to drop which would be
	strictly beneficial to $i$. Assume that retailer $i$ is considering dropping a link to
	upstream supplier $k \in \Nout_{1, i}$; the payoff and the corresponding neighborhood overlap
	after the link is dropped are denoted by $\widetilde{\pi}_{1, i}$ and $\widetilde{\uprho}^+_{1, i}$,
	respectively. Then,
	\begin{align*}
		\E[\widetilde{\pi}_{1, i} &- \pi_{1, i} ]\\[0.1in]
			&= \left[
				\lambda (1 - \lambda)^2 D^2 \frac{ 1 + (1 + \lambda)\widetilde{\uprho}^+_{1, i} }{ \dout_{1, i} - 1 } - c (\dout_{1, i} - 1)
			\right]
			-
			\left[
				\lambda (1 - \lambda)^2 D^2 \frac{ 1 + (1 + \lambda)\rout_{1, i} }{ \dout_{1, i} }
				- c \dout_{1, i}
			\right]\\[0.1in]
			&= \frac{ \lambda (1 - \lambda)^2 D^2 }{ \dout_{1, i}(\dout_{1, i} - 1) }
				 - \lambda (1 - \lambda) (1 - \lambda^2) D^2 \left(
				 	\frac{\widetilde{\uprho}^+_{1, i}}{\dout_{1, i} - 1}
				 	-
				 	\frac{\rout_{1, i}}{\dout_{1, i}}
				 \right) + c\\[0.1in]
			&=\lambda (1 - \lambda)^2 D^2 \frac{ 1 + (1 + \lambda) ( \dout_{1, i} (\widetilde{\uprho}^+_{1, i} - \rout_{1, i}) + \rout_{1, i} ) }{ \dout_{1, i}(\dout_{1, i} - 1) } + c.
	\end{align*}
	Let us take a closer look at one component of the obtained expression:
	\begin{align*}
		\dout_{1, i} & (\widetilde{\uprho}^+_{1, i} - \rout_{1, i}) + \rout_{1, i}
			= \text{(from Lemma~\ref{thm:rho-f-lemma})}
			= \dout_{1, i} (
					\sum\limits_{j \in \Nout_{1, i} \setminus \{k\}}{
						F_{2, j}^{-i}
					}  - \rout_{1, i}
				) + \rout_{1, i}\\
			&= \dout_{1, i} (
				(\rout_{1, i} - F_{2, k}^{-i}) - \rout_{1, i}
				) + \rout_{1, i}
				= \rout_{1, i} - F_{2, k}^{-i} \dout_{1, i}
				= \dout_{1, i} \left(
					\frac{\rout_{1, i}}{\dout_{1, i}} - F_{2, k}^{-i}
				\right)\\
			&= \text{(from Lemma~\ref{thm:rho-f-lemma})}
				= \dout_{1, i} \left(
					\frac{ \sum_{j \in \Nout_{1, i}}{
						F_{2, j}^{-i}
					} }{\dout_{1, i}} - F_{2, k}^{-i}
				\right)
				= \dout_{1, i}( \langle
						F_{2, j}^{-i}
					\rangle_{j \in \Nout_{1, i}}
					- F_{2, k}^{-i} ),
	\end{align*}
	where $F_{2, j}^{-i} = \sum_{i' \in \Nin_{2, j} \setminus \{i\}}{1 / \dout_{1, i'}}$.
	Taking into account that $k \in \Nout_{1, i}$, in the obtained expression we are subtracting
	one $F_{2, k}^{-i}$ from its arithmetic average per retailer $i$. It is clear that we can always pick
	$k = \argmin_{j}{F_{2, j}^{-i}}$ to make the obtained expression non-negative. Thus,
	$\dout_{1, i} (\widetilde{\uprho}^+_{1, i} - \rout_{1, i}) + \rout_{1, i} \geq 0$, and
	consequently, $\E[\widetilde{\pi}_{1, i} - \pi_{1, i} ] > 0$ for such $k$. In other words, an arbitrarily
	picked retailer with out-degree exceeding $1$ has a strict incentive  to drop a
	link to its ``least useful'' supplier. Hence, for any non-cone network there is a sequence of strictly improving unilateral deviations that terminate in a cone network.
	\qed
\endproof

 Theorem~\ref{thm:two-tier-equil-no-congest} generalizes easily
to the $T$-tier case and is summarized in
Corollary~\ref{thm:multi-tier-equil-no-congest}.

\begin{figure}[ht!]
	\centering
	\includegraphics[width=0.85\linewidth]{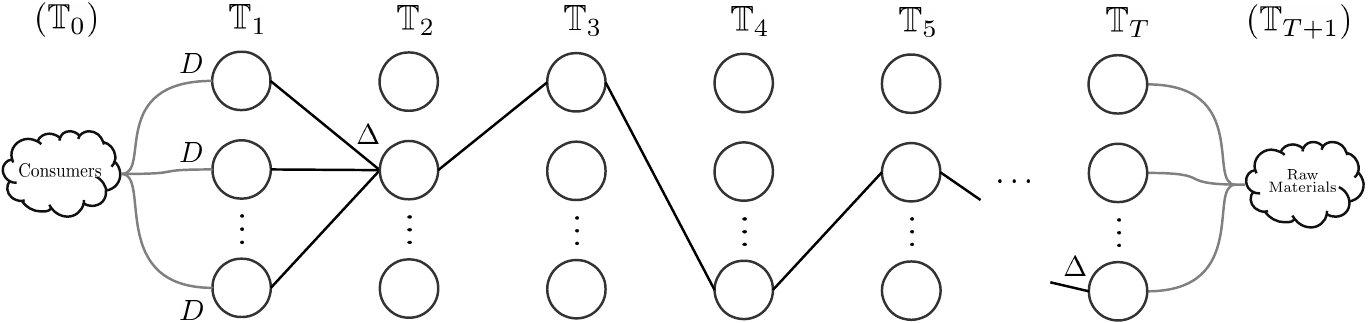}
	\caption{
		Pure strategy Nash equilibrium in multi-tier supply chain network formation game
		without congestion.
	}
	\label{fig:multi-tier-equil-no-congest}
\end{figure}

\begin{corollary}[Non-Empty Equilibria in $T$-tier Supply Chain]
	In a $T$-tier supply chain network formation game without congestion,  the networks illustrated in
	Fig.~\ref{fig:multi-tier-equil-no-congest}---in which $\T_1$ retailers concentrate links,
	and in each subsequent tier $\T_t$, $t \in \{2, \dots, T\}$, only one active supplier
	maintains a single link---are the unique non-empty pure strategy Nash equilibria.
	\label{thm:multi-tier-equil-no-congest}
\end{corollary}
Thus, in the absence of congestion equilibrium supply chains are almost chains, up to the
linkage between the first two tiers $\T_1$ and $\T_2$. From this corollary, it is clear why in~\cite{bimpikis2019supply}, among the exogenously given complete
$k$-partite supply chain networks, the inverted pyramid-shaped networks---with the number of
agents per tier decreasing along the supply chain---appear as optimal.

\begin{changed}
  Prior work considering production uncertainty,
    focuses on strategically setting order quantities and prices~\citep{cachon1999capacity,cachon2003supply,cho2013advance}
    rather than strategically choosing suppliers, as in our case. One may wonder whether
    Theorem~\ref{thm:two-tier-equil-no-congest} and Corollary~\ref{thm:multi-tier-equil-no-congest}
    hold if we also allow the agents to strategically set order quantities, under the
    assumption that the quantity that an agent can sell downstream does not exceed its downstream
    demand\footnote{\begin{changed}This assumption is encoded in the payoff expression~\eqref{eq:payoff} implicitly. When an agent does not strategically set order quantities, her output cannot exceed her
    demand. If we allow order quantities to vary, we need to truncate the quantity
    an agent sells downstream, changing the payoff expression~\eqref{eq:payoff} to $\pi_{t, i} = \min\{D_{t, i}, S_{t, i}\} \cdot p_t - R_{t, i} \cdot p_{t + 1} - c \cdot \dout_{t, i}$.\end{changed}}.
    The answer is \emph{yes}: without
    congestion, retailers' linking and order quantity setting decisions are orthogonal,
    so retailers would form a cone network (or the chain-like network from Corollary~\ref{fig:multi-tier-equil-no-congest}) at equilibrium, with the equilibrium
    order quantities being above downstream demand to hedge against the supplier's
    under-delivery. Due to our focus on strategic linking, computing these quantities
    is beyond the scope of this paper.
\end{changed}

\subsection{Discussion of the Model Without Congestion}
\label{sec:model-discussion-no-congest}

The analysis in Sec.~\ref{sec:analysis-no-congest} of the model of Sec.~\ref{sec:model-no-congest}
produced cone-shaped non-empty equilibria networks, in which the retailers, \begin{changed}having\end{changed} one link each,
point to the same supplier upstream. The sparsity of such networks as well as the link
concentration behavior are surprising. Intuitively, one would expect  a resilient / efficient network would have some link redundancy.
It is also surprising that too high production reliability, i.e., $\lambda$ large, can actually hurt retailers;
in particular, there is no way retailers can make a profit if production never fails. We discuss
these observations below.

\subsubsection{Sparsity}
\label{sec:model-discussion-no-congest-sparsity}
\emph{Equilibrium networks are sparse because there are no bounds on supplies}. Each upstream supplier can produce as much
product as requested (conditional upon production success at that supplier, and, possibly,
at other higher-level suppliers in a multi-tier model).
The consequence of this is easy to see in
a simplified model where we remove price formation and focus only
\begin{changed}on\end{changed}
 quantities: assume that a single retailer needs to satisfy a demand of $D$ units, and
has an option to source it from $\dout_{1, i}$ upstream suppliers ($\dout_{1, i} / D$ units from each)
each of whom successfully delivers the requested quantity with a fixed probability $\lambda$. Ignoring the cost of link formation, the expected payoff of the retailer is $\lambda D$. Hence,  it does not matter through how many links to source product, even
in the presence of failure. As soon as we introduce a positive linking cost, the retailer prefers to
source product via a single link.  
If, instead, each upstream supplier had a hard
cap on its production output strictly lower that a retailer's demand $D$,  retailers would be
forced to multi-source. Our model with congestion---described in Sec.~\ref{sec:supchain-formation-w-congestion}---incorporates a soft cap via a
congestion penalty.

\subsubsection{Link Concentration} \emph{Retailers favor link concentration as sourcing from
a single supplier allows them to buy at a low upstream price (conditional upon that supplier
successfully producing).} The positive effect of link concentration by the retailers is best
illustrated with a simple example.
Consider a supply chain, with two retailers $1$ and $2$ and two suppliers $A$ and $B$,
in which we are concerned with the expected payoff of retailer $1$. For simplicity, suppose
$D = 1$ and $c = 0$ (the introduction of linking costs does not affect the conclusion). Now, let us compare the link concentration scenario (\emph{cone network}) with the
scenario when the retailers source from separate suppliers (\emph{network with parallel links}).
\begin{itemize}
	\item{\emph{Cone Network:}
		If both retailers source from the same supplier, say, $A$, the upstream price
		is always low ($0$ in this example) when upstream production succeeds, and the positive
		expected payoff of retailer $1$ is obtained entirely from the case when both retailer $1$
		and the supplier $A$ succeed, while retailer $2$ fails, thereby, creating a 1 unit gap 
		between upstream and downstream prices, generating a payoff $\pi_{1, 1} = 1$,
		with probability $\lambda^2 (1 - \lambda)$.
	}
	\item{\emph{Network with Parallel Links:}
		If the retailers source from different suppliers, then, the expected payoff of retailer $1$
		has two components. One when $1$ succeeds, its peer $2$
		fails, and both upstream suppliers succeed establishing a low upstream price---which
		happens with probability $\lambda^3(1- \lambda)$, and in which retailer $1$ has a payoff
		$\pi_{1, 1} = 1$. The other case is when $1$ fails, yet, its supplier $A$ succeeds (so $1$ does
		not sell to consumers, yet has to buy from $A$), and $B$ fails (so $1$ buys at a high upstream price of $1$), which
		happens with likelihood $\lambda (1 - \lambda)^2$ and corresponds to $1$'s payoff
		$\pi_{1, 1} = -1$. Hence, the expected payoff of retailer $1$ in the network with parallel
		links is $\lambda^3(1- \lambda) - \lambda (1 - \lambda)^2 = \lambda (1 - \lambda) (\lambda^2 + \lambda - 1) < \lambda^2 (1 - \lambda)$
		for all $\lambda \in (0, 1)$. 
	}
\end{itemize}

The benefit that retailers enjoy from supply variance is related to~\citet{weitzman1974prices} who compares 
the benefit of controlling a system through quantities rather than prices when production costs are uncertain. In the first case, quantities are fixed and prices adjust to clear the market. In the second case, prices are fixed and quantity adjusts to clear the market.
Weitzman argued that
control through quantities is superior to control via prices, and the
advantage of such control for the system scales with
the variance of the (component of the) production cost. Thus, higher production
cost variance makes quantity control more advantageous for the producer. Our
results provide a complementary perspective: If the buyer can choose to source from distinct supply chains that are quantity controlled, it may prefer to source from the chain having higher
output uncertainty. Consequently, while, according to~\citet{weitzman1974prices},
higher cost variance encourages control through quantities, our results suggest
that competition among quantity controlled supply chains will increase output uncertainty.

\subsubsection{Retailers' Welfare vs. Production Failure} 
\begin{figure}[ht!]
	\centering
	\includegraphics[width=0.5\linewidth]{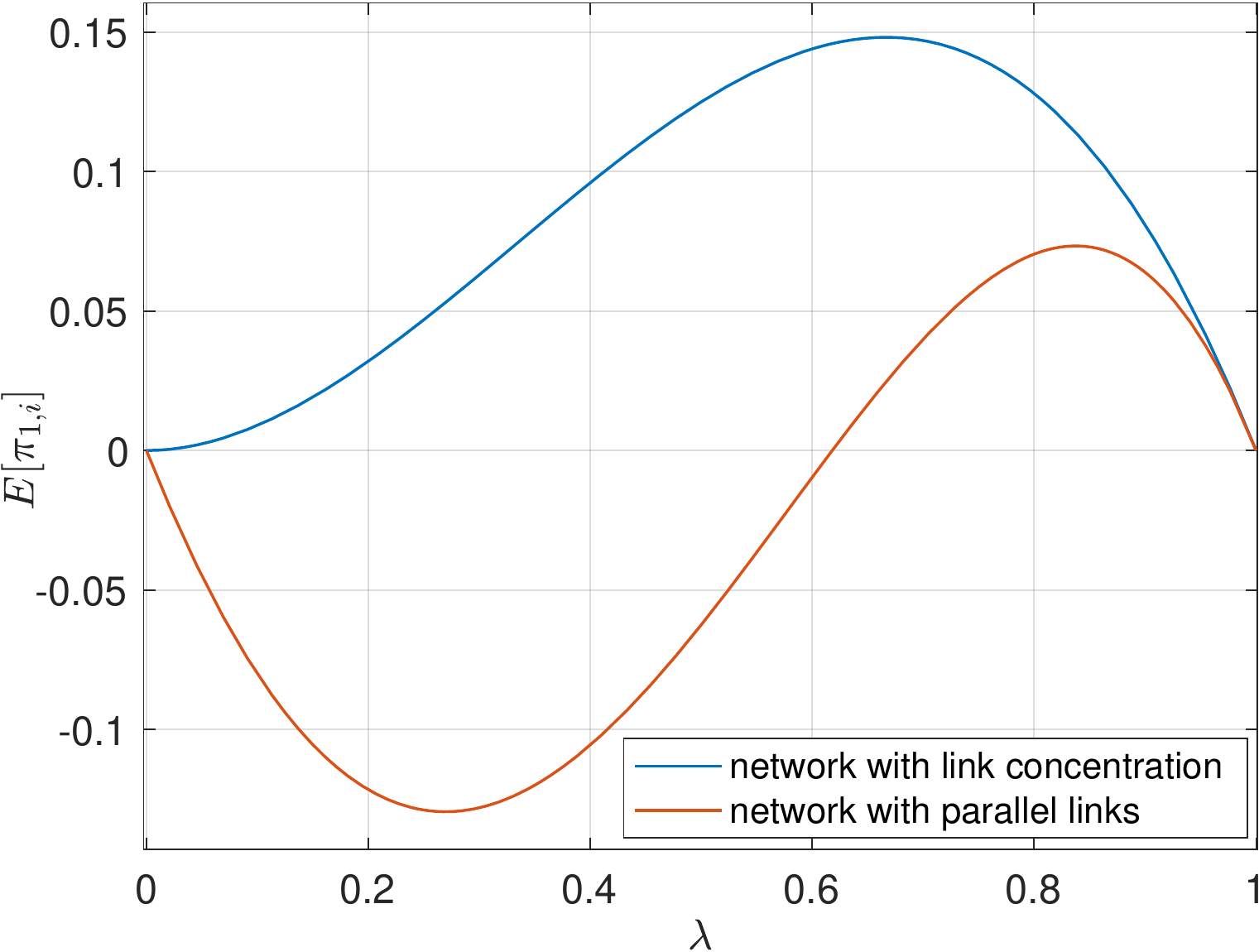}
	\caption{
		Dependency of a retailer's expected payoff in a small supply chain in a network where both retailers source from the same supplier and a network where they source from two different suppliers.
	}
	\label{fig:exp-payoff-vs-lambda-no-congest}
\end{figure}

\emph{Retailers' welfare suffers when
$\lambda$ is close to $1$ because a small number of failures among
a retailer's peers cannot result in a large enough gap between upstream and downstream prices,
to guarantee the retailer a positive expected payoff.} If we resort to the same simple supply chain
example from the discussion of link concentration behavior, with two retailers and two
suppliers, the above mentioned effect clearly manifests itself in Fig.~\ref{fig:exp-payoff-vs-lambda-no-congest}
when we look at the dependency of a retailer's expected payoff upon productivity $\lambda$.

The non-monotonic dependence of a retailer's expected
payoff upon $\lambda$ is valid only when we vary reliability of
\emph{every} agent in the system. If we admit a heterogeneous environment, where every retailer
and supplier $i \in \T_t$ had its own reliability parameter $\lambda_{t, i}$, then, in the model
without congestion, from the point of view of retailer $i \in \T_1$, a perfect situation would
be if every retailer including $i$ were linking to the most reliable supplier, $i$ itself would also
have maximal $\lambda_i$, while its peers' $\lambda_{i'}$ were minimal---this way, retailer
$i$ could guarantee itself both a large product quantity to sell, and a large gap between upstream
and downstream prices.

\section{Strategic Formation of Supply Chain Networks With Congestion}
\label{sec:supchain-formation-w-congestion}

In this section, we extend the previous model by incorporating a congestion penalty, modeling
either limited supply or a delay in supply delivery---based upon the total product quantity being
produced by a supplier---and show that the congestion effect changes the formed supply chains
qualitatively.

\subsection{Model}
\label{sec:general-model-def}
The payoff function~\eqref{eq:payoff-ex} for each agent $i \in \T_t$ is an extension of the payoff function~\eqref{eq:payoff} that incorporates a congestion penalty term $L_{t, i}$:

\begin{align}
	\pi_{t, i} &=
		\underbrace{ S_{t, i} \cdot p_t }_{ \substack{\text{selling}\\\text{downstream}} }
		\underbrace{-~R_{t, i} \cdot p_{t + 1}}_{ \substack{\text{buying}\\\text{upstream}} }
		\underbrace{-~c \cdot \dout_{t, i}}_{ \substack{\text{linking}\\\text{cost}} }
		\underbrace{-~L_{t, i}}_{ \substack{\text{congestion}\\\text{penalty}} }, \label{eq:payoff-ex} \\[0.1in]
	\pi_{T, i} &= S_{T, i} \cdot p_T - R_{T, i} \cdot p_{T + 1}, \notag
\end{align}
where
\begin{align}
	L_{t, i} = \frac{1}{\dout_{t, i}} \sum\limits_{j \in \Nout_{t, i}}{
		\ell(S_{t + 1, j})
	}
	= \frac{1}{\dout_{t, i}} \sum\limits_{j \in \Nout_{t, i}}{
		\frac{\gamma}{2} (S_{t + 1, j})^2
	}.
\label{eq:latency}
\end{align}

On can interpret the penalty as a soft constraint on supplies.
Another way is to treat it as a delay or latency in product
delivery---the larger the amount of product being in production at a supplier, the longer a retailer
would wait, on average, for the delivery of goods by that supplier. The congestion function
$\ell(x)$ depends upon the amount of product actually
produced by a supplier, though demand-dependent $\ell$ may also be a viable option\footnote{Dependency of the congestion function $\ell(x)$ upon either
the requested or produced amount of product is meaningful, depending on when a requesting
party learns about the upstream failure. For example, if a failure occurs due to a natural disaster
or a union strike---both of which are publicly observed---the congestion penalty would depend
on the amount delivered; if, however, a supplier reaches the deadline having not managed to
produced any output and having not timely informed its clients about it, then the congestion penalty's
dependency upon (non-realized) demand may be more appropriate.
We chose a supply-dependent congestion function to ensure that, if an upstream supplier
fails, a retailer sourcing from that supplier does not incur additional penalties associated with
the failed product delivery.}.

\begin{changed}
    We use congestion function $\ell(x) = \tfrac{\gamma}{2} x^2$ due to its simplicity and
    strict convexity. The rationale for strict convexity is to model decreasing returns to
    scale in production. We assume that the production process is already well-established and, at best,
    scales linearly in the product quantity to be produced, yet, delivering higher product
    quantities may incur extra delays, potentially resulting from the need for sequential production due to the
    limited output capacity of manufacturing equipment, or transportation of extra product
    from additional warehouses, or the need to service production equipment due to its
    amortization.

    There is no consensus in literature on the form of the congestion function $\ell(x)$.
    For example, in~\citet{song2000contract}, the ``tardiness penalty'' depends linearly
    on a product of excess wait time and product quantity, making the penalty function
    super-linear in its input; while in~\citet{wang2009wait}, the lead time-related penalty
    is a linear function of the order fulfilment excess wait time. \citet{akan2012congestion}
    use a convex-concave lead time cost function: it is initially convex, modeling customers'
    increasing impatience prior to a deadline, and concave past the deadline. One can imagine
    a purely concave congestion function, modeling a supplier's gradual learning of more
    efficient methods of production while executing large orders.
    However, in environments where the congestion function models
    delays---or, equivalently, lead time uncertainty---convexity is a natural
    choice (see \citet[Sec.~4]{johari2010investment} and references therein).
\end{changed}

Next, we update the expression for the expected payoff.
\begin{proposition}[Retailer's Expected Payoff in the Model With Congestion]
	For an active retailer \begin{changed}$i \in \T_1^a$\end{changed},
\begin{align}
	\E[ \pi_{1, i} ] &= \E[ S_{1, i} \cdot p_1 - R_{1, i} \cdot p_2 - c \cdot \dout_{1, i} - L_{1, i} ] \notag \\[0.1in]
		&= \lambda (1 - \lambda) D (\lambda D ((1 + \lambda)n_1^a - \lambda) - \Delta)
			+ \frac{\lambda D^2}{\dout_{1, i}} \left( (1 - \lambda)^2 - \frac{\gamma}{2 \dout_{1, i}} \right) \notag\\
			&\phantom{==}- c \dout_{1, i} + \frac{\lambda D^2}{\dout_{1, i}} \sum\limits_{j \in \Nout_{1, i}}	
				{
					F_{2, j}^{-i} \left( (1 - \lambda)(1 - \lambda^2) - \frac{\gamma}{\dout_{1, i}} - \frac{\gamma}{2} F_{2, j}^{-i} \right)
				} \label{eq:expected-retailer-payoff-ex1}\\[0.1in]
		&= \lambda (1 - \lambda) D (\lambda D ((1 + \lambda)n_1^a - \lambda) - \Delta) \notag\\[0.1in]
			&\phantom{==}+ \frac{\lambda D^2}{\dout_{t, i}} \left( (1 - \lambda)^2 - \frac{\gamma}{2 \dout_{1, i}} + \left( (1 - \lambda)(1 - \lambda^2) - \frac{\gamma}{\dout_{1, i}} \right) \rout_{1, i} \right) \notag\\[0.1in]
			&\phantom{==}- c \dout_{1, i} - \frac{\lambda D^2 \gamma}{2\dout_{1, i}} \sum\limits_{j \in \Nout_{1, i}}	
				{
					{\Big(F_{2, j}^{-i}\Big)}^2
				}. \label{eq:expected-retailer-payoff-ex2}
\end{align}
Here $n_1^a \leq n$ is the number of active retailers, having at least one out-link. $\rout_{1, i} = \sum_{i' \in \T_1^a \setminus \{i\}}{\dout_{1, i' \cap i}~/~\dout_{1, i'}} = \sum_{i' \in \T_1^a \setminus \{i\}}{|\Nout_{1, i} \cap \Nout_{1, i'}|~/~\dout_{1, i'}}$ is the aggregate relative overlap of out-neighborhoods of active retailers with the out-neighborhood of $i$, and $F^{-i}_{2, j} = \sum_{ i' \in \Nin_{2, j} \setminus \{ i \} }{ 1 / \dout_{1,i'} }$
is the congestion at supplier $j \in \T_2$ excluding the contribution of retailer
$i \in \T_1$. If retailer $i$ is inactive, then, $\dout_{1, i} = 0$ and $\E[ \pi_{1, i} ] = 0$.
\label{thm:expected-payoff-ex}
\end{proposition}
\proof{Proof of Proposition~\ref{thm:expected-payoff-ex}}
	The proof is easily obtained by substituting the
	congestion penalty~\eqref{eq:latency}, the expression for supply~\eqref{eq:supply}, and the
	expected payoff for the model without congestion given in Proposition~\ref{thm:expected-retailer-payoff} into the above expression for $\E[\pi_{1, i}]$.
	\qed
\endproof

\subsection{\texorpdfstring{Analysis of the Model With $n = 2$ and $m = 2$}{Analysis of the Model With n = 2 and m = 2}}
\label{sec:general-model-w-congest-2x2}

\begin{figure}[ht]
	\centering
	% \vsa\vsa
	\includegraphics[width=3.5in]{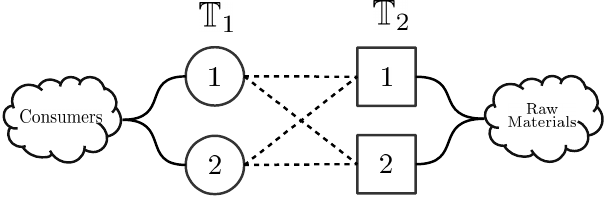}
	\caption{
		A small two-tier supply chain with two retailers $\T_1$, $|\T_1| = n = 2$, and two
		suppliers $\T_2$, $|\T_2| = m = 2$. The link that the retailers may create are
		displayed dashed.
	}
	\label{fig:small-supchain}
\end{figure}

In this section, we provide the analysis of the supply chain formation model with congestion,
described in Sec.~\ref{sec:general-model-def} in the case of two tiers, having two retailers
and two suppliers, as illustrated in Fig.~\ref{fig:small-supchain}. More general results for the
model with congestion appear in Sec.~\ref{sec:general-model-w-congest-analysis}.

We will be interested in which of the networks shown in Fig.~\ref{fig:small-supchain-equi-cands}
are equilibrium networks.
\newcommand{\equilcandfigwidth}[0]{0.15\linewidth}
\begin{figure}[ht]
	\centering
	\vsa\vsa
	\subfloat[empty]{\includegraphics[width=\equilcandfigwidth]{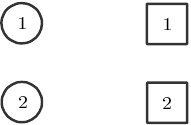}}
	\qquad
	\subfloat[cone]{\includegraphics[width=\equilcandfigwidth]{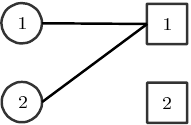}}
	\qquad
	\subfloat[parallel]{\includegraphics[width=\equilcandfigwidth]{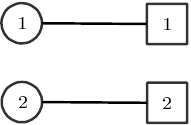}}
	\qquad
	\subfloat[zee]{\includegraphics[width=\equilcandfigwidth]{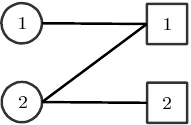}}
	\qquad
	\subfloat[full]{\includegraphics[width=\equilcandfigwidth]{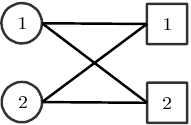}}
	\vspace{0.1in}
	\caption{
		Equilibrium network candidates. The empty network is always an equilibrium; other networks may
		be equilibria (potentially, simultaneously) in different regions of the parameter space.
	}
	\label{fig:small-supchain-equi-cands}
\end{figure}
The networks in Fig.~\ref{fig:small-supchain-equi-cands} exhaust the set of
equilibrium network candidates, up to agent labeling: (i) an empty network is always an equilibrium
for the same reason as in the case of the model without congestion (see
Theorem~\ref{thm:two-tier-empty-equil-no-congest})---even in the absence of the congestion penalty,
it is strictly preferred to the networks that can be obtained from it via unilateral deviations, so (ii) the
latter networks where only one retailer has links cannot be \begin{changed}equilibria\end{changed}; (iii) the zee-shaped network is
unique up to retailer labeling, and the cone network is unique up to supplier labeling.

\subsubsection{Payoffs}
We use
Proposition~\ref{thm:expected-payoff-ex} to compute a retailer's expected payoff in the candidate equilibrium networks.
\begin{proposition}[Retailer's Expected Payoff in 2x2 Candidate Networks]
	Retailers have the following expected payoffs in each of the candidate networks
	from Fig.~\ref{fig:small-supchain-equi-cands}:
	\begin{align*}
		\E[\pi^*_{1, i} \ecsempty ] &= 0,\\[0.1in]
		\E[\pi^*_{1, i} \ecscone ] &=
			\lambda (\lambda^3 - 2 \lambda^2 + 1 - \tfrac{7}{2} \gamma) D^2 - c,\\[0.1in]
		\E[\pi^*_{1, i} \ecspara ] &=
			\lambda(-\lambda^3 + 2\lambda - 1 - \tfrac{1}{2} \gamma) D^2 - c,\\[0.1in]
		\E[\pi^*_{1, 1} \ecszee ] &=
			\tfrac{1}{2} \lambda (-\lambda^3 - \lambda^2 + 3 \lambda - 1 - \tfrac{9}{4} \gamma) D^2 - c,\\[0.1in]
		\E[\pi^*_{1, 2} \ecszee ] &=
			\tfrac{1}{2} \lambda (-\lambda^3 - 2\lambda^2 + 5 \lambda - 2 - \tfrac{5}{4} \gamma) D^2 - 2c,\\[0.1in]
		\E[\pi^*_{1, i} \ecsfull ] &=
			\tfrac{1}{2} \lambda (-\lambda^3 - 2\lambda^2 + 5 \lambda - 2 - \gamma) D^2 - 2c.
	\end{align*}
	where $i \in \{1, 2\}$, $\lambda \in (0, 1)$ is the production success likelihood,
	$D$ is the consumer demand per retailer, and
	$c$ and $\gamma$ are the linking and the congestion costs, respectively.
\label{thm:small-equil-cands-expected-payoffs}
\end{proposition}
\proof{Proof of Proposition~\ref{thm:small-equil-cands-expected-payoffs}}
	The expressions for the payoffs as functions of consumer demand $D$ per retailer are
	obtained directly by specializing expression~\eqref{eq:expected-retailer-payoff-ex1} of a retailer's
	expected payoff from Proposition~\ref{thm:expected-payoff-ex} to the case of 2 retailers and 2 suppliers.
	\qed
\endproof

\subsubsection{Bounding Costs}
First, we determine the model parameters for which each of the candidate networks yields  non-negative expected payoffs for each retailer.

\begin{proposition}[Retailer Network Feasibility]
	For each of the candidate networks, retailer
	$i \in \{1, 2\} = \T_1$ enjoys non-negative expected payoff only within the
	following model parameter ranges:

	\hspace{0.15\linewidth}	
	\begin{minipage}{0.15\linewidth}
		\vspace{0.35in}
		\includegraphics[width=0.8\linewidth]{fig-2x2-chain-equil-cand-2-cone.pdf}
		\hspace{0.2\linewidth}
	\end{minipage}
	\begin{minipage}[c]{0.4\linewidth}
		\begin{align*}
			\begin{cases}
			\gamma~<~\gamma^{max}\ecsscone = \tfrac{2}{7} (1 - \lambda) (\tfrac{\sqrt{5} + 1}{2} - \lambda) (\tfrac{\sqrt{5} - 1}{2} + \lambda), \\[0.05in]
			\lambda~\in~(0, 1), \\[0.05in]
			c~\leq~c^{max}\ecsscone = \lambda (\lambda^3 - 2 \lambda^2 + 1 - \tfrac{7}{2} \gamma) D^2;
			\end{cases}
		\end{align*}
	\end{minipage}
	
	\hspace{0.15\linewidth}	
	\begin{minipage}{0.15\linewidth}
		\vspace{0.35in}
		\includegraphics[width=0.8\linewidth]{fig-2x2-chain-equil-cand-3-para.pdf}
		\hspace{0.2\linewidth}
	\end{minipage}
	\begin{minipage}[c]{0.4\linewidth}
		\begin{align*}
			\begin{cases}
			\gamma~<~\gamma^{max}\ecsspara = 2 (1 - \lambda) (\lambda + \tfrac{\sqrt{5} + 1}{2}) (\lambda - \tfrac{\sqrt{5} - 1}{2}), \\[0.05in]
			\lambda~\in~(\tfrac{\sqrt{5} - 1}{2}, 1) = (\lambda^{min}\ecsspara, 1), \\[0.05in]
			c~\leq~c^{max}\ecsspara = \lambda (-\lambda^3 + 2 \lambda^2 - 1 - \tfrac{1}{2} \gamma) D^2;
			\end{cases}
		\end{align*}
	\end{minipage}
	
	\hspace{0.15\linewidth}	
	\begin{minipage}{0.15\linewidth}
		\vspace{0.35in}
		\includegraphics[width=0.8\linewidth]{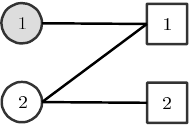}
		\hspace{0.2\linewidth}
	\end{minipage}
	\begin{minipage}[c]{0.4\linewidth}
		\begin{align*}
			\begin{cases}
			\gamma~<~\gamma^{max}\ecsszeeone = \tfrac{4}{9} (1 - \lambda) (\lambda + (\sqrt{2} + 1)) (\lambda - (\sqrt{2} - 1)), \\[0.05in]
			\lambda~\in~(\sqrt{2} - 1, 1) = (\lambda^{min}\ecsszeeone, 1), \\[0.05in]
			c~\leq~c^{max}\ecsszeeone = \tfrac{1}{2} \lambda (-\lambda^3 - \lambda^2 + 3 \lambda - 1 - \tfrac{9}{4} \gamma) D^2;
			\end{cases}
		\end{align*}
	\end{minipage}
	
	\hspace{0.15\linewidth}	
	\begin{minipage}{0.15\linewidth}
		\vspace{0.35in}
		\includegraphics[width=0.8\linewidth]{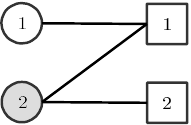}
		\hspace{0.2\linewidth}
	\end{minipage}
	\begin{minipage}[c]{0.4\linewidth}
		\begin{align*}
			\begin{cases}
			\gamma~<~\gamma^{max}\ecsszeetwo = \tfrac{4}{5} (1 - \lambda) (\lambda + \tfrac{\sqrt{17} + 3}{2}) (\lambda - \tfrac{\sqrt{17} - 3}{2}), \\[0.05in]
			\lambda~\in~(\tfrac{\sqrt{17} - 3}{2}, 1) = (\lambda^{min}\ecssfull, 1), \\[0.05in]
			c~\leq~c^{max}\ecssfull = \tfrac{1}{4} \lambda (-\lambda^3 - 2 \lambda^2 + 5 \lambda - 2 - \tfrac{5}{4} \gamma) D^2;
			\end{cases}
		\end{align*}
	\end{minipage}
	
	\hspace{0.15\linewidth}	
	\begin{minipage}{0.15\linewidth}
		\vspace{0.35in}
		\includegraphics[width=0.8\linewidth]{fig-2x2-chain-equil-cand-5-full.pdf}
		\hspace{0.2\linewidth}
	\end{minipage}
	\begin{minipage}[c]{0.4\linewidth}
		\begin{align*}
			\begin{cases}
			\gamma~<~\gamma^{max}\ecssfull = (1 - \lambda) (\lambda + \tfrac{\sqrt{17} + 3}{2}) (\lambda - \tfrac{\sqrt{17} - 3}{2}), \\[0.05in]
			\lambda~\in~(\tfrac{\sqrt{17} - 3}{2}, 1) = (\lambda^{min}\ecsszee, 1), \\[0.05in]
			c~\leq~c^{max}\ecsszeetwo = \tfrac{1}{4} \lambda (-\lambda^3 - 2 \lambda^2 + 5 \lambda - 2 - \gamma) D^2;
			\end{cases}
		\end{align*}
	\end{minipage}
\label{thm:equil-retail-payoff-and-cand-net-feasib}
\end{proposition}
\proof{Proof of Proposition~\ref{thm:equil-retail-payoff-and-cand-net-feasib}}		
	We verify the proposition for the cone network the proof for the other networks is similar.

	From Proposition~\ref{thm:small-equil-cands-expected-payoffs}, we have
	$$
	    \E[\pi^*_{1, i} \ecscone ] =
			\lambda (\lambda^3 - 2 \lambda^2 + 1 - \tfrac{7}{2} \gamma) D^2 - c.
	$$
	In this expression, if the coefficient of $D^2$ is non-positive, then
	$\E[\pi_{1, i} \ecscone ] < 0$ because we assumed that the consumer demand $D$ per
	retailer is positive. In the latter case, the cone network would not be a best response,
	as a retailer would prefer to drop its links, increasing its expected payoffs to zero.
	Hence, for a retailer to get non-negative expected payoff, that coefficient of $D^2$
	must be positive, resulting in
	$$
		\gamma < \tfrac{2}{7}(\lambda^3 - 2 \lambda^2 - 2 \lambda + 3)
			= \tfrac{2}{7} (1 - \lambda) (\tfrac{\sqrt{5} + 1}{2} - \lambda)(\tfrac{\sqrt{5} - 1}{2} + \lambda).
	$$
	(For the obtained upper bound to be well-defined, we \begin{changed}do not\end{changed} need additional restrictions on $\lambda \in (0, 1)$, though, in the proof for the other
	networks we must lower-bound $\lambda$ to ensure that the upper bound of
	$\gamma$ is non-negative.)
	
	Additionally, for the cone network to be an equilibrium candidate, we require
	$\E[\pi^*_{1, i} \ecscone ] \geq 0$, resulting in the upper bound for $c$
	$$
		c \leq \lambda (\lambda^3 - 2 \lambda^2 + 1 - \tfrac{7}{2} \gamma) D^2
		    = c^{max}_{\text{cone}}.
	$$
	\qed
\endproof

Similarly to how it was done for the model without congestion, we will assume that the parameters
of the model with congestion are such that the network with parallel links is feasible, that is, the
retailers are getting positive payoffs in such a network. The cost bounds in the following 
assumption follow directly from Proposition~\ref{thm:equil-retail-payoff-and-cand-net-feasib}.

\begin{assumption}[Bounding Costs in 2x2 Model With Congestion]
	Assume that the parameters of the supply chain network formation model with congestion
	are such, that, if the number of suppliers were at least as large as the number of retailers, 
	then, in the network with parallel links, every retailer would have a positive expected payoff.
	From Proposition~\ref{thm:equil-retail-payoff-and-cand-net-feasib}, this assumption holds \begin{changed}if and only if\end{changed}
	\begin{align*}
		\gamma &< \gamma^{max}\ecsspara = 2 (1 - \lambda) (\lambda + \tfrac{\sqrt{5} + 1}{2}) (\lambda - \tfrac{\sqrt{5} - 1}{2}) \qquad \text{and} \qquad% \\[0.05in]
		\lambda > \tfrac{\sqrt{5} - 1}{2} = \lambda^{min}\ecsspara.
	\end{align*}
\label{asm:parallel-network-feasibility-small-network}
\end{assumption}

\subsubsection{\texorpdfstring{Nash Equilibria When $\gamma > 0$ and $c = 0$}{Nash Equilibria When gamma > 0 and c is 0}}

In what follows, we analyze the small supply chain network formation model
with congestion assuming a negligible linking cost $c$. For now, we focus on how different combinations of $(\lambda, \gamma)$ affect retailers'
behavior.

\begin{figure}[ht]
	\centering
	\includegraphics[width=5in]{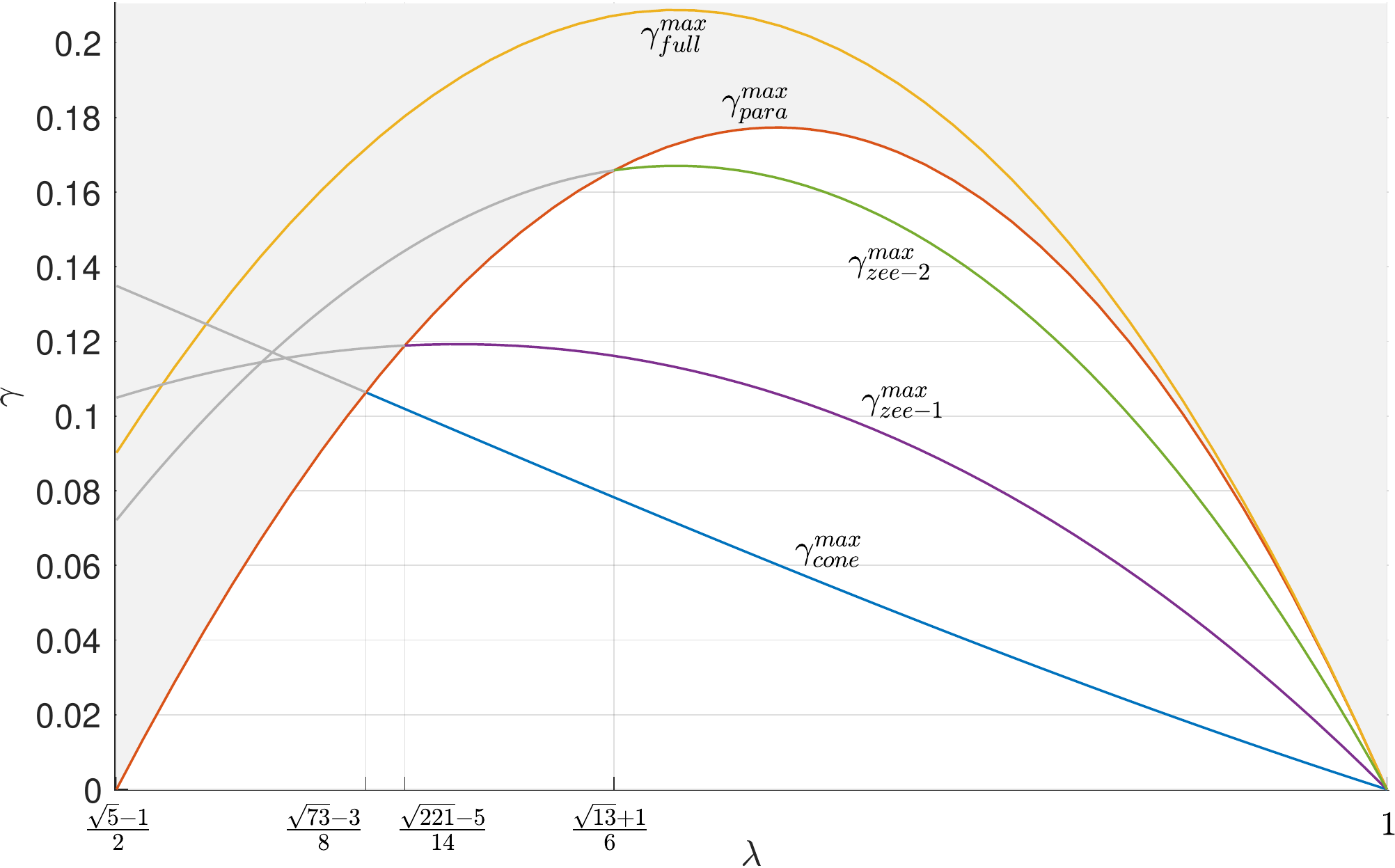}
	\caption{
		Parameter space for the model with congestion having 2 retailers and 2 suppliers.
		The feasibility regions of each equilibrium candidate network (for zee-shaped
		network---from the points of view of both retailers)---in which the corresponding
		retailers have positive expected payoffs---are enclosed between the horizontal axis,
		strictly below the curve $\gamma^{max}_{para}$ and the curve for the corresponding
		candidate network.
	}
	\label{fig:small-chain-param-space}
\end{figure}

In the light of Proposition~\ref{thm:equil-retail-payoff-and-cand-net-feasib} and
Assumption~\ref{asm:parallel-network-feasibility-small-network}, the relevant space of parameters is depicted in Fig.~\ref{fig:small-chain-param-space}. Due
to Assumption~\ref{asm:parallel-network-feasibility-small-network}, we are interested
only in the part of the parameter space under the curve $\gamma^{max}_{para}$.

\begin{figure}[ht]
	\centering
	\includegraphics[width=5in]{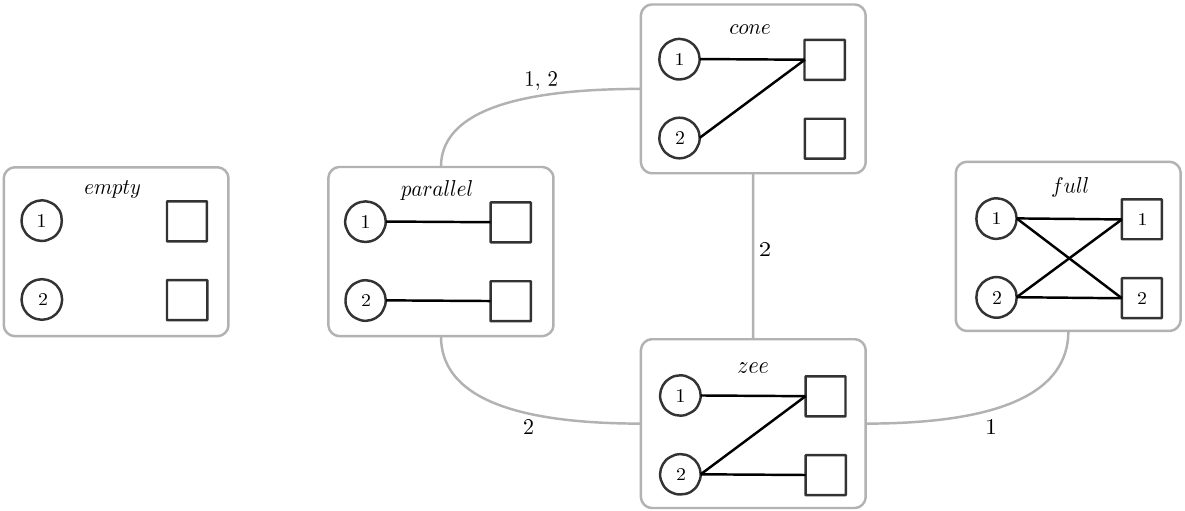}
	\caption{
		Possible unilateral deviations in a supply chain with 2 retailers and 2 suppliers.
		Links between networks indicate a possibility of a unilateral deviation by the retailers
		whose indices label that link.
	}
	\label{fig:small-chain-unilat-devs}
\end{figure}

In order to reason about when each of the candidate networks is an equilibrium,
let us outline the possible unilateral deviations in Fig.~\ref{fig:small-chain-unilat-devs}.
From Theorem~\ref{thm:two-tier-empty-equil-no-congest} we know that no unilateral deviation from an empty network can provide a non-negative
expected payoff to a retailer, so the empty network is isolated in Fig.~\ref{fig:small-chain-unilat-devs}, and, hence, is always an equilibrium. Other candidate networks may or may not be
equilibria depending on which of them are preferred by the retailers performing the corresponding
unilateral deviations. These latter preferences vary across the parameter space, as the following
proposition establishes.

\begin{proposition}[Retailers' Preference Over Equilibrium Network Candidates]
	Let us assume that linking cost $c$ is negligibly small, and define
	\begin{align*}
		\widehat{\gamma}_{fz1} &= \tfrac{4}{5} (1 - \lambda)^2, \quad % \\[0.1in]
		\widehat{\gamma}_{z2c} = \tfrac{4}{23} (1 - \lambda)^2 (3 \lambda + 4), \quad % \\[0.1in]
		\widehat{\gamma}_{pc} = \tfrac{2}{3} (1 - \lambda)^2 (\lambda + 1), \quad % \\[0.1in]
		\widehat{\gamma}_{pz2} = \lambda (1 - \lambda)^2.
	\end{align*}
	Then, for all $\lambda \in (\tfrac{\sqrt{5} - 1}{2}, 1)$,
	$
		\widehat{\gamma}_{fz1}(\lambda) < \widehat{\gamma}_{z2c}(\lambda)
			< \widehat{\gamma}_{pc}(\lambda) < \widehat{\gamma}_{pz2}(\lambda),
	$
	and for the different ranges of $\gamma$, retailers' preferences over networks are
	as shown in Fig.~\ref{fig:small-chain-net-pref-diag},
	\begin{figure}[h!]
		\centering
		\includegraphics[width=0.9\linewidth]{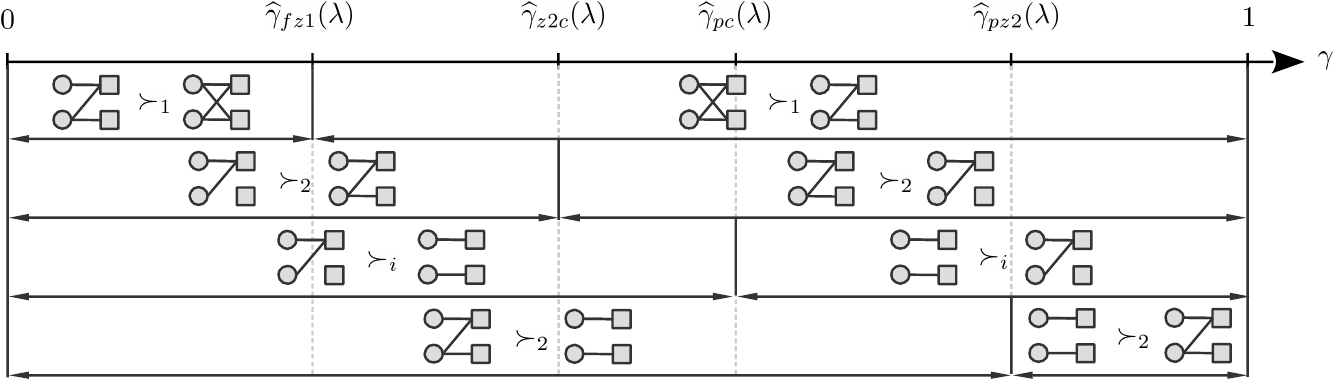}
		\caption{Retailers' preferences over equilibrium network candidates
		    for different congestion costs.}
		\label{fig:small-chain-net-pref-diag}
	\end{figure}

\noindent where $A \succ_i B$ indicates that $\E[\pi_{1, i} \big\rvert_{A}] > \E[\pi_{1, i} \big\rvert_{B}]$, that is, retailer $i \in \{1, 2\}$ strictly 	prefers network $B$ to network $A$, and there
	is a unilateral deviation via which $i$ can switch between $A$ and $B$.
	Non-strict preference $\succeq_i$ is defined analogously.
\label{thm:small-chain-pref-gamma-ranges}
\end{proposition}
\proof{Proof of Proposition~\ref{thm:small-chain-pref-gamma-ranges}}
	Expressions for
		$\widehat{\gamma}_{fz1}$,
		$\widehat{\gamma}_{z2c}$,
		$\widehat{\gamma}_{pc}$, and
		$\widehat{\gamma}_{pz2}$
	are obtained by solving equations
		$\E[\pi_{1, i} \ecsfull] = \E[\pi_{1, 1} \ecszee ]$,
		$\E[\pi_{1, 2} \ecszee] = \E[\pi_{1, i} \ecsfull ]$,
		$\E[\pi_{1, i} \ecspara] = \E[\pi_{1, i} \ecscone ]$, and
		$\E[\pi_{1, i} \ecspara] = \E[\pi_{1, 2} \ecszee ]$,
	respectively, under the assumption that linking cost $c$ can be dropped. In these equations,
	the expected payoffs are given in
	Proposition~\ref{thm:equil-retail-payoff-and-cand-net-feasib}. The rest is straightforward.
	\qed
\endproof

\begin{figure}[ht]
	\centering
	\includegraphics[width=5in]{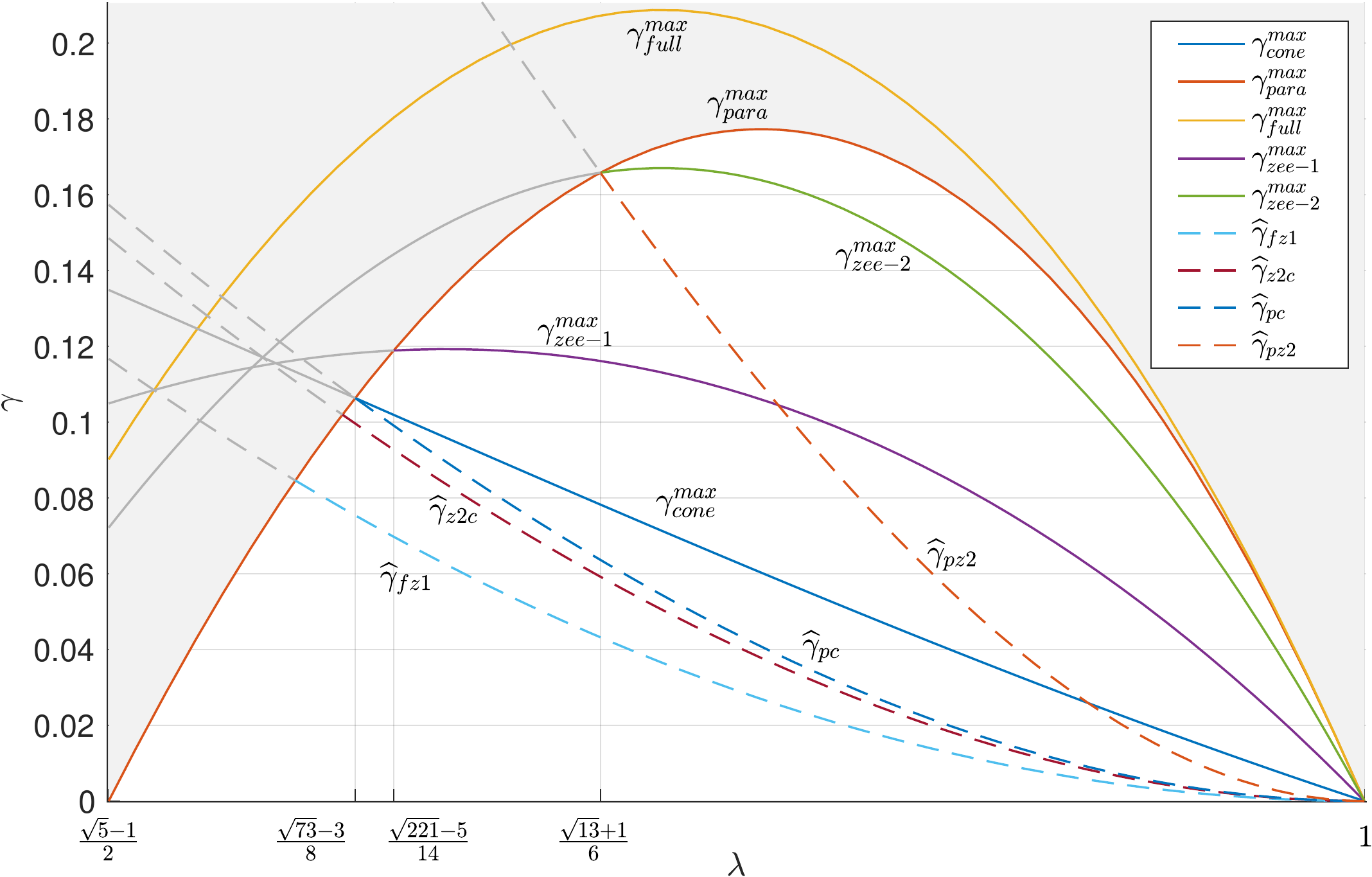}
	\caption{
		Parameter space for the model with congestion having 2 retailers and 2 suppliers.
		In addition to the upper bounds on $\gamma$ necessary for candidate network
		feasibility, we show thresholds $\widehat{\gamma}$ that affect the retailers'
		preferences over equilibrium candidate networks.
	}
	\label{fig:small-chain-param-space-w-gamma-hats}
\end{figure}

We, now, can augment the parameter space in~Fig.~\ref{fig:small-chain-param-space} with the
obtained thresholds $\widehat{\gamma}$. The result is shown in Fig.~\ref{fig:small-chain-param-space-w-gamma-hats}. 

Having information about both candidate network feasibility and retailers' preference over
them in different regions of the parameter space, we can  characterize equilibrium networks,
as we do in Theorem~\ref{thm:small-model-w-congest-equils} and
Fig.~\ref{fig:small-chain-param-space-w-equils}.

\begin{theorem}[Equilibria in 2x2 Model With Congestion and $c = 0$]
	Assume that the linking cost $c$ is negligibly small, and $\gamma^{max}_*$ and
	$\widehat{\gamma}_*$ are defined in
	Propositions~\ref{thm:equil-retail-payoff-and-cand-net-feasib} and
	Proposition~\ref{fig:small-chain-param-space-w-gamma-hats}, respectively.
	Then, in the supply chain network formation game with congestion (Definition~\ref{def:game})
	with 2 retailers and 2 suppliers, the following holds for the pure strategy Nash equilibria
	networks:
	\begin{enumerate}
		\item{
			An empty network is always an equilibrium.
		}
		\item{
			If $\tfrac{\sqrt{5} - 1}{2} < \lambda < 1$, and $0 \leq \gamma < \gamma^{max}_{para}$,
			and
			\begin{enumerate}
			\item $\gamma < \widehat{\gamma}_{fz1}(\lambda)$, then the \emph{cone} network is the unique non-empty
				equilibrium;
			\item $\widehat{\gamma}_{fz1}(\lambda) \leq \gamma \leq \widehat{\gamma}_{z2c}(\lambda)$,
				then the \emph{cone} and \emph{full} networks are the only non-empty equilibria;
			\item $\widehat{\gamma}_{z2c}(\lambda) < \gamma < \widehat{\gamma}_{pz2}(\lambda)$,
				then the \emph{full} networks is the unique non-empty equilibrium;
			\item $\gamma \geq \widehat{\gamma}_{pz2}(\lambda)$, then
				the \emph{parallel} and the \emph{full} networks are the only non-empty equilibria.
			\end{enumerate}
		}
		\item{
			If none of the above conditions is met, then the empty network is the only equilibrium.
		}
	\end{enumerate}
	\begin{figure}[ht]
		\centering
		\includegraphics[width=5in]{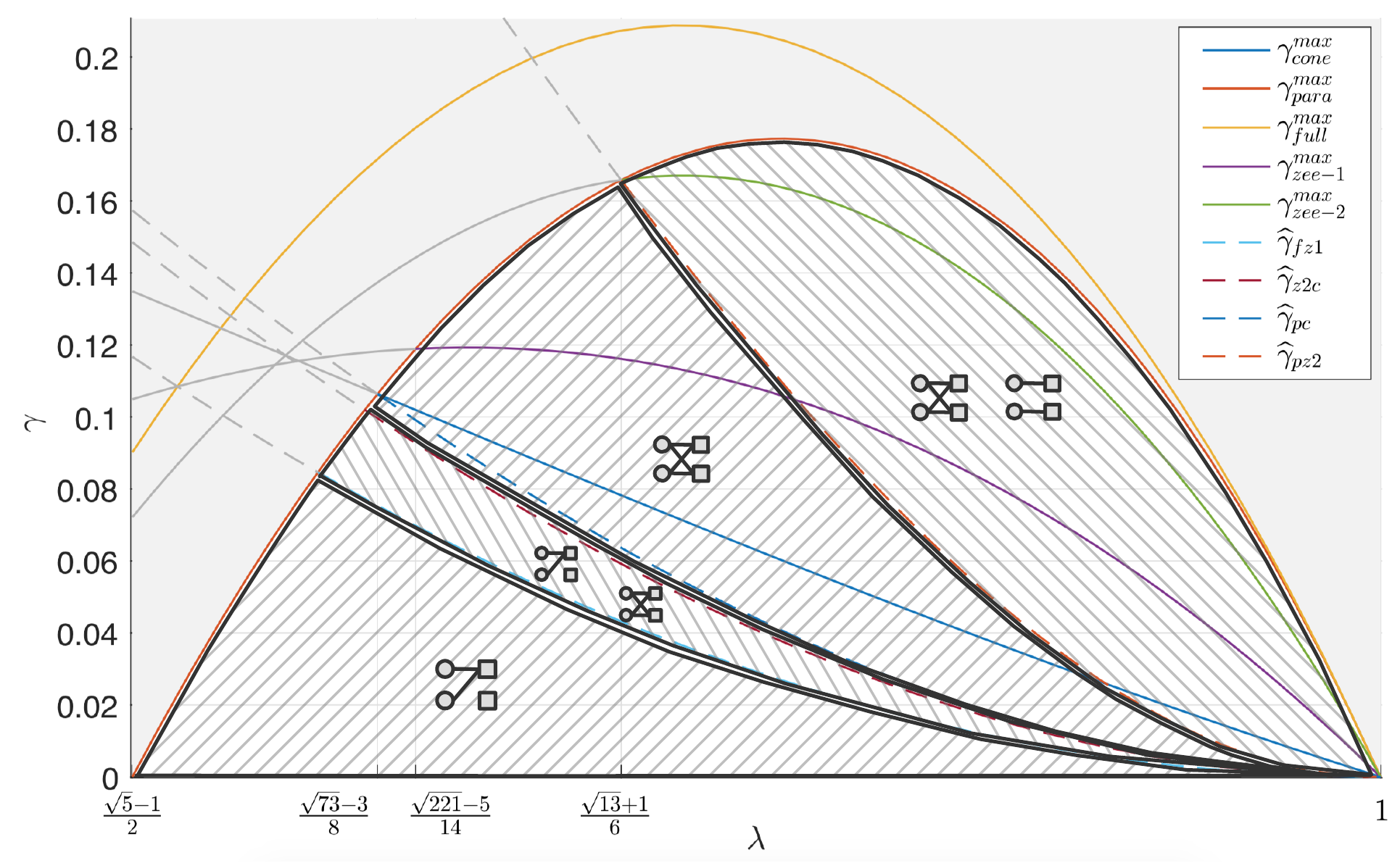}
		\caption{
			Equilibria networks in different parts of the parameter space for the model with congestion with 2 retailers and 2 suppliers, and a negligible linking cost $c \approx 0$.
		}
		\label{fig:small-chain-param-space-w-equils}
	\end{figure}
	\label{thm:small-model-w-congest-equils}
\end{theorem}
\proof{Proof of Theorem~\ref{thm:small-model-w-congest-equils}}
	To characterize equilibria networks in different parts of the model parameter space---shown in
	Fig.~\ref{fig:small-chain-param-space-w-gamma-hats}---we will rely on
	Proposition~\ref{thm:equil-retail-payoff-and-cand-net-feasib} that
	provides us with retailers' expected payoffs in a best response as well as the conditions
	for when that payoff is non-negative, as well as on Proposition~\ref{thm:small-chain-pref-gamma-ranges} that
	establishes the retailers' preference over equilibrium network candidates.
	
	According to Assumption~\ref{asm:parallel-network-feasibility-small-network}, and Proposition~\ref{thm:equil-retail-payoff-and-cand-net-feasib} characterizing the necessary condition for the assumption to hold, we are interested only in the
	region of the parameter space strictly\footnote{The case $\gamma = 0$ corresponds to the model without
	congestion, which is the topic of study in Sec.~\ref{sec:supchain-formation}.} above the horizontal axis
	and strictly below curve $\gamma^{max}_{\text{para}}(\lambda)$, which also implies a lower bound on
	$\lambda$:
	\begin{align*}
			0 <~&\gamma < 2 (1 - \lambda) (\lambda + \tfrac{\sqrt{5} + 1}{2}) (\lambda - \tfrac{\sqrt{5} - 1}{2}), \\[0.05in]
			\tfrac{\sqrt{5} - 1}{2} <~&\lambda < 1.
	\end{align*}
	
\begin{figure}[ht!]
			\centering
			\includegraphics[width=3in]{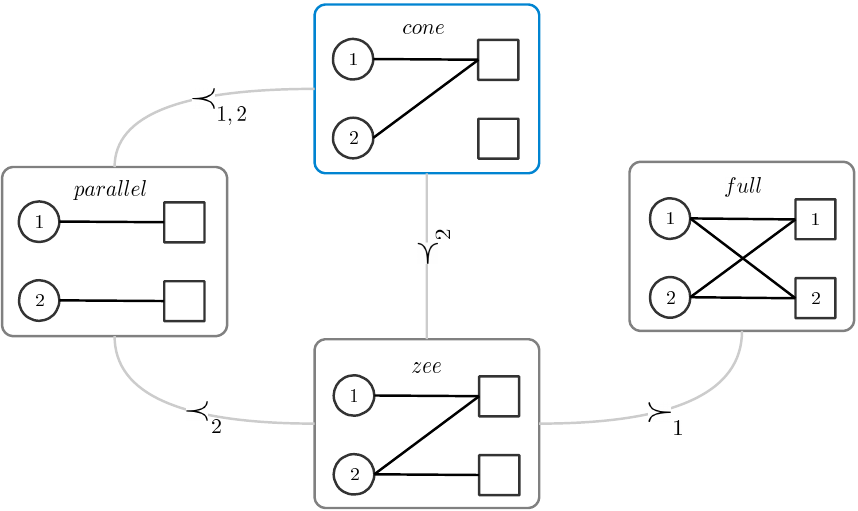}
			\caption{
				Retailers' preference over networks when $\gamma < \widehat{\gamma}_{fz1}(\lambda)$.
				$A \succ_i B$ indicates that in network $A$, 
				retailer $i$ has a strictly larger expected payoff than in network $B$.
				Each network is feasible (retailers have non-negative expected payoffs in each of them).
				The cone network is an equilibrium.
			}
			\label{fig:net-preference-diag-slice-one}
		\end{figure}

	In the above defined region of the parameter space, we will focus on 5 parts that curves
	$\widehat{\gamma}_{fz1}$, $\widehat{\gamma}_{z2c}$, $\widehat{\gamma}_{pc}$, and
	$\widehat{\gamma}_{pz2}$ slice the region into:
	
	\vspace{0.08in}
	\noindent \emph{1) $\gamma < \widehat{\gamma}_{fz1}(\lambda)$:} According to Proposition~\ref{thm:equil-retail-payoff-and-cand-net-feasib},
		all the non-empty candidate networks are feasible here (the retailers are getting a non-negative payoff), and,
		based on Proposition~\ref{thm:small-chain-pref-gamma-ranges}, the retailers' preferences over networks
		are as follows:
		\begin{align*}
			 \eczee~\succ_2~\ecpara, \qquad
			 \eczee~\succ_1~\ecfull, \qquad
			 \eccone~\succ_{1,2}~\ecpara, \qquad
			 \eccone~\succ_{2}~\eczee
		\end{align*}
		The same relationships summarized as a diagram in Fig.~\ref{fig:net-preference-diag-slice-one}.
		Thus, the only non-empty network from which no retailer wants to unilaterally deviate
		is the cone network, making it the unique non-empty equilibrium network up to supplier labeling.
		
\begin{figure}[h!]
			\centering
			\includegraphics[width=3in]{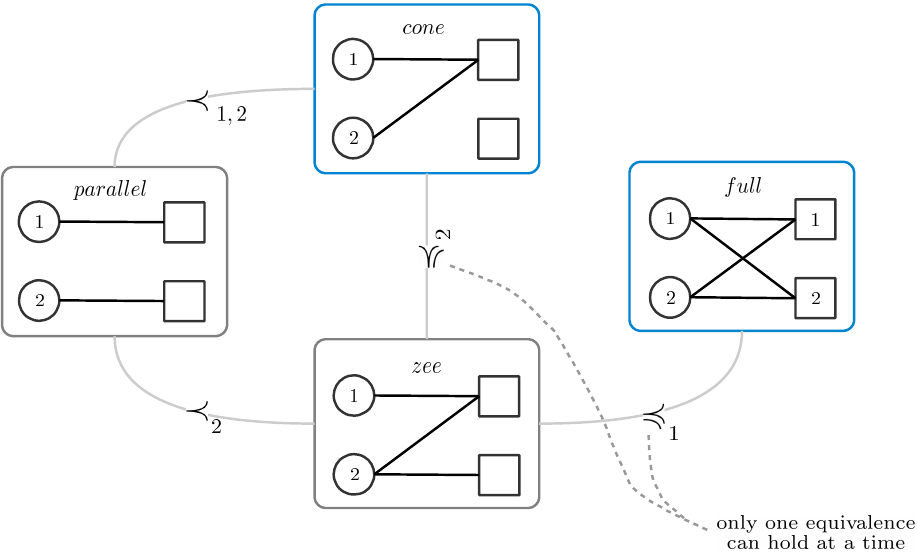}
			\caption{
				Retailers' preference over networks when
				$\widehat{\gamma}_{fz1}(\lambda) \leq \gamma \leq \widehat{\gamma}_{z2c}(\lambda)$.
				$A \succ_i B$ indicates that in network $A$, 
				retailer $i$ has a strictly larger expected payoff than in network $B$,
				and $A \succcurlyeq_i B$ allows for expected payoff equality. Among the two
				$\succcurlyeq_i$ relationships, equality can hold only for one of them at a
				time. All networks are feasible, and the cone and full networks are equilibria.
			}
			\label{fig:net-preference-diag-slice2}
		\end{figure}

	\vspace{0.08in}
	\noindent \emph{2) $\widehat{\gamma}_{fz1} \leq \gamma \leq \widehat{\gamma}_{z2c}$:} Proceeding similarly
		to the previous case, we end up with the relationships between the candidate networks shown
		in Fig.~\ref{fig:net-preference-diag-slice2}.
		While all networks are feasible, only in the cone and the full networks, retailers prefer
		not to unilaterally deviate. Notice that, out of two non-strict preference
		relations---$\eccone~\succcurlyeq_{2}~\eczee$ and
		$\ecfull~\succcurlyeq_{1}~\eczee$---equivalence can hold in one of them at
		a time (either when $\gamma = \widehat{\gamma}_{fz1}(\lambda)$ or when
		$\gamma = \widehat{\gamma}_{z2c}(\lambda)$, which cannot hold simultaneously
		in $\lambda \in (\tfrac{\sqrt{5} - 1}{2}, 1)$), which is why zee-shaped network is
		not an equilibrium.

\begin{figure}[h!]
			\centering
			\includegraphics[width=3in]{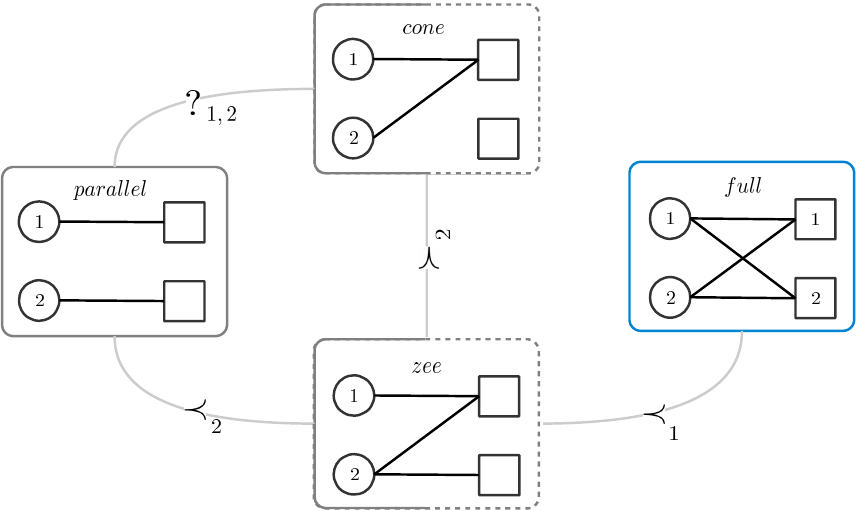}
			\caption{
				Retailers' preference over networks when
				$\widehat{\gamma}_{z2c}(\lambda) < \gamma < \widehat{\gamma}_{pz2}(\lambda)$.
				The retailers' expected payoffs are non-negative in the cone network only in the
				part of the region, and so is the expected payoff of retailer $1$ in the zee-shaped
				network; also the relationship between the cone and the parallel networks change
				within the region.
			}
			\label{fig:net-preference-diag-slice34}
		\end{figure}

	\vspace{0.08in}
	\noindent \emph{3) $\widehat{\gamma}_{z2c}(\lambda) < \gamma < \widehat{\gamma}_{pz2}(\lambda)$}:
		The retailers' preferences over the candidate networks are shown in Fig.~\ref{fig:net-preference-diag-slice34}.
		Here, the feasibility of networks varies across the region, and the cone network's
		relationship with the parallel network also varies. However, this does not affect the
		full network's being the only non-empty equilibrium in this region of the parameter
		space.

\begin{figure}[h!]
			\centering
			\includegraphics[width=3in]{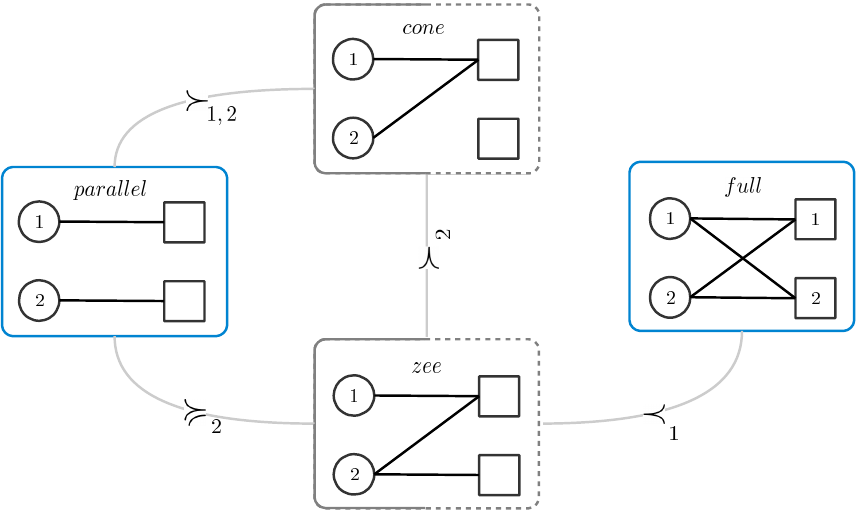}
			\caption{
				Retailers' preference over networks when
				$\widehat{\gamma}_{pz2}(\lambda) \leq \gamma < \gamma^{max}_{\text{para}}(\lambda)$. Feasibility of zee and cone networks varies across the region. Parallel and full networks
				are unique non-empty equilibria.
			}
			\label{fig:net-preference-diag-slice5}
		\end{figure}

	\vspace{0.08in}
	\noindent \emph{4)
	$\widehat{\gamma}_{pz2}(\lambda) \leq \gamma < \gamma^{max}_{\text{para}}(\lambda)$:}
		In the last slice of the parameter space partition, the retailers' preferences over
		the candidate networks are shown in Fig.~\ref{fig:net-preference-diag-slice5}.
		Here, the parallel and the full network are the only non-empty equilibria.

	\qed
\endproof

\subsubsection{\texorpdfstring{Nash Equilibria When $\gamma > 0$ and $c > 0$}{Nash Equilibria When gamma > 0 and c > 0}}
\begin{figure}[ht]
	\centering
	\subfloat[$c = 0$]{\includegraphics[width=0.46\linewidth]{fig-2x2-chain-param-space-w-gamma-hats/fig-2x2-chain-param-space-w-gamma-hats-v1.pdf}}
	\quad
	\subfloat[$c > 0$]{\includegraphics[width=0.46\linewidth]{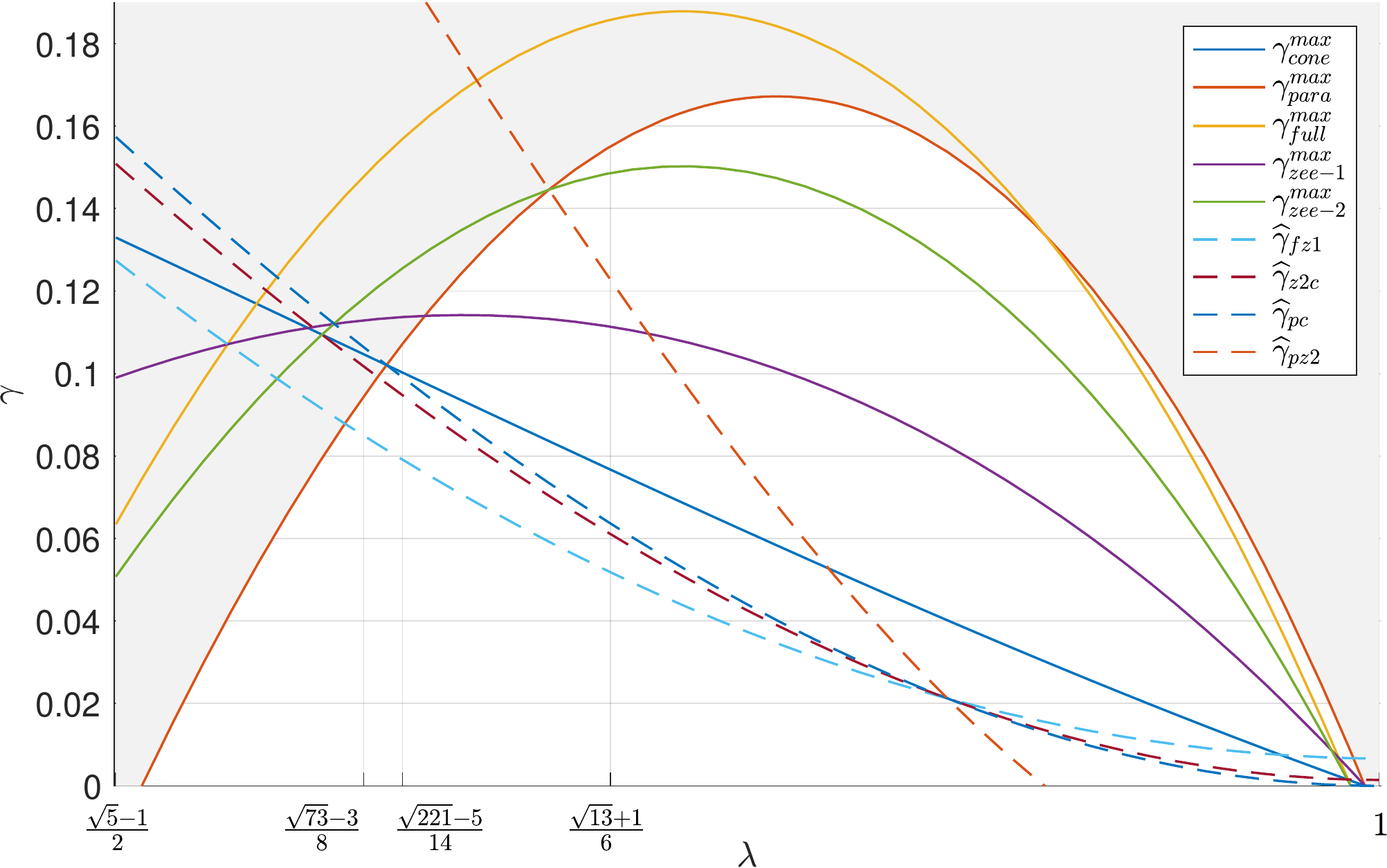}}
	\caption{
		Parameter space of the model with congestion with two retailers and suppliers,
		with zero and positive linking cost.
	}
	\label{fig:small-chain-param-space-w-gamma-hats-pos-c}
\end{figure}

\begin{figure}[ht]
	\centering
	\includegraphics[width=0.5\linewidth]{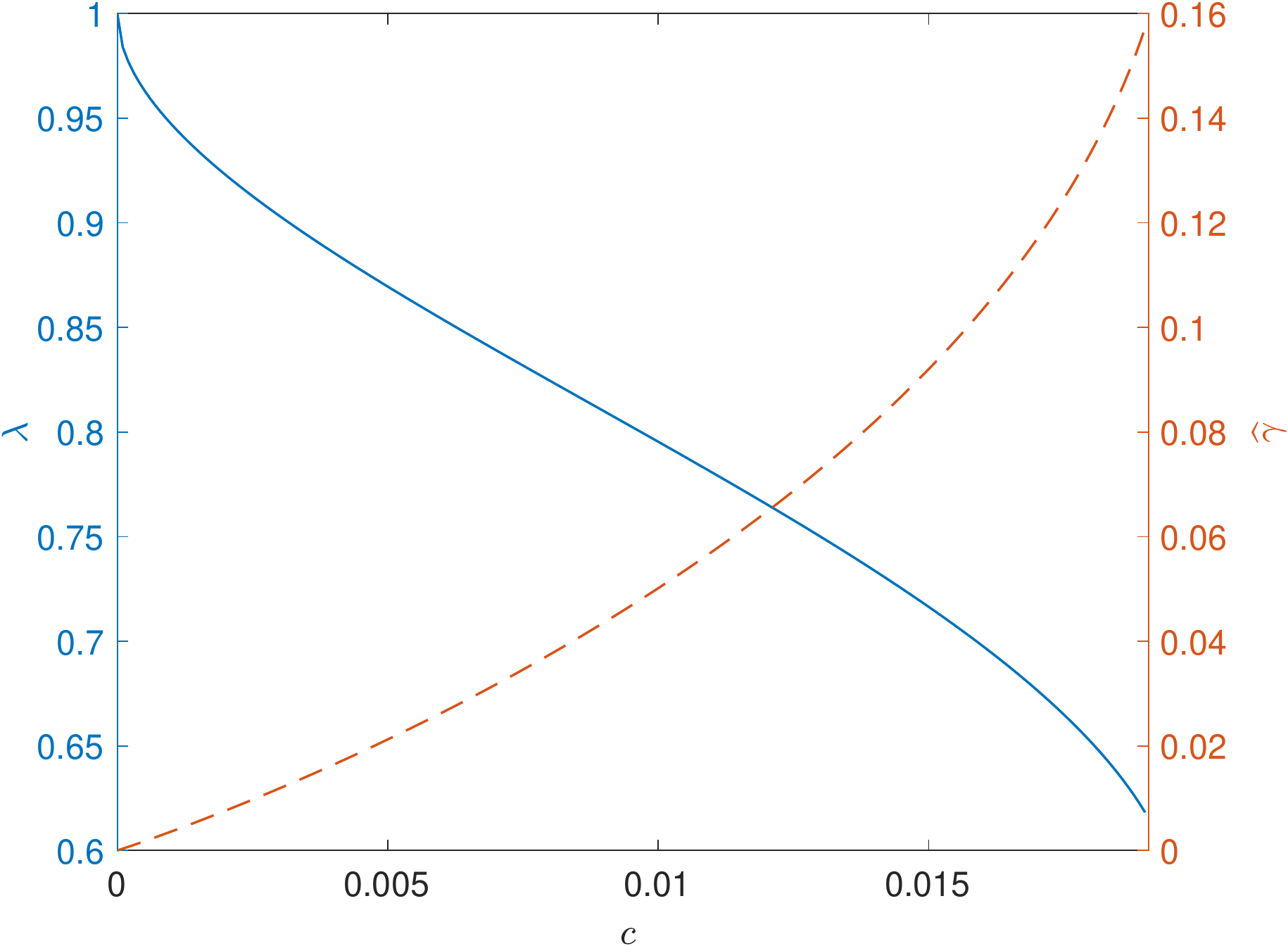}
	\caption{
		The location of the intersection point of the curves
		$\widehat{\gamma}_{fz1}(\lambda, c)$, $\widehat{\gamma}_{z2c}(\lambda, c)$,
		$\widehat{\gamma}_{pc}(\lambda, c)$, and $\widehat{\gamma}_{pz2}(\lambda, c)$.
	}
	\label{fig:2x2-chain-gamma-hat-intersect-location}
\end{figure}

\begin{figure}[ht]
	\centering
	\includegraphics[width=0.75\linewidth]{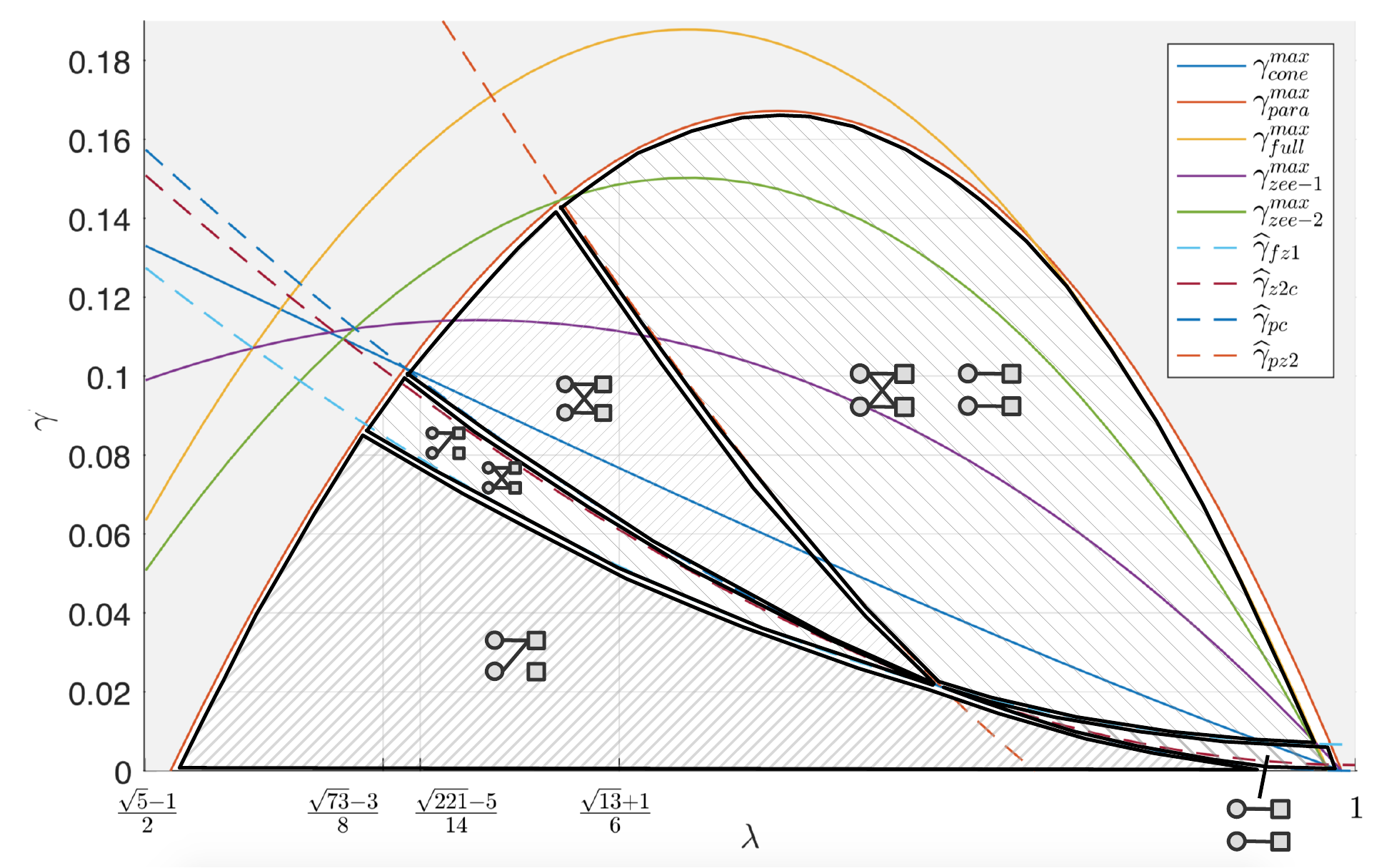}
	\caption{
		Equilibrium networks in different parts of the parameter space for the model with congestion
		having 2 retailers and 2 suppliers, and a \emph{small} positive linking cost $c$.
	}
	\label{fig:small-chain-param-space-w-equils-pos-c}
\end{figure}

 In this section, we characterize equilibrium networks
when $c > 0$. The qualitative changes in the parameter space partitioning are
depicted in Fig.~\ref{fig:small-chain-param-space-w-gamma-hats-pos-c}.
It is easy to see that the introduction of positive linking cost $c$ results in a shift of curves $\widehat{\gamma}_{z2c}(\lambda)$ and $\widehat{\gamma}_{pz2}(\lambda)$. However, there are two critical
things to notice:
\begin{enumerate}
	\item[(i)]{
		all such curves intersect at the same point, whose location in the parameter space is described
		in Fig.~\ref{fig:2x2-chain-gamma-hat-intersect-location}; and
	}
	\item[(ii)]{
		the order of the curves $\widehat{\gamma}$ on each side of that intersection point
		is the same, regardless of the value of $c$.
	}
\end{enumerate}
Thus, following the reasoning from the proof of
Theorem~\ref{thm:small-model-w-congest-equils}, it is easy to generalize the latter's statement
to the case of positive linking cost $c$. To aid understanding, we provide this generalization
here informally: Fig.~\ref{fig:small-chain-param-space-w-equils-pos-c} shows how equilibrium
networks change when the linking cost $c$ is positive but not large.

\begin{figure}[ht]
	\centering
	\includegraphics[width=0.75\linewidth]{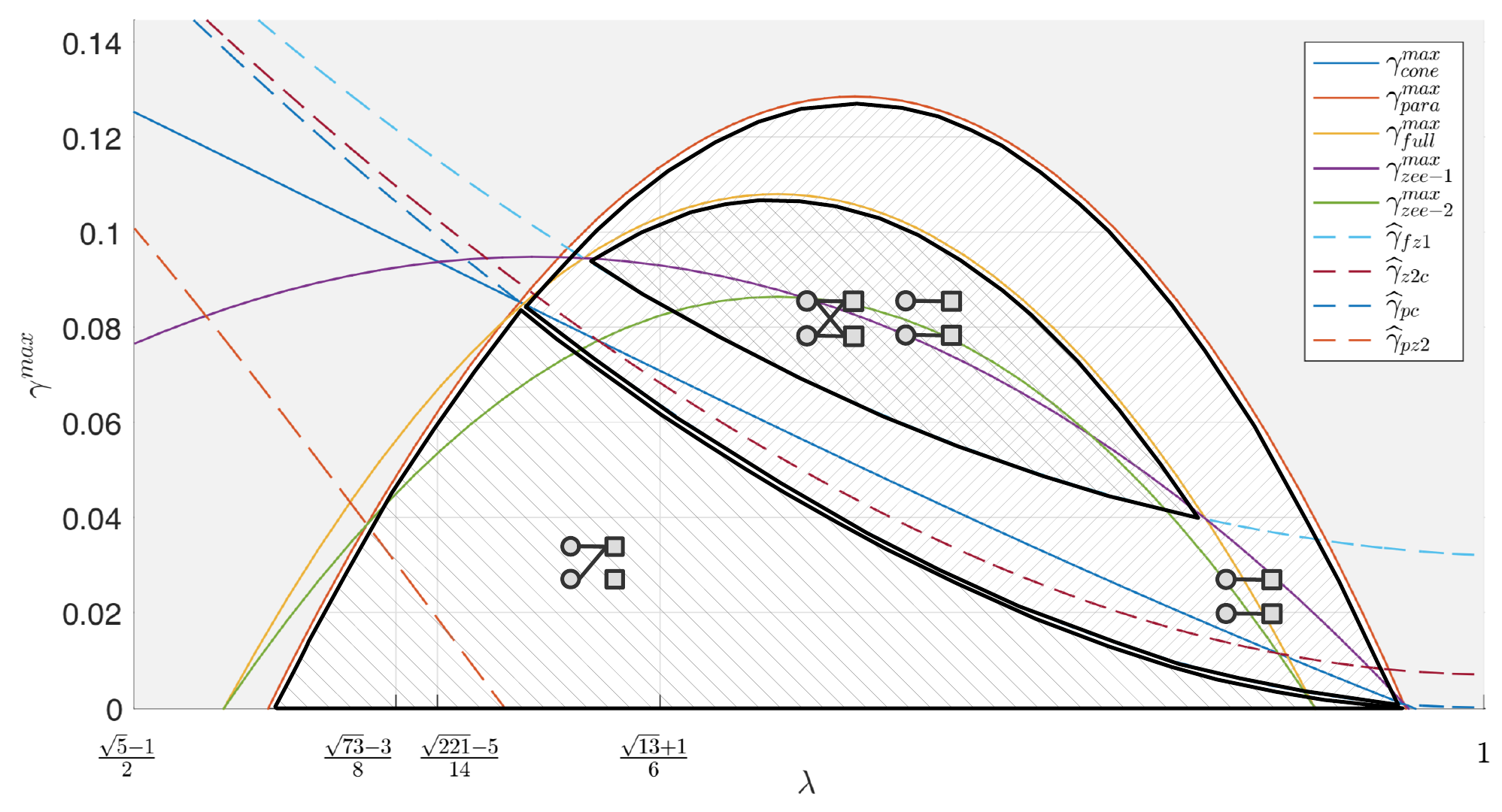}
	\caption{
		Equilibria networks in different parts of the parameter space for the model with congestion
		having 2 retailers and 2 suppliers, and a positive linking cost $c$ of \emph{larger}
		magnitude.
	}
	\label{fig:small-chain-param-space-w-equils-pos-c-large}
\end{figure}

Besides $\widehat{\gamma}$ curves' shifting with growing $c$, the network feasibility regions---outlined
by curves $\gamma^{max}$---shrink, and the feasibility regions of denser networks (only
the full network for the case of 2 retailers and suppliers) shrink faster than those of sparser
networks.
As a result, when $c$ grows further, the equilibria network distribution over the parameter space
changes as shown in Fig.~\ref{fig:small-chain-param-space-w-equils-pos-c-large}.

As $c$ grows even larger, first, the island region in which the full network is an equilibrium gradually
disappears; then the region where the parallel network is feasible shrinks to a point (it does not
disappear because of
Assumption~\ref{asm:parallel-network-feasibility-small-network}).

We conclude the discussion of the model with 2 retailers and suppliers with
Table~\ref{tbl:small-chain-payoffs-at-equil} showing the values of expected payoff of a retailer
at equilibrium for the candidate networks when the linking cost $c$ grows. We see that the
positive values of the expected payoffs outline the earlier defined regions within which different
networks may be equilibria.

\newcommand{\xotext}[1]{\footnotesize #1}
\newcommand{\payatequilfigw}[0]{0.27\linewidth}
\begin{table}[ht]
	\setlength{\fboxsep}{4pt}
	\setlength{\fboxrule}{0pt}
	\centering
	\begin{tabular}{|c|c|c|c|}
	 	\hline
	 	$c$ & \framebox{\ecconebig} & \framebox{\ecfullbig} & \framebox{\ecparabig}\\ \hline
		\raisebox{0.5in}{$0.000$} &
		\framebox{\vspace{0.2in}\includegraphics[width=\payatequilfigw]{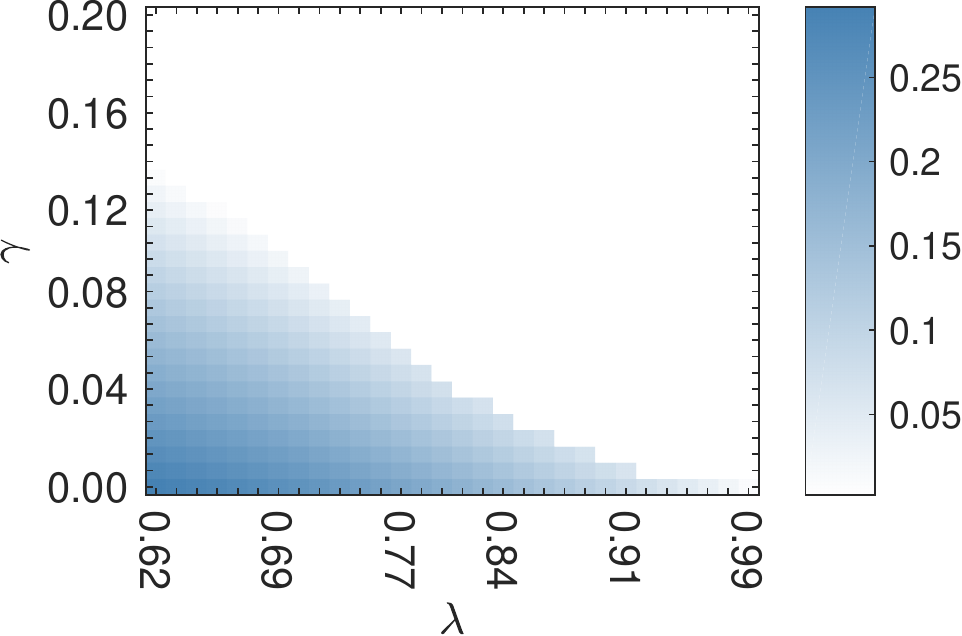}} &
		\framebox{\vspace{0.2in}\includegraphics[width=\payatequilfigw]{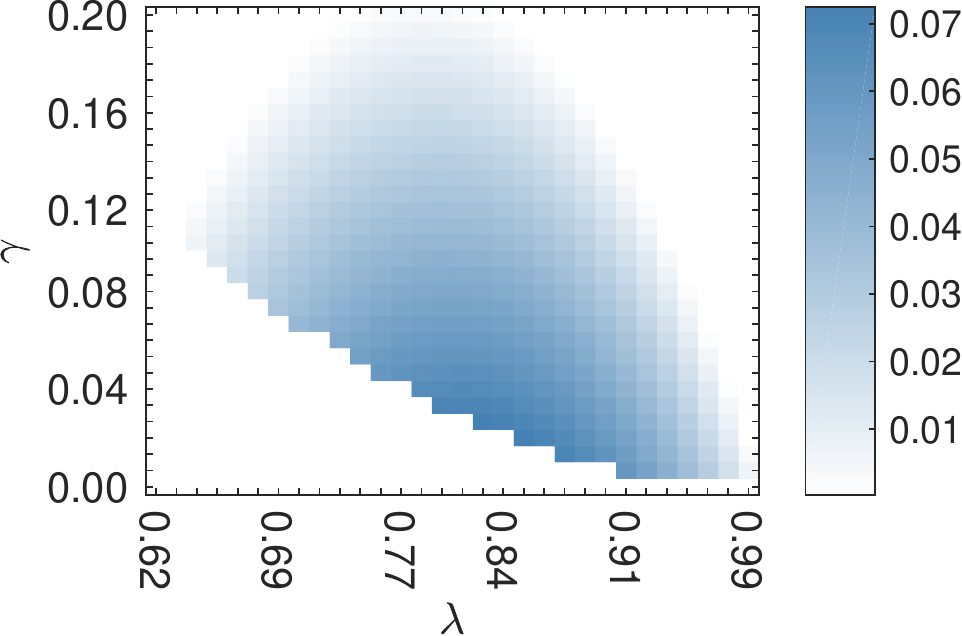}} &
		\framebox{\vspace{0.2in}\includegraphics[width=\payatequilfigw]{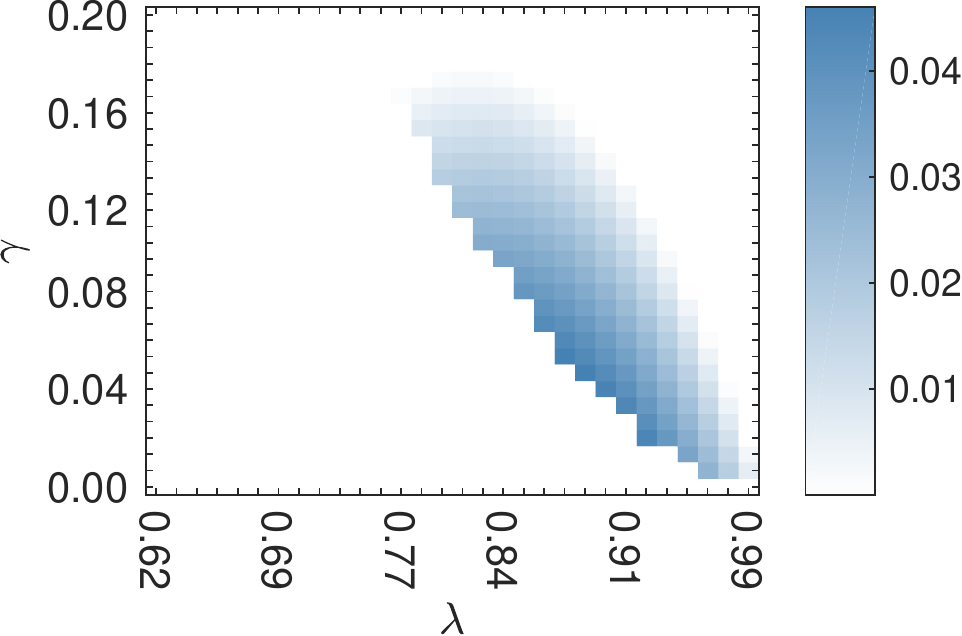}} \\ \hline
		\raisebox{0.5in}{$0.002$} &
		\framebox{\vspace{0.2in}\includegraphics[width=\payatequilfigw]{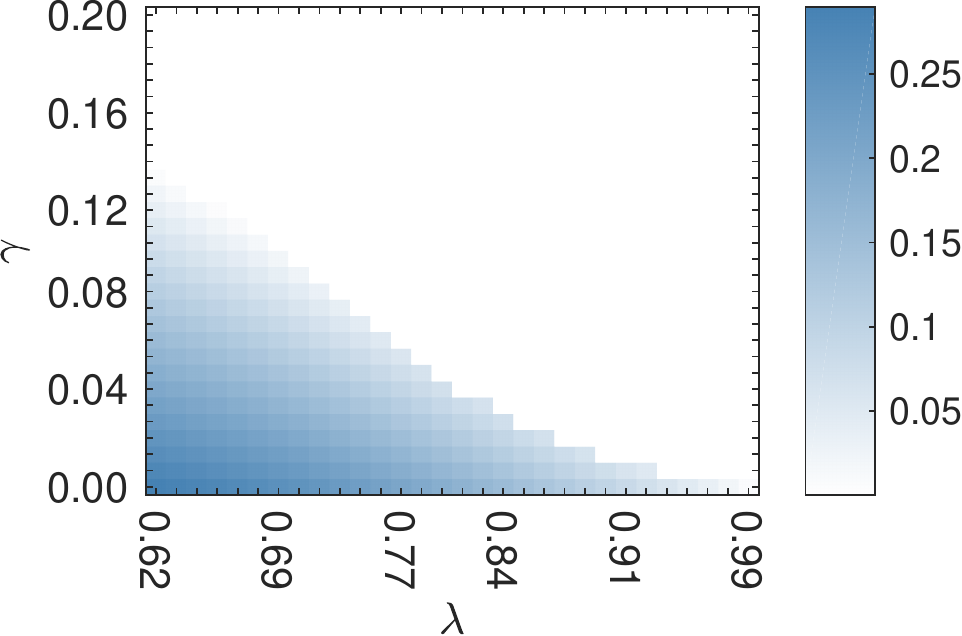}} &
		\framebox{\vspace{0.2in}\includegraphics[width=\payatequilfigw]{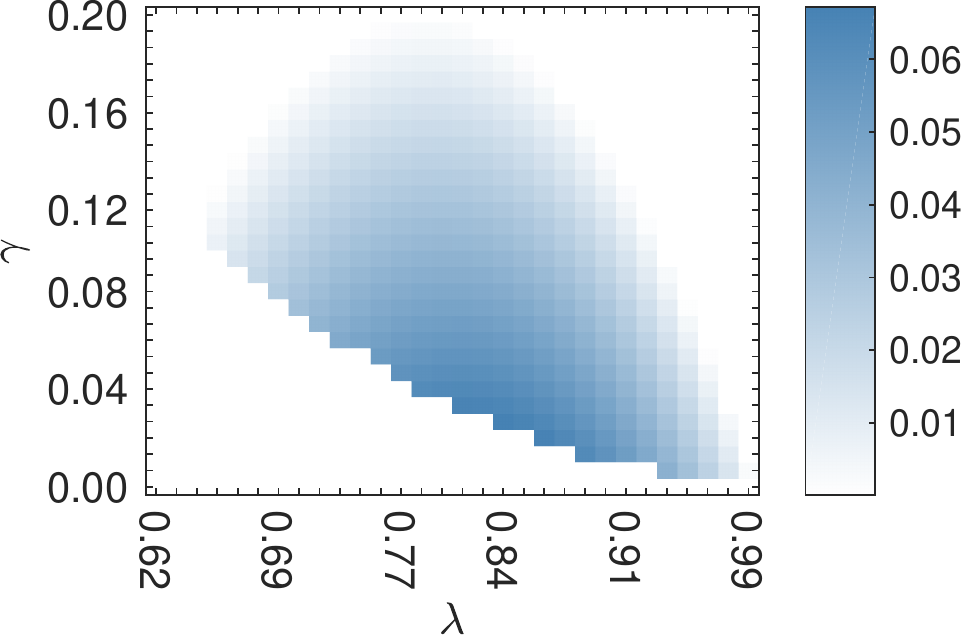}} &
		\framebox{\vspace{0.2in}\includegraphics[width=\payatequilfigw]{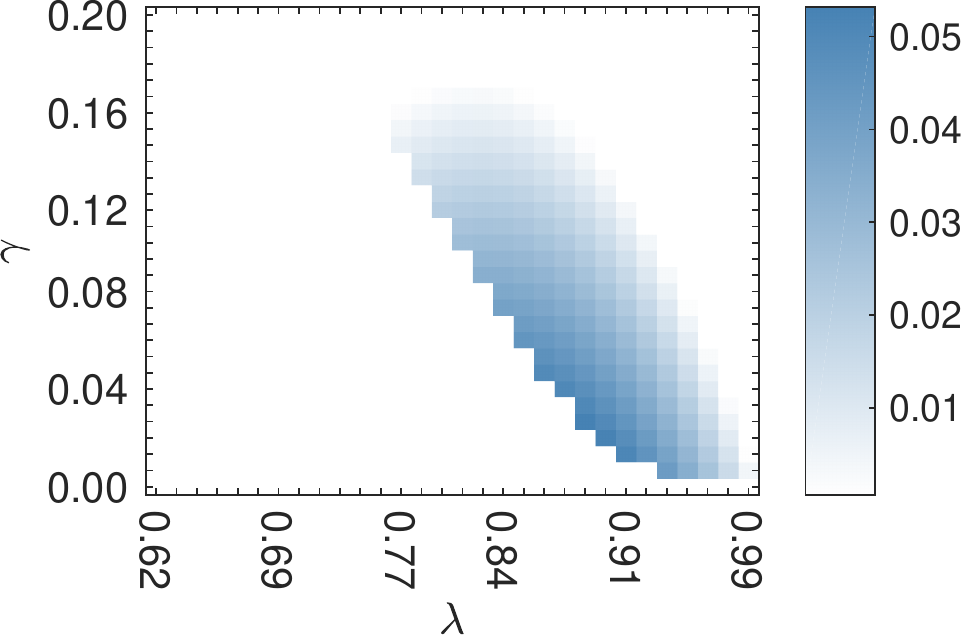}} \\ \hline
		\raisebox{0.5in}{$0.010$} &
		\framebox{\vspace{0.2in}\includegraphics[width=\payatequilfigw]{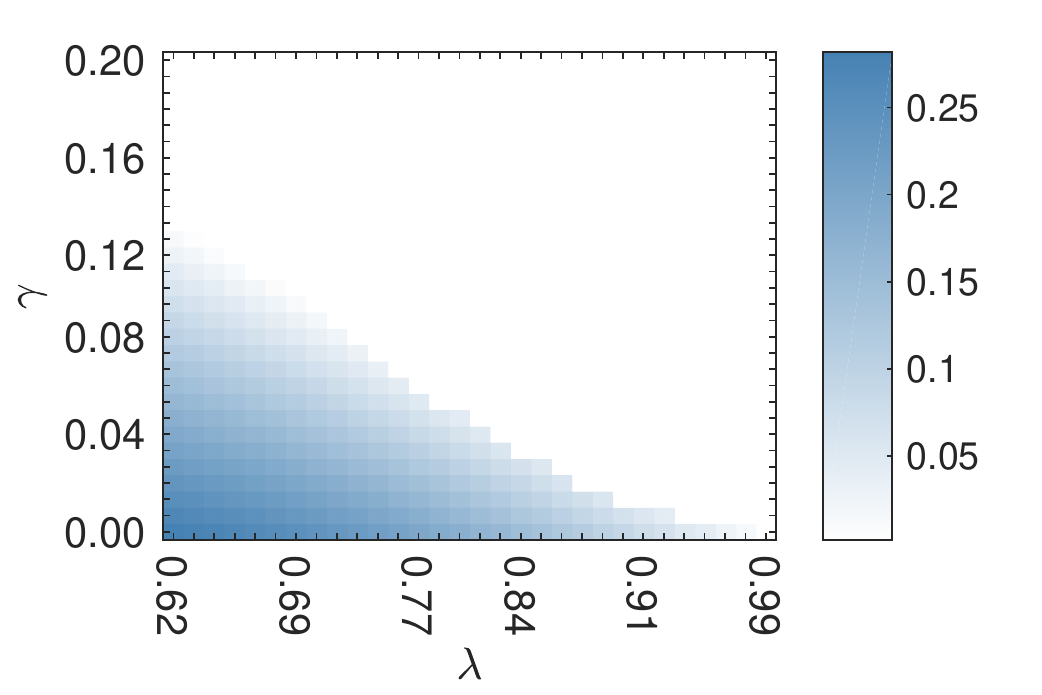}} &
		\framebox{\vspace{0.2in}\includegraphics[width=\payatequilfigw]{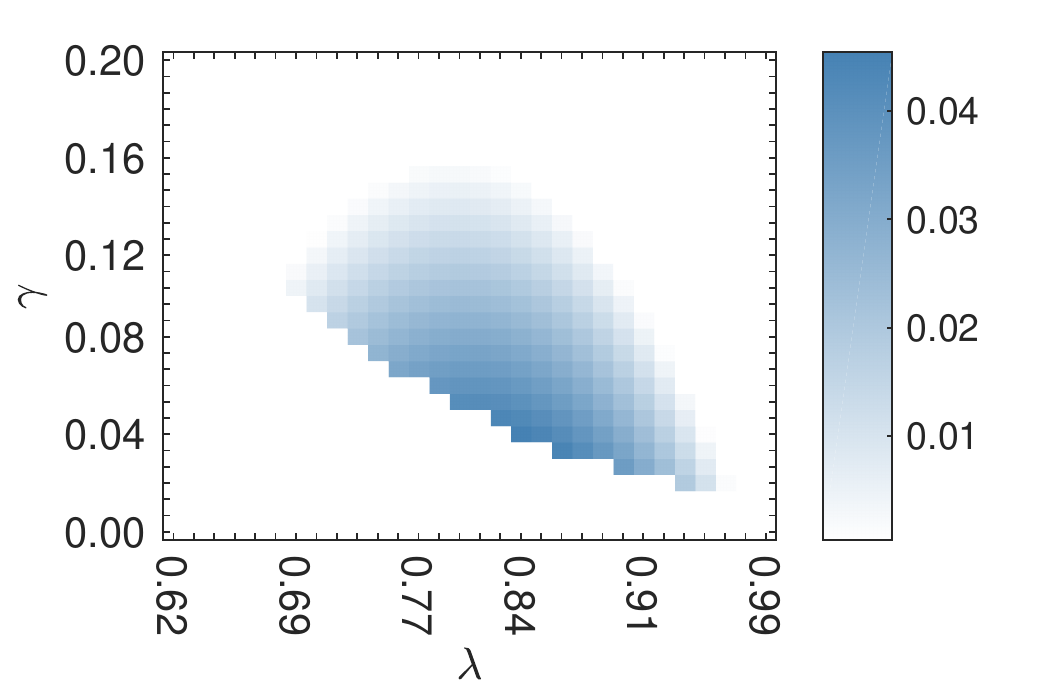}} &
		\framebox{\vspace{0.2in}\includegraphics[width=\payatequilfigw]{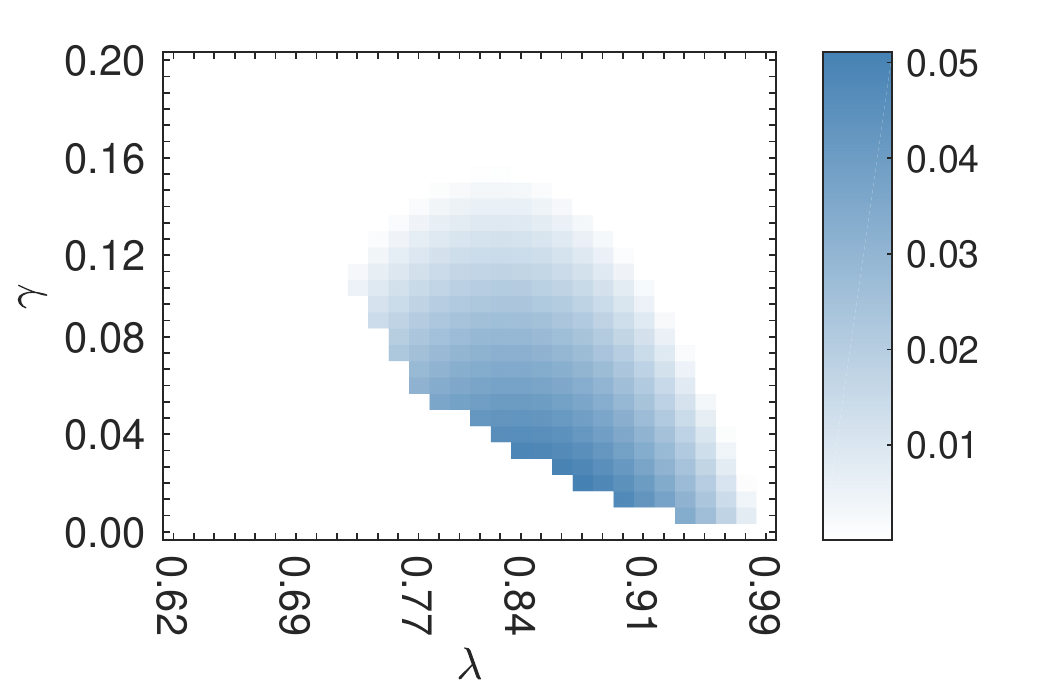}} \\ \hline
		\raisebox{0.5in}{$0.015$} &
		\framebox{\vspace{0.2in}\includegraphics[width=\payatequilfigw]{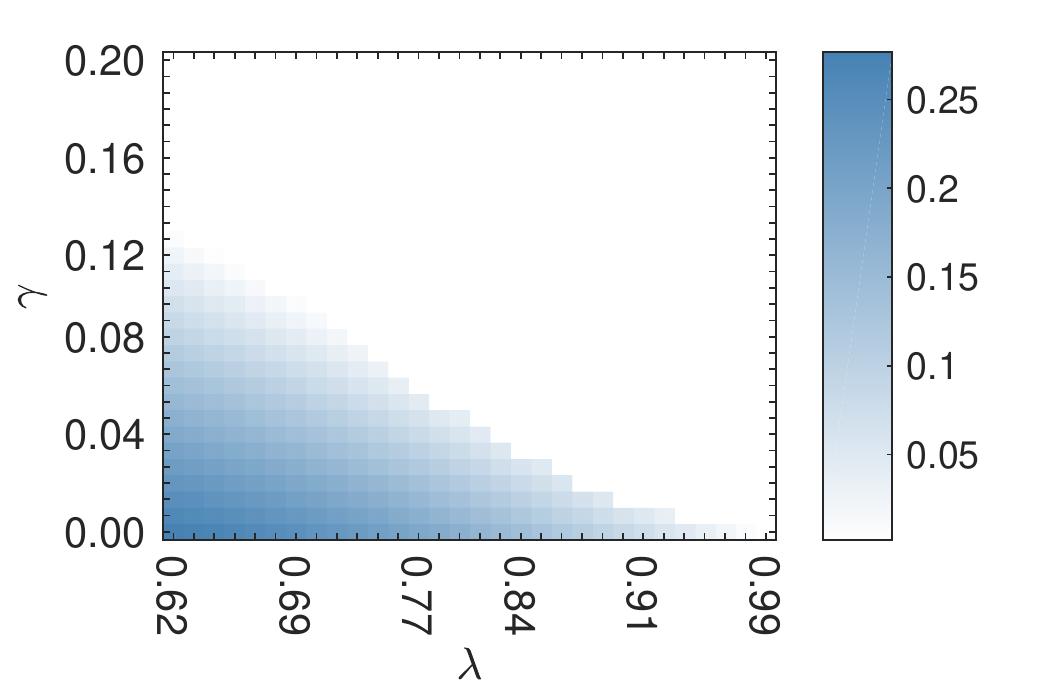}} &
		\framebox{\vspace{0.2in}\includegraphics[width=\payatequilfigw]{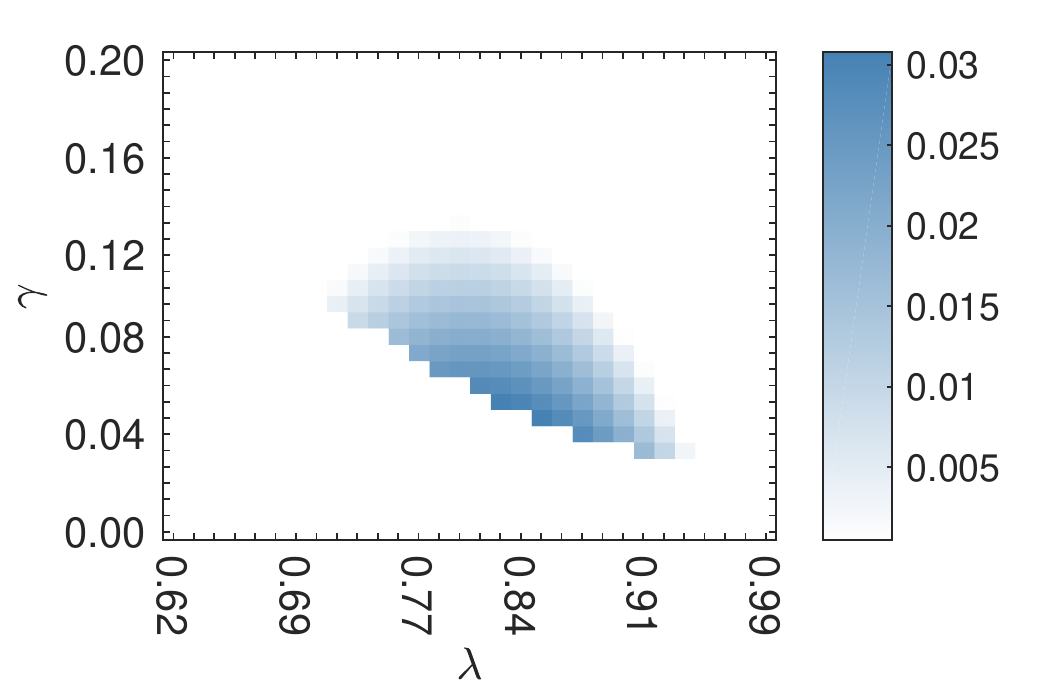}} &
		\framebox{\vspace{0.2in}\includegraphics[width=\payatequilfigw]{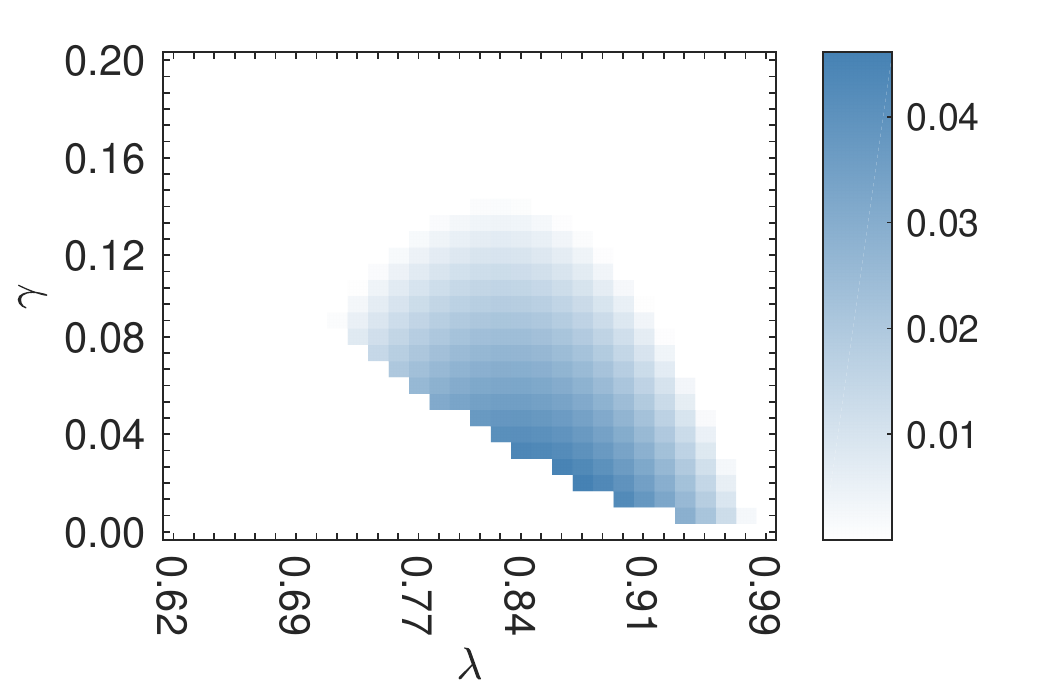}} \\ \hline
		\raisebox{0.5in}{$0.020$} &
		\framebox{\vspace{0.2in}\includegraphics[width=\payatequilfigw]{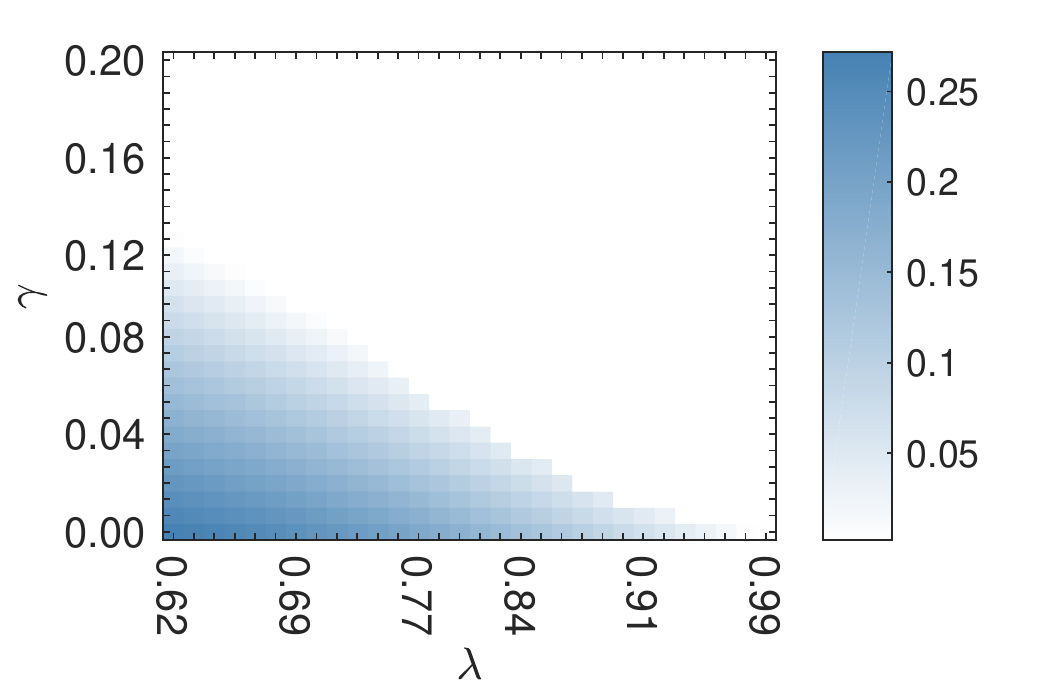}} &
		\framebox{\vspace{0.2in}\includegraphics[width=\payatequilfigw]{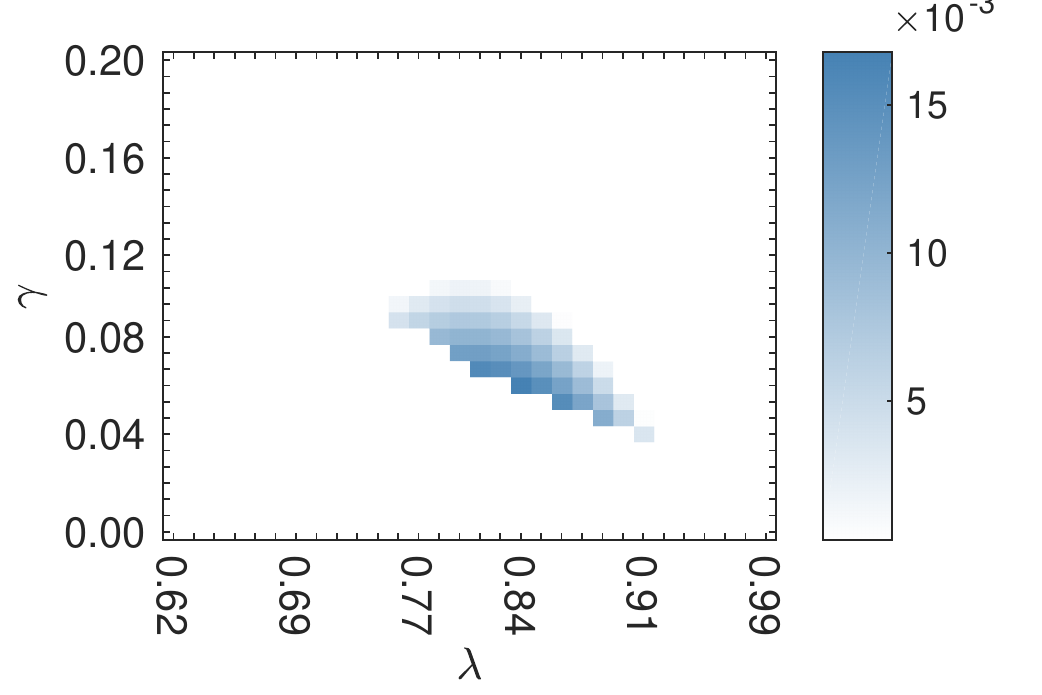}} &
		\framebox{\vspace{0.2in}\includegraphics[width=\payatequilfigw]{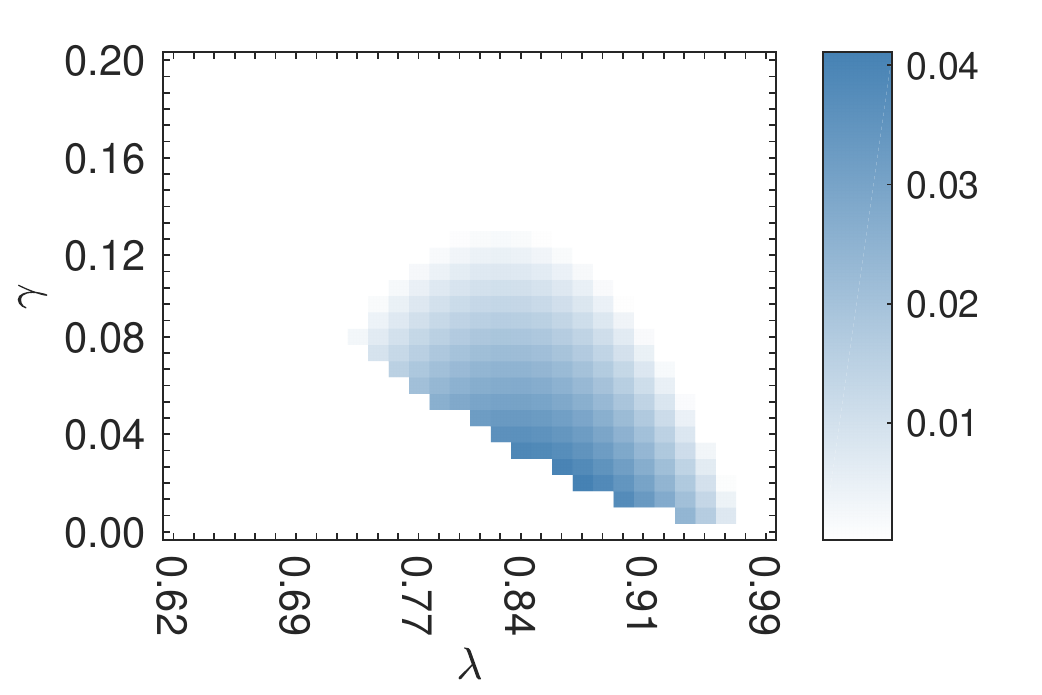}} \\ \hline
		\raisebox{0.5in}{$0.030$} &
		\framebox{\vspace{0.2in}\includegraphics[width=\payatequilfigw]{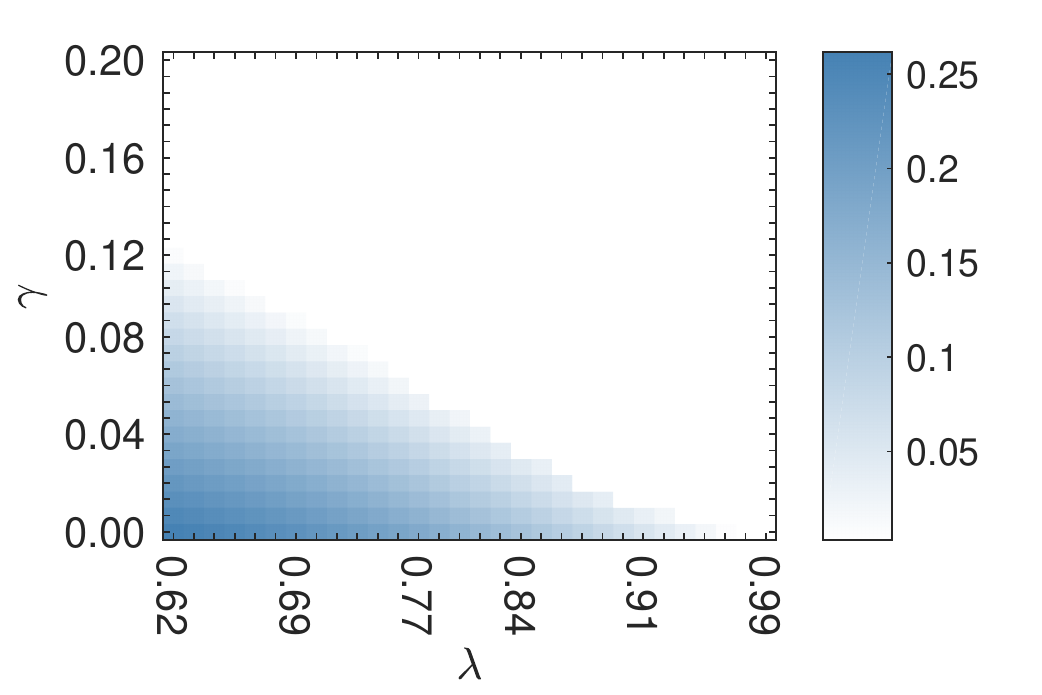}} &
		\framebox{\vspace{0.2in}\includegraphics[width=\payatequilfigw]{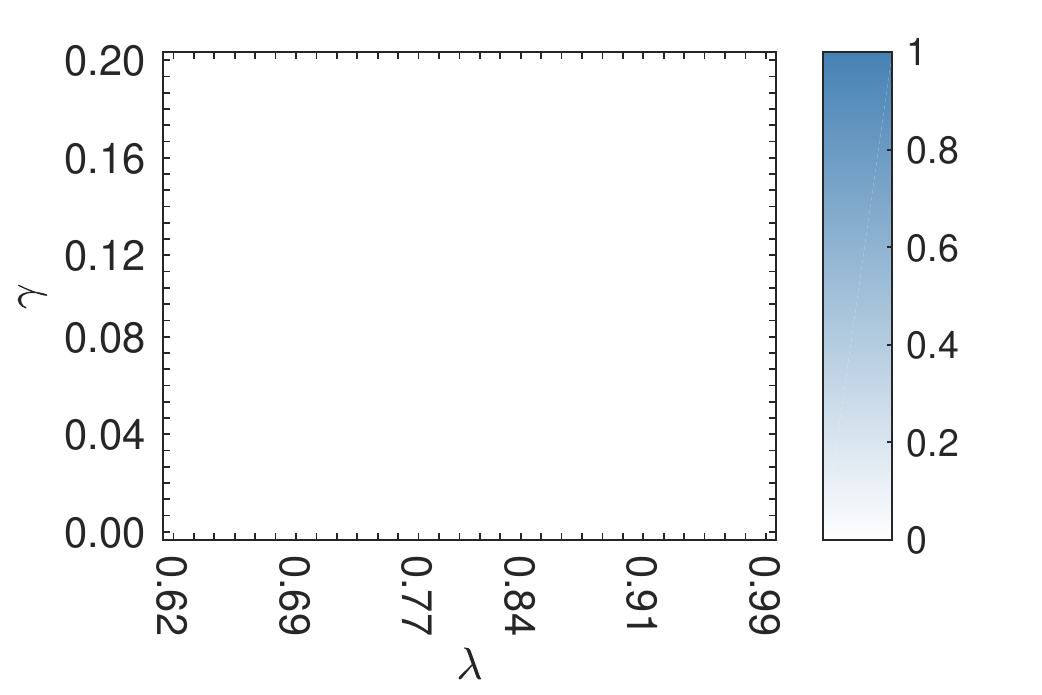}} &
		\framebox{\vspace{0.2in}\includegraphics[width=\payatequilfigw]{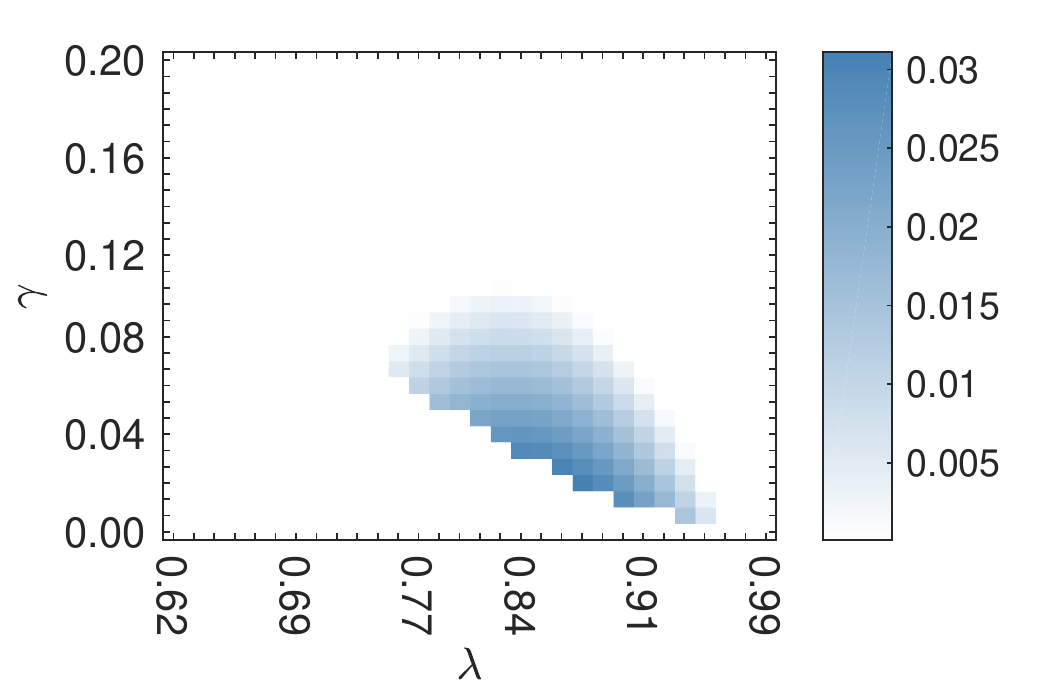}} \\ \hline
	\end{tabular}
	\caption{
		Expected payoff $\E[\pi^*_{1, i} \rvert_{\bullet} ]$ of a retailer at equilibrium when linking cost
		$c$ varies, and the consumer demand per retailer is normalized to $D = 1$.
		The $0$ values in these figures indicate that the corresponding network is not an
		equilibrium for the given combination of model parameter values $(\lambda, c, \gamma)$.
	}
	\label{tbl:small-chain-payoffs-at-equil}
\end{table}

\subsection{Analysis of the General Model}
\label{sec:general-model-w-congest-analysis}

In this section, we will provide a limited set of theoretical results for the general model with congestion defined in Sec.~\ref{sec:general-model-def}, putting no constraints on the number
of retailers and suppliers. These results will be proved under the following assumptions.

\begin{assumption}[Cost Magnitude in General Model]
	Let us assume that the linking cost $c$ and the congestion cost $\gamma$ in the general model
	with congestion are not too high, so that
	\begin{enumerate}
		\item{
			every retailer can be active---having $\dout_{1, i} > 0$ and $\E[\pi_{1, i}] > 0$; and
		}
		\item{
			there is a positive value of consumer demand $D$ per supplier, under which
			the retailers have non-negative expected payoffs $\E[\pi_{1, i}] \geq 0$.
		}
	\end{enumerate}
	\label{asm:assumption-for-general-analysis}
\end{assumption}

The second item in Assumption \ref{asm:assumption-for-general-analysis} states that there is a network in which retailers
enjoy non-negative profit. The first item in the assumption restricts attention
to those networks where every retailer links to some suppliers (there may be equilibria
networks in which some retailers have no links).

First, let us define
\begin{align}
	\douthat &= \frac{\lambda (1- \lambda) D}{ \sqrt{c} },  \label{eq:douthat}\\
	\Fjmihat &= \max\Big\{0, \frac{(1 - \lambda)(1 - \lambda^2)}{\gamma} - \frac{1}{\dout_{1, i}}\Big\}, \label{eq:Fjmihat}
\end{align}
and assume that $\douthat \in \{1, \dots, m\}$.

\begin{theorem}[Equilibria in Low Congestion Cost Regime]
	Given that Assumption~\ref{asm:assumption-for-general-analysis} holds, in a
	\emph{low congestion cost regime}, when
	$$
		\gamma < \frac{(1 - \lambda)(1 - \lambda^2)}{n}
	$$
	the behavior of the general model with congestion is qualitatively identical to that of the model without
	congestion---characterized earlier in Theorem~\ref{thm:two-tier-equil-no-congest}---so there is only
	one unique (up to supplier labeling) non-empty equilibrium network where every retailer
	maintains a single link, and all the retailers source from the same supplier.
	\label{thm:general-equil-low-congest-regime}
\end{theorem}

\begin{theorem}[Symmetric Equilibrium in Higher Congestion Cost Regime]
	Given that Assumption~\ref{asm:assumption-for-general-analysis} holds, in a \emph{high congestion cost
	regime}, where, from the perspective of every retailer $i$, the congestion at every supplier $F_{2, j}^{-i}$ can
	reach $\Fjmihat$, let us ignore parity and divisibility issues and assume existence of a regular network, in
	which all retailers have out-degree $\dout_{1, i} = \douthat$, all the congestion at every
	supplier is $\Fjmihat \vert_{\dout_{1, i} = \douthat}$.
	Then, this network is a pure strategy Nash equilibrium of the general model with congestion.
	\label{thm:best-equil-higher-congest-regime}
\end{theorem}

\proof{Proof of Theorems~\ref{thm:general-equil-low-congest-regime} and~\ref{thm:best-equil-higher-congest-regime}}

From~\eqref{eq:expected-retailer-payoff-ex1}, we have the expression for the expected
payoff of a retailer in the general model with congestion
\begin{align*}
	\E[ \pi_{1, i} ]
		= &\lambda (1 - \lambda) D (\lambda D ((1 + \lambda)n_1^a - \lambda) - \Delta)
			+ \frac{\lambda D^2}{\dout_{1, i}} \left( (1 - \lambda)^2 - \frac{\gamma}{2 \dout_{1, i}} \right) \notag\\
			&- c \dout_{1, i} + \frac{\lambda D^2}{\dout_{1, i}} \sum\limits_{j \in \Nout_{1, i}}	
				{
					F_{2, j}^{-i} \left( (1 - \lambda)(1 - \lambda^2) - \frac{\gamma}{\dout_{1, i}} - \frac{\gamma}{2} F_{2, j}^{-i} \right)
				}.
\end{align*}
From Assumption~\ref{asm:assumption-for-general-analysis}, we know that at an equilibrium, $n_1^a = n$.

1) \emph{Low congestion cost regime:} Based on the above expression for $\E[ \pi_{1, i} ]$, when a best-responding
retailer $i$ decides which $\dout_{1, i}$ suppliers to link to, it assesses each supplier with respect to the value of
$$
	f(F_{2, j}^{-i}) = F_{2, j}^{-i} \left( (1 - \lambda)(1 - \lambda^2) - \frac{\gamma}{\dout_{1, i}} - \frac{\gamma}{2} F_{2, j}^{-i} \right),
$$
which is retailer dependent. $f(F_{2, j}^{-i})$ is a quadratic function, reaching its maximal value on $F_{2, j}^{-i} \in \mathbb{R}_+$ at
$$
	F_{2, j}^{-i} = \Fjmihat = \max\Big\{0, \frac{(1 - \lambda)(1 - \lambda^2)}{\gamma} - \frac{1}{\dout_{1, i}}\Big\}.
$$
In a low congestion regime, when $\gamma < \tfrac{(1 - \lambda)(1 - \lambda^2)}{n}$, for any retailer $i$ and
supplier $j$, $\Fjmihat > n - \tfrac{1}{\dout_{1, i}} \geq n - 1$ is so large, that
$$
	F_{2, j}^{-i} <  \Fjmihat.
$$
Taking into account that, under the same constraint imposed upon $\gamma$, the following term in the
expected payoff expression is non-negative
$$
	(1 - \lambda)^2 - \frac{\gamma}{2 \dout_{1, i}} > (1 - \lambda)^2 - \frac{(1 - \lambda) (1 - \lambda^2)}{2 n \dout_{1, i}}
	= (1 - \lambda)^2 \Big[ 1 - \frac{1 + \lambda}{2 n \dout_{1, i}}\Big] \geq 0,
$$
we see that congestion in the low congestion cost regime is unambiguously good from the perspective of a retailer, and the latter also has no incentive to create more than one link as the expected payoff
is a decreasing function of $\dout_{1, i}$. As a result, the behavior of the model is the same as that of the
model without congestion, as described in Theorem~\ref{thm:two-tier-equil-no-congest}.

2) \emph{Higher congestion cost regime:} In this regime, and unlike the previous case of low $\gamma$,
congestion $F_{2, j}^{-i}$ at every supplier can actually reach its optimal value $\Fjmihat$. Let us consider
the network in which retailers have identical out-degrees $\dout_{1, i} = \tau \geq 1$, and suppliers
have identical congestion values $F_{2, j}^{-i} = \Fjmihat$. This network's existence is assumed in
Theorem~\ref{thm:best-equil-higher-congest-regime}. In this network, in terms of retailers' link distribution
over suppliers, the retailers are best-responding, as all the suppliers have identical value $f(F_{2, j}^{-i} )$,
and it does not matter which $\dout_{1, i}$ of them to link to. Thus, we are interested in the question of
whether any retailer would like to unilaterally change its outdegree $\dout_{1, i}$ improving its expected
payoff. Let us pick the best possible out-degree from which a retailer would not have an incentive to
deviate.
$$
	\E[\pi_{1, i}] \rvert_{F_{2, j}^{-i} = \Fjmihat} \xrightarrow{\dout_{1, i}} \max.
$$
The considered objective function, as a function of $\dout_{1, i}$, reaches its maximum at
$$
	\dout_{1, i} = \douthat = \frac{\lambda (1 - \lambda) D}{\sqrt{c}}.
$$
If $\forall i \in \T_1: \dout_{1, i} = \douthat$ and $\forall i \in \T_1, j \in \T_2: F_{2,j}^{-i} = \Fjmihat$,
does a retailer $i$ have an incentive to unilaterally deviate? If it decides to reduce its outdegree $\dout_{1, i}$,
then no changes happen to $i$'s preferences on which suppliers to link to, as its $\Fjmihat$ decreased,
and by simply dropping any of existing links it is already doing its best at reducing congestion $F_{2, j}^{-i}$
at its suppliers. Hence, as $\dout_{1, i}$ is already at its optimal value, a link drop cannot improve $i$'s
expected payoff. Similarly, if we consider link addition, it does not matter whom to create an extra link
to, as all the suppliers have the same value, and a retailer's out-degree is already at its optimal value. Thus,
the considered network is an equilibrium.

\qed
\endproof

We list several
observations obtained from computing equilibria in specific instances of the
general model with congestion:
\vspace{0.1in}
\begin{itemize}
	\item{
		We conjecture that, under moderate linking and congestion costs, a non-empty equilibrium network always exists.
	}
	\vspace{0.1in}
	\item{
		There are irregular equilibrium networks (not left- or right-regular).
	}
	\vspace{0.1in}
	\item{
		In some equilibrium networks, a fraction of retailers have no links.
	}
	\vspace{0.1in}
	\item{
		While the symmetric equilibrium network of
		Theorem~\ref{thm:best-equil-higher-congest-regime}
		need not exist in general, we often observe existence of an equilibrium network
		structurally similar to it.
	}
\end{itemize}

\subsection{Supplier Heterogeneity and Incentives for Reliability Improvement}
\label{sec:hetero-suppliers}

The general model with congestion assumed a homogeneous set of
suppliers, having identical production success likelihoods $\lambda_i = \lambda$ and
congestion costs $\gamma_i = \gamma$. Here we are interested in whether
suppliers are incentivized to invest in improving their
reliability through either reducing their congestion costs $\gamma_i$ or increasing
their production success likelihood $\lambda_i$. Furthermore, if there are two
options for a supplier---either to invest in production reliability, or to invest congestion reduction---which should they pick?

\begin{theorem}[Investment in Quality by Heterogeneous Strategic Suppliers]
	Assume that the linking and congestion costs are moderate, so that non-empty
	equilibria exist. Then, the following holds in our two-tier models:
	\begin{enumerate}
		\item{ Absent congestion, suppliers are always incentivized to maximize their
	production success likelihood $\lambda_i$.
		}
		\item{
	With congestion, \begin{changed}while\end{changed} a reduction in congestion cost
	$\gamma_i$
	is unambiguously good for supplier $i$, increasing its production success likelihood $\lambda_i$
	is \emph{not} always profitable---retailers may prefer to source from suppliers that are
	\emph{less} likely to succeed in production. 
		}
	\end{enumerate}
	\label{thm:hetero-sup-improv}
\end{theorem}

\proof{Proof of Theorem~\ref{thm:hetero-sup-improv}}

Let us, now, extend the general model with congestion with heterogeneous reliability parameters.
Since we are mostly interested in the strategic behavior of suppliers competing for retailers,
we will assume that retailers are all equally reliable, having the likelihood of production success
$\lambda_{1, i} = \lambda_r$ for all $i \in \T_1$. Reliability of suppliers, however, varies:
supplier $j \in \T_2$ has production success likelihood $\lambda_{2, j} = \lambda_j$, and
congestion cost $\gamma_{2, j} = \gamma_j$.

First, we must modify the expected payoffs
of the general model with congestion to this environment.

\emph{1) Payoffs:} The general expressions~\eqref{eq:payoff-ex} for agent payoff still holds
in the heterogeneous model:
\begin{align*}
	\pi_{1, i} &= S_{1, i} p_1 - R_{1, i} p_2 - c \dout_{1, i} - L_{1, i},\\[0.1in]
	\pi_{2, j} &= S_{2, j} p_2 - R_{2, j} p_3,
\end{align*}
where, as before, $S_{t, i}$ is the supply of $i \in \T_t$, $R_{t, i}$ is the same agent's realized
demand, $p_t$ is the price at which tier $t$ trades product downstream, and
$$
	L_{1, i} = \frac{1}{\dout_{1, i}} \sum_{j \in \Nout_{1, i}}{\frac{\gamma}{2} (S_{2, j})^2}
$$
is the
penalty incurred by agent $i \in \T_t$ due to congestion at the upstream suppliers $\Nout_{t, i}$
it is linked to. Notice that, as suppliers in $\T_2$ do not strategically create links---all
of them are assumed to have access to the raw material market---they do not suffer penalties
associated with linking or upstream congestion.

Now, we need to derive expected payoffs. This derivation will go along the lines of the proof
of Proposition~\ref{thm:expected-payoff-ex}, with the difference that, suppliers are
heterogeneous.
\begin{align*}
	\E[\pi_{1, i}] &= \E[S_{1, i} p_1 - R_{1, i} p_2 - c \dout_{1, i} - L_{1, i}] = \text{(from proof of Proposition~\ref{thm:expected-payoff-ex})}\\[0.1in]
		&= \E\Big[ R_{1, i} \big( \sum_{i' \in \T_1^a}{ (1 - \omega_{1, i} \omega_{1, i'}) R_{1, i'} } - (1 - \omega_{1, i} \Delta \big) - c \dout_{1, i} - L_{1, i} \Big].
\end{align*}
At first, focus on the terms other than the congestion penalty $L_{1, i}$; we deal with
the later in the last part of the proof.
\begin{align*}
	\E[R_{1, i}] &= \text{(from~\eqref{eq:realized-retailer-demand})} = \E\Big[ D \frac{\eout_{1, i}}{\dout_{1, i}} \Big]
		= \E\Bigg[ D \frac{\sum_{j \in \Nout_{1, i}}{\omega_{2, j}}}{\dout_{1, i}} \Bigg]\\
		&= \frac{D}{\dout_{1, i}} \sum_{j \in \Nout_{1, i}}{\E[\omega_{2, j}]}
		= \frac{D}{\dout_{1, i}} \sum_{j \in \Nout_{1, i}}{\lambda_j}
		= \frac{D}{\dout_{1, i}} \lout_{1, i} = D \lbarout_{1, i},
\end{align*}
where
$$
	\lout_{1, i} = \sum_{j \in \Nout_{1, i}}{\lambda_j}
	\quad\text{and}\quad
	\lbarout_{1, i} = \frac{\lout_{1, i}}{\dout_{1, i}}.
$$
Within the scope of this proof, we will be
using some extra notation to handle supplier heterogeneity; in this notation, the plus superscript
indicates relation to the out-neighborhood of an agent specified in the subscript, while the
bar indicates averaging over that out-neighborhood.

\begin{align*}
	\E[R_{1, i} R_{1, i'}] &= \text{(from~\eqref{eq:realized-retailer-demand})}
		= \E\Big[ \Big(D \frac{\eout_{1, i}}{\dout_{1, i}}\Big) \Big(D \frac{\eout_{1, i'}}{\dout_{1, i'}}\Big) \Big]
		= \frac{D^2}{\dout_{1, i} \dout_{1, i'}} \E\Big[
			\Big( \sum_{j \in \Nout_{1, i}}{\omega_{2, j}} \Big)
			\Big( \sum_{j' \in \Nout_{1, i'}}{\omega_{2, j'}} \Big)
		\Big]\\
		&= \frac{D^2}{\dout_{1, i} \dout_{1, i'}} \Big(
			\sum_{j \in \Nout_{1, i}}{ \sum_{j' \in \Nout_{1, i'}}{
				\E[\omega_{2, j}] \E[\omega_{2, j'}]
			}}
			+
			\sum_{j \in \Nout_{1, i'} \cap \Nout_{1, i'}}{
				\Var[\omega_{2, j}]
			}
		\Big)\\
		&= \frac{D^2}{\dout_{1, i} \dout_{1, i'}} \Big(
			\sum_{j \in \Nout_{1, i}}{ \sum_{j' \in \Nout_{1, i'}}{
				\lambda_j \lambda_{j'}
			}}
			+
			\sum_{j \in \Nout_{1, i'} \cap \Nout_{1, i'}}{
				\lambda_j (1 - \lambda_j)
			}
		\Big)\\
		&= D^2 \Bigg(
			\frac{\sum_{j \in \Nout_{1, i}}{ \lambda_j }}{\dout_{1, i}} \cdot
			\frac{\sum_{j' \in \Nout_{1, i'}}{ \lambda_j' }}{\dout_{1, i'}}
			+
			\frac{\sum_{j \in \Nout_{1, i'} \cap \Nout_{1, i'}}{ \lambda_j (1 - \lambda_j) }}{
				\dout_{1, i} \dout_{1, i'}
			}
		\Bigg)
		= D^2 \Big(
			\lbarout_{1, i} \lbarout_{1, i'}
			+
			\sigout_{1, i \cap i'}
		\Big),
\end{align*}
where
$$
	\sigout_{1, i \cap i'} =
		\frac{\sum_{j \in \Nout_{1, i'} \cap \Nout_{1, i'}}{ \lambda_j (1 - \lambda_j) }}{
				\dout_{1, i} \dout_{1, i'}
			}.
$$
Substituting the obtained expressions involving realized demands into the expression for
the expected payoff of a retailer, we end up with
\begin{align*}
	\E[\pi_{1, i}] &=
		\E\Big[ R_{1, i} \big( \sum_{i' \in \T_1^a}{ (1 - \omega_{1, i} \omega_{1, i'}) R_{1, i'} } - (1 - \omega_{1, i} \Delta \big) - c \dout_{1, i} - L_{1, i} \Big]\\[0.1in]
		&= \sum\limits_{i' \in \T_1^a}{
			(1 - \E[\omega_{1, i}]\E[\omega_{1, i'}]) \E[R_{1, i} R_{1, i'}]
		} - \Delta (1 - \E[\omega_{1, i}]) \E[R_{1, i}] - c \dout_{1, i} - \E[L_{1, i}]\\[0.1in]
		&= D^2 (1 - \lambda_r^2) \sum\limits_{i' \in \T_1^a}{
			 (\lbarout_{1, i} \lbarout_{1, i'} + \sigout_{1, i \cap i'})
		} - \Delta (1 - \lambda_r) D \lbarout_{1, i} - c \dout_{1, i} - \E[L_{1, i}].
\end{align*}
The expected payoff $\E[\pi_{2, j}]$ of a supplier has the following form:
\begin{align*}
	\E[\pi_{2, j}] &= \E[ S_{2, j} p_2 - R_{2, j} p_3 ]
		= \E[ \omega_{2, j} R_{2, j} (\Delta - S_2) - R_{2, j} (\Delta - S_3) ]\\[0.1in]
		&=\text{(as raw material production never fails)}\\[0.1in]
		&= \E[ \omega_{2, j} D_{2, j} (\Delta - \sum_{j' \in \T_2^a}{ S_{2, j'} }) - D_{2, j} (\Delta - n_1^a D) ]\\
		&= D_{2, j} \E[ \omega_{2, j} (\Delta - \sum_{j' \in \T_2^a}{ \omega_{2, j'} D_{2, j'} }) - (\Delta - n_1^a D) ]\\
		&= D_{2, j} \big( \lambda_j (\Delta - D_{2, j} - \sum_{j' \in \T_2^a \setminus \{j\}}{ \lambda_{j'} D_{2, j'} }) - (\Delta - n_1^a D) \big).
\end{align*}

Now, we use the obtained expected payoffs to analyze the behavior of retailers and suppliers,
first, in the model without congestion, and, then, with congestion.

\bigskip
\emph{2) Supplier Behavior Without Congestion:} Expected payoff of a retailer without congestion is
$$
\E[\pi_{1, i}] =
	D^2 (1 - \lambda_r^2) \sum\limits_{i' \in \T_1^a}{
			 (\lbarout_{1, i} \lbarout_{1, i'} + \sigout_{1, i \cap i'})
		} - \Delta (1 - \lambda_r) D \lbarout_{1, i} - c \dout_{1, i}.
$$
From the above expression and the definition of
$$
	\sigout_{1, i \cap i'} =
		\frac{\sum_{j \in \Nout_{1, i'} \cap \Nout_{1, i'}}{ \lambda_j (1 - \lambda_j) }}{
				\dout_{1, i} \dout_{1, i'}
			},
$$
it is clear that the link concentration behavior by retailers in the model
without congestion transfers persists with heterogeneous suppliers. Thus, at an
equilibrium, there will be only one supplier, say $s \in \T_2$, to which all the active
retailers link. Thus, at an equilibrium,
\begin{align*}
	\lbarout_{1, i} = \frac{ \lout_{1, i} }{\dout_{1, i}} = \lout_{1, i} = \sum_{j \in \Nout_{1, i}}{\lambda_j} = \lambda_s \quad \text{and} \quad
	\sigout_{1, i' \cap i} = \lambda_s (1 - \lambda_s).
\end{align*}
We can also repeat the reasoning from the analysis of equilibria for the model without
congestion, and conclude that, at an equilibrium,
\begin{align*}
	\E[\pi_{1, i}] &=
		D^2 (1 - \lambda_r^2) \sum\limits_{i' \in \T_1^a}{
			 (\lbarout_{1, i} \lbarout_{1, i'} + \sigout_{1, i \cap i'})
		} - \Delta (1 - \lambda_r) D \lbarout_{1, i} - c \dout_{1, i}\\
		&= D^2 (1 - \lambda_r^2) \sum\limits_{i' \in \T_1^a}{
			 (\lambda_s^2 + \lambda_s (1 - \lambda_s))
		} - \Delta (1 - \lambda_r) D \lambda_s - c\\
		&= D^2 (1 - \lambda_r^2) n_1^a \lambda_s
		- n D (1 - \lambda_r) D \lambda_s - c\\[0.1in]
		&= D^2 (1 - \lambda_r) (n_1^a (1 + \lambda_r) - n) \lambda_s - c.
\end{align*}
As we assumed that non-empty equilibria exist, the factor in front of $\lambda_s$ in the
obtained expression must be non-negative. Hence, a best-responding retailer maximizes
$\lambda_s$.

At the same time, at the equilibrium where retailers concentrate links, supplier expected
payoff is
$$
	\E[\pi_{1, j}] = D_{2, j} (\lambda_j (\Delta - D_{2, j}) - (\Delta - n_1^a D))
$$
if $j = s$ and $\E[\pi_{1, j}] = 0$ otherwise. Notice that, generally, $\Delta > D_{2, j}$, so,
unsurprisingly, suppliers are incentivized to attract more demand from retailers.

Consequently, if suppliers are strategic about choosing their production success likelihoods
$\lambda_j$, and taking into account that, in the absence of congestion, only one supplier
gets links, suppliers are unconditionally incentivized to maximize their reliability to get a
positive expected payoff.

\bigskip
\emph{3) Supplier Behavior With Congestion:} To analyze the model with congestion, let
us, first, update the corresponding expression for the expected retailer payoff, and, in 
particular, get the expected value of the congestion penalty.
\begin{align*}
	L_{1, i} = \frac{1}{\dout_{1, i}} \sum_{j \in \Nout_{1, i}}{\frac{\gamma}{2} (S_{2, j})^2},
	\qquad
	\E[L_{1, i}] = \frac{\gamma}{2 \dout_{1, i}} \sum_{j \in \Nout_{1, i}}{ \lambda_j (D_{2, j})^2 }
	 = \frac{\gamma D^2}{2 \dout_{1, i}} \sum_{j \in \Nout_{1, i}}{ \lambda_j (F_{2, j})^2 },
\end{align*}
where $F_{2, j} = \sum_{i' \in \Nin_{2, j}}{1 / \dout_{1, i'}}$ describes how congested supplier
$j \in \T_2$ is. Consequently, a retailer's expected payoff in the model with congestion
and heterogeneous suppliers is
\begin{align*}
	\E[\pi_{1, i}] =
	D^2 (1 - \lambda_r^2) \sum\limits_{i' \in \T_1^a}{
			 (\lbarout_{1, i} \lbarout_{1, i'} + \sigout_{1, i \cap i'})
		} - \Delta (1 - \lambda_r) D \lbarout_{1, i} - c \dout_{1, i}
		-  \frac{\gamma D^2}{2 \dout_{1, i}} \sum_{j \in \Nout_{1, i}}{ \lambda_j (F_{2, j})^2 }.
\end{align*}
Already from the expression above we can see that, now, a retailer, besides aiming at
picking reliable suppliers due to the first two summands in the expected payoff
expression, may avoid linking to highly reliable suppliers, as the corresponding large
$\lambda_j$ would increase the congestion penalty---the last term in the expected payoff.

To get a better feeling for why retailers may prefer lower-reliability suppliers, let us look
at two specific equilibria that we have already encountered.

First, let us inspect the symmetric equilibrium of Theorem~\ref{thm:best-equil-higher-congest-regime}. In it,
\begin{align*}
	\dout_{1, i} = \tau, \quad \din_{2, j} = n \tau / m, \quad F_{2, j} = \sum_{i' \in \Nin_{2, j}}{1 / \dout_{1, i'}} = n / m, \quad \lbarout_{1, i} = \lout_{1, i} / \tau.
\end{align*}
Consequently, again, assuming that every retailer is active, that is, $n_1^a = n$, 
\begin{align*}
	\E[\pi_{1, i}] = D^2 \sum_{i' \in \T_{1}}{
		\Big(
			\lbarout_{1, i} \Big( (1 - \lambda_r^2) \lbarout_{1, i'} - (1 - \lambda_r) - \frac{\gamma n}{2 m^2} \Big) + (1 - \lambda_r^2) \sigout_{1, i \cap i'}
		\Big)
	} - c \tau.
\end{align*}
Notice that, in the obtained expression, the factor next to $\lbarout_{1, i}$ can be
positive or negative, depending on the balance between the sizes of supplier and retailer
sets, as well as the congestion penalty value. If this is the case, the first summand under
the sum would be such that a best-responding retailer would choose lower-reliability
suppliers to lower its $\lbarout_{1, i}$---the average reliability over the suppliers
retailer $i$ is linking to. Another term under the sum, involving
$\sigout_{1, i \cap i'}$ would still drive the retailers to link to ``mid-reliability'' suppliers.
Thus, the retailers would be driven to link to a few suppliers having an intermediate
value of production success likelihood $\lambda_j$.

Secondly, let us investigate a simpler equilibrium of Sec.~\ref{sec:general-model-w-congest-2x2}, when $n = m = 2$, and in the network, every retailer maintains a single
link, sourcing from a separate supplier (a network with parallel links). In that network,
assuming, w.l.o.g., that retailer $i \in \T_1$ links to supplier $i \in \T_2$,
\begin{align*}
	\dout_{1, i} = 1, \quad \din_{2, j} = 1, \quad F_{2, j} = 1, \quad \lbarout_{1, i} = \lambda_i,
	\quad \sigout_{1, i} = \lambda_i (1 - \lambda_i),
\end{align*}
which gives
\begin{align*}
	\E[\pi_{1, i}] = D^2 \big[
		-(1 - \lambda_r^2) \lambda_i^2 + ((1 - \lambda_r^2)(\lambda_1 + \lambda_2 + 1) - 2 (1 - \lambda_r) - \gamma / 2) \lambda_i
	\big] - c.
\end{align*}
Expected payoff is a quadratic function, whose maximum in $\lambda_i$ is attained at
$$
	\widehat{\lambda}_i = \lambda_1 + \lambda_2 + 1 - \frac{2}{1 + \lambda_r} - \frac{\gamma}{2 (1 - \lambda_r^2)}.
$$
In general, it is possible that $\widehat{\lambda}_i \in (0, 1)$, in which case, suppliers would
compete to drive their reliability $\lambda_i$ towards an intermediate value.

Hence, in the model with congestion, suppliers may not be incentivized to improve their
production reliability $\lambda_i$. At the same time, however, it is clear from the expression
for the expected payoff of a retailer that improvement of the congestion cost
$\gamma_i$ is unambiguously good and lets the corresponding supplier attract more demand (links).

\qed
\endproof

Theorem~\ref{thm:hetero-sup-improv} confirms something intuitive---there
is no reason why reduction of congestion cost $\gamma_j$ can hurt a supplier. Indeed, reduction
of delays in order fulfillment unequivocally makes the corresponding supplier more attractive
for retailers, boosting the supplier's demand and, consequently, payoff. Alternatively, if we 
interpret the congestion penalty as a soft cap on supply, then it is unsurprising that
suppliers prefer higher supply caps.

Surprisingly, higher production reliability,
 can actually harm a supplier. The intuition for
why this happens is as follows. In a model with congestion, there are two competing forces present.
One, coming from the base model without congestion, drives the retailers towards link
concentration to secure better upstream prices. Another force, present in the form of the
explicit congestion penalty term in the retailer's payoff in the model with congestion, drives
the retailers towards diversifying and spreading their supplier bases to avoid congestion or
long waits for their order fulfillment. When both these forces are present, their balance
results in some optimal value of congestion for the retailers, and to approach that optimal
congestion, the retailers are incentivized to link to ``medium congestion'' suppliers.
Such suppliers are characterized by lower demand or lower production success likelihood.

In the light of Theorem~\ref{thm:hetero-sup-improv}, we can conclude that, despite
the seeming similarity between production failures and production delays, these two types
of failures are qualitatively different.

\subsection{Discussion of the Model With Congestion}

The model with congestion of Sec.~\ref{sec:model-no-congest} was obtained by extending the model of Sec~\ref{sec:general-model-def}. Being an extension, the model with congestion inherited the retailers' drive towards creating sparse networks (as they are cheap from the linking cost perspective) are concentrating links (as it allows retailers to attain better upstream prices). However, the congestion penalty introduces a countervailing force, that drives the retailers to create redundant links and spread them to achieve a certain optimal supplier overlap with their peers.

While the equilibrium networks of the model without congestion---where agents concentrate
links---are absent in the supply chain literature, the equilibrium networks of the model
with congestion: sparse networks possessing a sufficient amount of redundancy---and, in particular, the symmetric equilibrium network of Theorem~\ref{thm:best-equil-higher-congest-regime}---resemble $k$-partite graph expanders, which have been argued to form resilient supply chains~\citep{chou2011process}. However, unlike this latter work, where
the network structure was exogenously imposed, our resilient networks are endogenously
formed by the agents in an uncoordinated fashion.

It is also surprising that, in a heterogeneous environment, according to Theorem~\ref{thm:hetero-sup-improv}, suppliers may not want to improve their production reliability,
while always being willing to improve their production delays via reducing congestion costs.
This behavior also stems from the balance between two forces present in the
system---network sparsification and link concentration versus redundancy creation.

\begin{changed}
    Allowing the agents to strategically set order quantities
    in addition to strategic linking does not change equilibrium networks in the
    model without congestion. One may wonder whether it affects the structure
    of equilibrium networks when congestion is present. Unlike the case without
    congestion, creation of redundant links and boosting order quantities 
    interact through the congestion function. Strategic
    order quantity setting may result in the shrinkage of the retailers'
    supplier bases at equilibrium. Characterization of the balance between the
    extent of supplier base
    diversification and over-ordering at equilibrium is \emph{an open problem}.
\end{changed}

\section{Conclusion}

This paper introduced a model of strategic formation of supply chain network in the
presence of yield uncertainty and congestion. We use it to derive three conclusions. First, in the absence of
congestion, retailers tend to create very sparse networks and concentrate links, which results
in chain-like multi-tier networks. Sparsity is the result of unconstrained supplies, while link
concentration lets retailers secure lower prices at the high yield upstream.

In the presence of congestion, retailers tend to
form expander-like networks, which are sparse, yet possessing sufficient redundancy.
Finally, we show the qualitative difference between yield uncertainty and congestion failures
in an environment with heterogeneous strategic suppliers: reducing congestion
costs is unambiguously beneficial. Improving production reliability, however, is beneficial only in the absence
of congestion. In the presence of congestion making production process more reliable
can actually hurt a supplier. That is, there can be too much production reliability
in a supply chain.

In this paper, we focused on multi-sourcing as the primary device for mitigating
production disruptions. Others
focus on varying ordered quantities to tackle uncertainty. Combining the two
in one model\begin{changed}---would not change the equilibrium networks that emerge in the
absence of congestion---is an important direction for future study of supply chain
network models with congestion.
\end{changed}

\ACKNOWLEDGMENT{The authors are grateful to Ozan Candogan and Ben Golub for constructive suggestions that have led to tangible improvements of this work.}

\bibliographystyle{informs2014}
\bibliography{biblio-alias,biblio}

\end{document}